\documentclass[twocolumn,prl,superscriptaddress,showpacs]{revtex4-1} 
\usepackage{amsmath,amssymb}
\usepackage{graphicx}
\usepackage{verbatim}
\usepackage{amsfonts}
\usepackage{subfigure}
\usepackage{dcolumn}
\usepackage{float}
\usepackage{bm}
\usepackage{color}
\usepackage[colorlinks,citecolor=blue]{hyperref}
\usepackage{framed} 
\usepackage{mathrsfs}
\usepackage{ntheorem}
\usepackage{tikz-cd}
\usepackage{longtable}
\usepackage{hyperref}
\usepackage{natbib}
\usepackage{empheq}
\usepackage{multirow}
\usepackage{enumitem}
\usepackage[normalem]{ulem} 
\useunder{\uline}{\ul}{}
\usetikzlibrary{cd}

\newcommand{\beq}{\begin{eqnarray} }
	\newcommand{\eeq}{\end{eqnarray} }
\newcommand{\Beq}{\begin{eqnarray*} }
	\newcommand{\Eeq}{\end{eqnarray*} }
\newcommand{\Bmat}{\left(\begin{matrix}}
	\newcommand{\Emat}{\end{matrix}\right)}

\newcommand{\bit}{\begin{itemize} }
\newcommand{\eit}{\end{itemize} }
\newcommand{\ben}{\begin{enumerate} }
\newcommand{\een}{\end{enumerate} }

\definecolor{lightblue}{rgb}{0, 0.439, 0.753}

\newcommand{\mi}{\mathrm{i}}

\begin{document}

\title{Spin-Charge Groups for Fermions in Fluids and Crystals: General Structures and Physical Consequences}

\author{Arist Zhenyuan Yang}
\affiliation{School of Physics and Beijing Key Laboratory of Opto-electronic Functional Materials and Micro-nano Devices, Renmin University of China, Beijing, 100872, China}
\affiliation{Key Laboratory of Quantum State Construction and Manipulation (Ministry of Education), Renmin University of China, Beijing, 100872, China}
\affiliation{Department of Physics and HK Institute of Quantum Science \& Technology, The University of Hong Kong, Pokfulam Road, Hong Kong, China}

\author{Zheng-Xin Liu}
\email{liuzxphys@ruc.edu.cn}
\affiliation{School of Physics and Beijing Key Laboratory of Opto-electronic Functional Materials and Micro-nano Devices, Renmin University of China, Beijing, 100872, China}
\affiliation{Key Laboratory of Quantum State Construction and Manipulation (Ministry of Education), Renmin University of China, Beijing, 100872, China}

\date{\today}

\begin{abstract}

Known symmetry groups are insufficient to describe the various couplings among spin, charge, and spatial degrees of freedom in fermionic systems. To address this problem, we introduce spin-charge groups (SCGs), which provide a unified framework for fermionic symmetries. SCGs incorporate spin and charge operations as `internal' symmetries, spatial and temporal operations as `external' symmetries, together with their couplings and projective twists.  After deriving the general group structure of SCGs, we explore their applications in concrete physical systems, including $^3$He superfluids, charge-4e superconductors, collinear magnets with spin-fluxes, and superconductors with coexisting magnetic orders. We show that SCGs can enforce additional band degeneracies, Chern numbers and cross spin-charge responses. Hence SCGs provide a symmetry-based route toward the classification and exploration of new phases of matter even when strong interactions are included. 

\end{abstract}
\maketitle

\paragraph*{Introduction.—} Symmetry plays a central role in describing physical laws and classifying phases of matter. For example, crystalline materials are organized into 230 classes according to their discretized spatial symmetries \cite{spacegroup}. Together with time-reversal symmetry, these space group symmetries protect a wide variety of topological insulators \cite{2DTI, 2DTI_exp, 3DTI, 3DTI_exp, crystaleynTI, CrystTI_prop, CrystTI_exp, TI_cohomology16, Sym_indicator17, CrystTI_Dong16, CrystTI_Fang18, SlagerPRX17, SlagerNP13, GuWenPRB14superchmlg}, semimetals \cite{Topo_semimetal, Weyl_semimetal, Dirac_semimetal, Dirac_semimetal_kane, Semimetal_FangCPB}, and superconductors \cite{TopoSC_Wire, TopoSC_Heter, Topo_Wang14, TopoRev_QiZhang11}. Their representations also provide a powerful framework for describing emergent quasiparticles without direct counterparts in high-energy physics \cite{Bernevig_quasiparticles, Quasipart_MSG, Quasipart_SSG, AZY2024_rep}. In magnetic systems, the 1651 magnetic space groups \cite{MSG} were introduced to classify different magnetic orders. Since magnetic moments are primarily carried by spins, weak spin-orbit coupling allows spin and lattice rotations to decouple, leading to a broad class of magnetic materials whose symmetries are described by spin space groups (SSGs) \cite{Brinkman1966, Litvin1974, SSG1, SSG2, SSG3,song2025constructions}, including recently discovered altermagnets \cite{AlterMag1, AlterMag2, AlterMag3}. Nevertheless, SSGs still do not exhaust the possible symmetries of magnetic systems. For example, spin fluxes in collinear magnets \cite{SpinFlux95_1, SpinFlux95_2, SpinFlux24, Wu_Spinflux} can generate projective twists in symmetry operations that lie beyond the conventional SSG framework.

For itinerant electrons in crystalline materials, including insulators and metals, symmetries are described by double space groups or double magnetic/spin space groups \cite{MSG, Quasipart_SSG}. The symmetry structure of superconductors (SCs) and superfluids is more subtle because particle-hole degrees of freedom give rise to an emergent \(SU(2)\) charge symmetry \cite{AndersonSU2_88, CNYang90}. Incorporating the coupling between charge and lattice operations in SCs, together with magnetic fluxes \cite{piFlux_PRX13} (later referred to as ``charge flux''), leads to projective symmetry groups (PSGs) \cite{Wen_PSG, Alicea_PSG14}, originally introduced in the context of parton descriptions of quantum spin liquids. However, in systems with coexisting superconducting and magnetic orders -- such as iron-based superconductors \cite{Coexist_MagSC09, Li2012PRB_Iron, Luo2012PRL_Iron} and heavy-fermion superconductors \cite{Hegger2000_heavy, Stockert2011_heavy, Weng2016_heavy} -- the possible couplings among spin, charge, and lattice degrees of freedom lie beyond the scope of existing symmetry frameworks. This points to a broader class of symmetries whose physical consequences remain largely unexplored. A complete description of such symmetries is challenging because of the intricate intertwining among different symmetry operations.

In the present work, we introduce spin-charge groups (SCGs) to incorporate these broader classes of symmetries. An SCG contains internal and spatial symmetry operations acting simultaneously on lattice, spin, and charge degrees of freedom. In addition, central spin and charge symmetries can projectively twist space-time operations, leading to distinct projective classes characterized by second group cohomology invariants. SCGs unify the description of many important electronic systems, including charge-\(4e\) superconductors, spin-twisted magnets, and superconductors coexisting with magnetic order. We further demonstrate that SCGs provide a framework for exploring new phases of matter and unconventional physical responses, such as manipulating pairing symmetries and spin supercurrents in superconductor.

\paragraph*{$SO(4)$ and internal symmetries.—} Electron operators admit a representation in terms of four Majorana fermions. Thus, the maximal connected group acting on the spin and charge degrees of freedom of electrons is $SO(4) = (SU(2)_s \times SU(2)_c)/\mathbb{Z}_2$\cite{CNYang90, AZY2024_rep}, where the spin rotation group $SU(2)_s$ and the charge rotation group $SU(2)_c$ are generated by Pauli matrices ${1\over2}\sigma_{x,y,z}$ and ${1\over2}\tau_{x,y,z}$, respectively. 
The SO(4) is composed of elements $\{e^{-\mi(\theta_s\bm{n}_s\cdot\bm{\sigma}+\theta_c\bm{n}_c\cdot\bm{\tau})}\}$ with a center $Z_2^f=\{E,P_f\}$ generated by the fermion-parity $P_f=e^{-\mi\pi\bm{n}_s\cdot\bm{\sigma}}=e^{-\mi\pi\bm{n}_c\cdot\bm{\tau}}$. For a given electron system, two 
types of internal symmetries can be distinguished: spin-only and charge-only. The corresponding groups, denoted as $\mathcal{S}_0$ and $\mathcal{C}_0$ respectively, consist of symmetry operations acting solely on the spin and charge degrees of freedom. For instance, in Bardeen-Cooper-Schrieffer type spin-singlet SCs\cite{BCS57_1, BCS57_2}, one has $\mathcal{S}_0 = SU(2)_s$ and $\mathcal{C}_0 = Z_2^f$.  
In the superfluid $A$-phase of $^3$He\cite{He3A72_exp, He3A72_thr}, $\mathcal{S}_0 = U(1)_s$ (with the spin-axis parallel to the $\pmb d$-vector) and $\mathcal{C}_0 = Z_2^f$. On the other hand, in the $A_1$-phase and $B$-phase of $^3$He, both $\mathcal{S}_0$ and $\mathcal{C}_0$ are trivial. In spin $S_z$-conserved charge-$4e$ superconductors,  one has $\mathcal{S}_0 = U(1)_s=\{e^{-\mi\sigma_z\theta},\theta\in[0,2\pi)\}$ and $\mathcal{C}_0 = Z_{4c}=\{E, e^{-\mi{\tau_z\over2}\pi}, P_f, e^{-\mi{\tau_z\over2}\pi}P_f\}$ with $\mathcal{C}_0$ generated by $e^{-\mi{\tau_z\over2}\pi}$. For the $\pi$-flux Dirac semimetal on square lattice at half filling, one has $\mathcal{S}_0 = SU(2)_s$ and $\mathcal{C}_0 = SU(2)_c$.

\begin{table}[t]
\caption{The internal symmetry groups $ O_f $.}\label{tab:Of}
\centering
\resizebox{0.48\textwidth}{!}{
\renewcommand{\arraystretch}{2}
\begin{tabular}{|c|c|c|c|c|}
\hline
$\mathcal{S}_0$ & $\mathcal{C}_0$ & ${\mathcal{S}_{O_f}\over \mathcal{S}_0}$ & $O_f$ & Physical systems\\ 
\hline
\multirow{5}{*}{$Z_2^f$} & \multirow{2}{*}{$Z_2^f$} &  $\mathbb Z_1$ & $Z_2^f,$ & $^3$He-$B$, 2e SC-I \\
\cline{3-5}
& & $U(1),\! ...$ & ${ U(1)_{sc}\times Z_2^f}, ...$ & $^3$He-$A_1$\!,\! 2e SC-II,\!\! ... \\
\cline{2-5}
& $Z_{4c}$  & $\mathbb Z_1,...$ &$Z_{4c},...$ &  4e SC-I, ...\\ 
\cline{2-5}
& \multirow{2}{*}{$U(1)_c$}  & $\mathbb Z_1$ &$U(1)_c$ &Insulators-I \\ 
\cline{3-5}
& & $\mathbb{Z}_2$ & $[ E+ e^{-\mi({\sigma_z\over2}+{\tau_x\over2})\pi}]U(1)_c$ & Insulator-II\\
\cline{2-5}
& $SU(2)_c$  & $\mathbb Z_1$ & $SU(2)_c$  & Insulator-III \\
\hline
\multirow{8}{*}{$U(1)_s$\!}& \multirow{2}{*}{$Z_2^f$} & $\mathbb Z_1$&$U(1)_s$ & 2e SC-III \\
\cline{3-5}
& & $\mathbb{Z}_2$ & $[E+e^{-\mi({\sigma_x\over2}+{\tau_{z}\over2})\pi}]U(1)_s$  & {$^3$He-$A$,} 2e SC-IV\\
\cline{2-5}
& \multirow{3}{*}{$Z_{4c}$}  & $\mathbb Z_1$&$U(1)_s\star Z_{4c}$ & 4e SC-II \\
\cline{3-5}
& & \multirow{2}{*}{$\mathbb{Z}_2$} &$[E+e^{-\mi({\sigma_x\over2}+{\tau_{z}\over4})\pi}]U(1)_s\!\star\! Z_{4c}$  &  4e SC-III \\
\cline{4-5}
&  & &$[E+e^{-\mi({\sigma_x\over2}+{\tau_{x}\over2})\pi}]U(1)_s\!\star\! Z_{4c}$ &  4e SC-IV\\
\cline{2-4}
\cline{5-5}
& \multirow{2}{*}{$U(1)_{c}$}  &$\mathbb Z_1$ & $U(1)_s\star U(1)_c$ & Insulators-IV\\
\cline{3-5}
&  & $\mathbb{Z}_2$ &$[E\!+\!e^{-\mi({\sigma_x\over2}\!+\!{\tau_{x}\over2})\pi}]U(1)_s\!\!\star\! U(1)_c$ & Insulators-V \\
\cline{2-5}
& $SU(2)_c$  & $\mathbb Z_1$ &$U(1)_s\star SU(2)_c$ & Insulators-VI\\
\hline
\multirow{4}{*}{$SU(2)_s$\!}& $Z_2^f$ & $\mathbb Z_1$&$SU(2)_s$ &  2e SC-V \\
\cline{2-5}
& $Z_{4c}$  & $\mathbb Z_1$&$SU(2)_s\star Z_{4c}$ & 4e SC-V \\
\cline{2-5}
& $U(1)_c$  &$\mathbb Z_1$&$SU(2)_s\star U(1)_c$ & Insulator-VII\\
\cline{2-5}
& $SU(2)_c$  & $\mathbb Z_1$&$SO(4)$ &Insulator-VIII\\
\hline
\end{tabular}}
\end{table}

Since $\mathcal S_0$ and $\mathcal C_0$ are commuting and share the common center $Z_2^f$, their `product' is given by
$(\mathcal S_0 \times \mathcal C_0)/\mathbb Z_2$, which will be denoted by $\mathcal S_0 \star \mathcal C_0 \equiv (\mathcal S_0 \times \mathcal C_0)/\mathbb Z_2$ later. In general, the group $\mathcal S_0 \star \mathcal C_0$ does not coincide with the full set of internal symmetry operations that act trivially on the lattice and leave the system invariant, since spin and charge operations may be intertwined. We denote the full internal symmetry group as $O_f$, which is a normal subgroup of the complete symmetry group $G_f$. Accordingly, one has $O_f \subseteq SO(4)$ and
$(\mathcal S_0 \star \mathcal C_0) \lhd O_f$, with the quotient group $ O_f / (\mathcal S_0 \star \mathcal C_0)$ characterizing the coupling between spin and charge operations. 
The following isomorphism (known as Goursat’s lemma) characterizes restriction of {\it spin-charge coupling} to the structure of $O_f$. \\
\textbf{Lemma}. Denoting the group formed by spin(charge) operations in $O_f$ as $\mathcal{S}_{O_f}$($\mathcal{C}_{O_f}$) with $\mathcal{S}_0\lhd \mathcal{S}_{O_f}, \mathcal{C}_0\lhd \mathcal{C}_{O_f}$, then
\begin{align}\label{Grousat_Lemma_Of}
\frac{\mathcal{S}_{O_f}}{\mathcal{S}_0}\cong \frac{\mathcal{C}_{O_f}}{\mathcal{C}_0} \cong \frac{O_f}{(\mathcal{S}_0 \star \mathcal{C}_0)}.
\end{align}

For given $\mathcal{S}_0$ and $\mathcal{C}_0$, solving isomorphism relation (\ref{Grousat_Lemma_Of}) yields all possible internal symmetry groups $O_f$. Specially, $O_f = \mathcal{S}_0 \star \mathcal{C}_0$ means the absence of spin-charge coupling. In the following, we illustrate the physical consequence of $O_f$ via two concrete examples. 

The first example is the superfluid $A_1$ phase of $^3$He \cite{He3A1_72, He3A1_73}. Since the Cooper pairs in the $A_1$ phase are spin-polarized by external magnetic field,
the internal symmetry of $A_1$ 
reads $O_f = U(1)_{sc}\times Z_2^f$, where $U(1)_{sc}=\{ e^{-\mi({\sigma_z\over2}+{\tau_z\over2})\theta} \mid \theta \in [0,2\pi)\}$. In contrast, the $B$ phase just has a minimal internal symmetry group $O_f=Z_2^f$ although $A_1$ and $B$ have the same $\mathcal S_0$ and $\mathcal C_0$.

Another example is provided by $S_z$-conserved charge-$4e$ SCs, with $\mathcal{S}_0 = U(1)_s$ and $\mathcal{C}_0 = \mathbb{Z}_{4c}$. In this setting, the quotient group $\mathcal{S}_{O_f}/\mathcal{S}_0$ takes only two possible forms, $\mathbb{Z}_1$ or $\mathbb{Z}_2$, as summarized in Table~\ref{tab:Of}. These possibilities correspond to three distinct types of $S_z$-conserved charge-$4e$ SCs. 
The possible quartet pairing Hamiltonians for the three charge-$4e$ SCs are listed below. 

\noindent\ (1) For 4e SC-II, $O_f \cong U(1)\times Z_4$. Letting $\Delta, \delta_{1,2,3,4}\in\mathbb{C}$ and denotting $C ^\dag = (c_\uparrow^\dag, c_\downarrow^\dag), C ^* = (c_\uparrow^\dag, c_\downarrow^\dag)^T$, we have
\begin{small}
\Beq
H\!=\!\!\sum_{ijmn}\!\!\Delta_{ijmn}\ \!\! C _i^\dag(\delta_1\sigma_x\!+\!\delta_2\sigma_y) C _j^{*}C _m^\dag(\delta_3\sigma_x\!+\!\delta_4\sigma_y) C _n^{*} + {\rm h.c.}, 
\Eeq
\end{small} 
(2) For 4e SC-III, $O_f \cong U(1)\rtimes Z_8$, and 
\begin{small}
$$H=\sum_{ijmn}\Delta_{ijmn}\; C _i^\dag(\delta_1\sigma_x)(C _j^\dag)^{T}C _m^\dag(\delta_2\sigma_y)(C _n^\dag)^{T}+{\rm h.c.} ;$$
\end{small}
(3) For 4e SC-IV, \!$O_f\! \cong U(1)\!\rtimes Q_8$. \!\!Let $\varepsilon,\lambda_{1,2,3,4}\!\in\!\mathbb{R}$, \!then
\begin{small}
$$ H\!=\!\!\sum_{ijmn}\!\!\varepsilon_{ijmn}\ \!\! C _i^\dag(\lambda_1\sigma_x\!+\!\mi\lambda_2\sigma_y) C _j^{*}C _m^\dag(\lambda_3\sigma_x\!+\!\mi\lambda_4\sigma_y) C _n^{*} \!+\! {\rm h.c.}$$
\end{small}


\begin{table}[t]
\caption{Number of SCGs (without gauge operation $\gamma_l$) for 3D fermionic superfluids and 2D  superconductors. $Z_2^T$ stands for time-reversal group and $Z_2^P$ means inversion group.
} \label{tab:scpg_4e}
\centering
\resizebox{0.49\textwidth}{!}{
\begin{tabular}{ c|c|c}
\hline\hline
  $O_f$ & $G_b$   & {Num.\ of SCG} \\
\hline
   {$Z_2^f$}  & $O(3)\times Z_2^T$\ \ (3D) & {\bf 36}\\
\hline
 $U(1)_{sc}\times Z_2^f$ &\multirow{2}{*}{$ SO(2)\times Z_2^P$\ (3D)} &\ {\bf $\mathbb{Z}$}  \\ 
\cline{1-1}\cline{3-3}
$[E+e^{-\mi({\sigma_x\over2}+{\tau_{z}\over2})\pi}]U(1)_s$ &  &\ {\bf$\mathbb{Z}$} \\ 
\hline
\hline
$U(1)_s$ &  \multirow{5}{*}{2D-MPGs}  &\ \ \ \ \ \  {\bf629} (2e)\\
\cline{1-1}\cline{3-3}
$[E+e^{-\mi({\sigma_x\over2}+{\tau_{z}\over2})\pi}]U(1)_s$& &\ \ \ \ \ \  {\bf266} (2e)\\
\cline{1-1}\cline{3-3}
$U(1)_s\!\star\! Z_{4c}$ & &\ \ \ \ \ \  {\bf1161}(4e)\\
\cline{1-1}\cline{3-3}
$[E+e^{-\mi({\sigma_x\over2}+{\tau_{z}\over4})\pi}]U(1)_s\!\star\! Z_{4c}$ & &\ \ \ \ \ \  {\bf266} (4e)\\
\cline{1-1}\cline{3-3}
$[E+e^{-\mi({\sigma_x\over2}+{\tau_{x}\over2})\pi}]U(1)_s\!\star\! Z_{4c}$ & &\ \ \ \ \ \  {\bf358} (4e)\\
 \hline\hline
\end{tabular}}
\end{table}

\paragraph*{Structure of Spin-Charge Groups.—} We now formulate the general structure of an SCG $G_f$. We first consider $G_f$ as unitary. Since $O_f\lhd G_f$, depending on the quotient group $G_b=G_f/O_f$, $G_f$ can be a spin-charge space group (SCSG), a spin-charge point group (SCPG), or a spin-charge Euclidian group (SCEG). $G_f$ is an extension of $G_b$ by $O_f$. Meanwhile, due to the nontrivial center $\mathcal Z(O_f)\supseteq Z_2^f$, an SCG contains two essential pieces of information \cite{eilenberg1947cohomology1,eilenberg1947cohomology,serre2016finite,SM}: (I) the obstruction-free global action of \( G_b \) on \( O_f \), and (II) the projective twist of \( G_b \) by \(2\)-cocycles valued in \( \mathcal Z(O_f) \).

We first focus on sector (I) and neglect the twist. For fermionic systems, the global action on $O_f$ is realized by associating $l\in G_b$ with element $\phi_l\in\mathscr{N}(O_f)$, where $\mathscr N(O_f)$ is the normalizer of $O_f$ within $SO(4)$, $\mathscr{N}(O_f) \subseteq SO(4)$. We write $G_f$ in forms of cosets of $O_f$, namely $G_f=\big\{ (O_f \phi_{l} \mid l);\; l\in G_b\big\}$, with $(\phi_l | l)$ a section of the projection map $\pi: G_f\to G_b$. The operation $\phi_l$ is determined by the group homomorphism $\tilde\phi: G_b\to \mathscr N(O_f)/O_f, l\mapsto O_f\phi_{l}$, with $\phi_{l}$ a coset representative. The multiplication of two elements in $G_f$ then reads:
\begin{align}\label{Multi-I}
(  x   \phi_{l_1} \mid l_1 )  (y  \phi_{l_2} \mid l_2 ) = (xy^{l_1}  \alpha(l_1, l_2) \phi_{l_1l_2} \mid l_1l_2),
\end{align}
where $x,y\in O_f$, $y^{l_1}=\phi_{l_1}y \phi_{l_1}^{-1}$, and $\alpha(l_1, l_2)=\phi_{l_1}\phi_{l_2}\phi_{l_1l_2}^{-1} \in O_f$. The variables $\alpha(l_1,l_2)$ satisfy the associativity relation $\alpha(l_1,l_2)\alpha(l_1l_2,l_3) = [\phi_{l_1}\cdot\alpha(l_2,l_3)]\alpha(l_1,l_2l_3)$ under the action $\phi_{l_1}\cdot\alpha(l_2,l_3)=\phi_{l_1}\alpha(l_2,l_3)\phi_{l_1}^{-1}$.  When $O_f$ is Abelian, $\alpha(l_1, l_2)$ forms a 2-cocycle. For instance, in the case $O_f=Z_2^f$, the 2-cocycle $\alpha(l_1, l_2)$ appears in the double-valued representation of MSGs or SSGs\cite{Quasipart_MSG, Quasipart_SSG}.  Two different choice of global actions (or equivalently, different homomorphisms $\tilde{\phi}$,$\tilde{\phi}'$) generally give rise to two different group structures, except that they are conjugate to each other via an element $g\in SO(4)$ and an element $l\in \operatorname{Isom}(\mathbb{R}^3)$. In the latter case, the corresponding SCGs $G_f, G_f'$ are also mutual-conjugate, namely $G_f'= (g\mid l) G_f  (g\mid l)^{-1}$.

We then turn to sector (II) -- the projective twists. 
Such twists can be realized by  introducing site-dependent $SO(4)$ operations $\gamma_{l}({r})$ for each $l\in G_b$, which amounts to introducing background $SO(4)$ gauge field. Meanwhile, since the global action has been implemented by $\phi_l$, one can simply require the `gauge' sector $\gamma_l( r)$ to be commuting with $O_f$. Hence, $\gamma_l( r) \in \mathscr X(O_f)$ with $\mathscr X(O_f)\subset \mathscr N(O_f)$ the centralizer of $O_f$. The resulting SCG reads $G_f=\left\{ \big(\gamma_{l}({r})O_f \phi_{l}\mid l\big);\; l\in G_b\right\}$ with the  multiplication of two elements
\beq\label{multiply}
&& ( \gamma_{l_1}({r}) x  \phi_{l_1} |  l_1 )  (\gamma_{l_2}({r}) y \phi_{l_2} |  l_2 ) \notag\\
&&\ \ \ \ \ \ \ \  = \omega_2(l_1, l_2)\Big(\gamma_{l_1l_2}({r}) xy^{l_1}\alpha(l_1, l_2)\phi_{l_1l_2} | {l_1l_2}\Big),
\eeq
twisted by the factor system (site-independent) $\omega_2(l_1, l_2) = \gamma_{l_1}({r}) \gamma'_{l_2}({r})\gamma_{l_1l_2}^{-1}({r})\in \mathcal{Z}(O_f)$, 
where $\gamma'_{l_2}({r}) = \phi_{l_1}\gamma_{l_2}(l_1^{-1}{r})\phi_{l_1}^{-1}$. 
Here $\omega_2(l_1, l_2) $  is classified by the second group cohomology $H^2_\varphi(G_b, \mathcal{Z}(O_f))$, where $\varphi$ denotes the global action of $G_b$ on $\mathcal Z(O_f)$ determined by $\tilde\phi$, namely $\varphi(l_1)\cdot\omega = \phi_{l_1}\omega \phi_{l_1}^{-1}$ for $\omega\in\mathcal Z(O_f)$. 

When $G_f$ is anti-unitary, $G_b$ is essentially a magnetic group. The internal symmetry group $O_f \subseteq SO(4)$ is chosen to be unitary (which is not true in SSG) such that most general projective twist is allowed. We first define the time reversal operation as $T=M(T)K$ with $M(T)=e^{-i({\sigma_y\over2}+{\tau_y\over2})\pi}$ such that it commutes with $SO(4)$ and $G_b$ simulaneously. Then for a general anti-unitary operation $l\in G_b$, the associated spin-charge operation $\phi_l\in \mathscr N(O_f)$ is defined on top of $M(T)$, hence $l$ acts on $y\in O_f$ as $y^{l} = \phi_{l}M(T)K y K M(T)^{-1}\phi_{l}^{-1}=\phi_l y \phi_l^{-1}$, and accordingly $\varphi({l})\cdot \omega = \phi_{l}\omega \phi_{l}^{-1}$. In this way, the relation (\ref{multiply}) holds for both unitary and anti-unitary elements. An alternative convention $T''=K$ and $\phi''_l = \phi_l M(T)$ is also used, where $K$ acts on both lattice and internal degrees of freedom, and $\phi''_l$ acts on internal degrees of freedom only. In this case $y^{l} = \phi''_{l} y^*\phi_{l}^{''-1}$ and  $\varphi''({l})\cdot \omega = \phi''_{l}\omega^* \phi_{l}^{''-1}$.

Here we illustrate SCGs through representative examples. For three-dimensional chiral fermionic superfluids, including the \(^{3}\)He \(A\) and \(A_1\) phases, the spatial symmetry group \(G_b\) is given by \(SO(2)\times \mathbb{Z}_2^{P}\), where \(\mathbb{Z}_2^{P}\) denotes inversion symmetry. In this case, infinitely many maps \(\tilde{\phi}\) arise (see Table~\ref{tab:scpg_4e}). The corresponding \(\mathbb Z\) index labels the angular momentum \(m\) of the Cooper pairs and characterizes the \((p_x+ip_y)^m\) pairing symmetry. For two-dimensional magnetic point groups \(G_b\), a large number of maps \(\tilde{\phi}\) associated with charge-\(2e\) and charge-\(4e\) SCs are counted in Table~\ref{tab:scpg_4e}. The derivation of these maps is presented in Sec.~S3 of the Supplemental Material (SM) \cite{SM}, while the corresponding projective twists, invariants, and projective representations are summarized in Sec.~S4 \cite{SM}. Below we discuss several  physical consequences of SCGs.

\paragraph*{Projective twists in the spin sector.—} 
If \(S_z\) is conserved, i.e., if \(U(1)_s\subset O_f\), the extension of \(G_b\) can contain projective twists in the spin sector that are not captured by SSGs \cite{SSG1, SSG2, SSG3}. Such spin twists can be interpreted as background \(U(1)_s\) spin gauge fields carrying spin fluxes \cite{Wu_Spinflux}. Two types of spin fluxes can then be distinguished: ferromagnetic (FM) fluxes, which vary continuously in analogy with charge fluxes, and antiferromagnetic (AFM) fluxes, which are quantized and exhibit more exotic properties.

\begin{figure}[b]
  \centering
\includegraphics[scale=0.35]{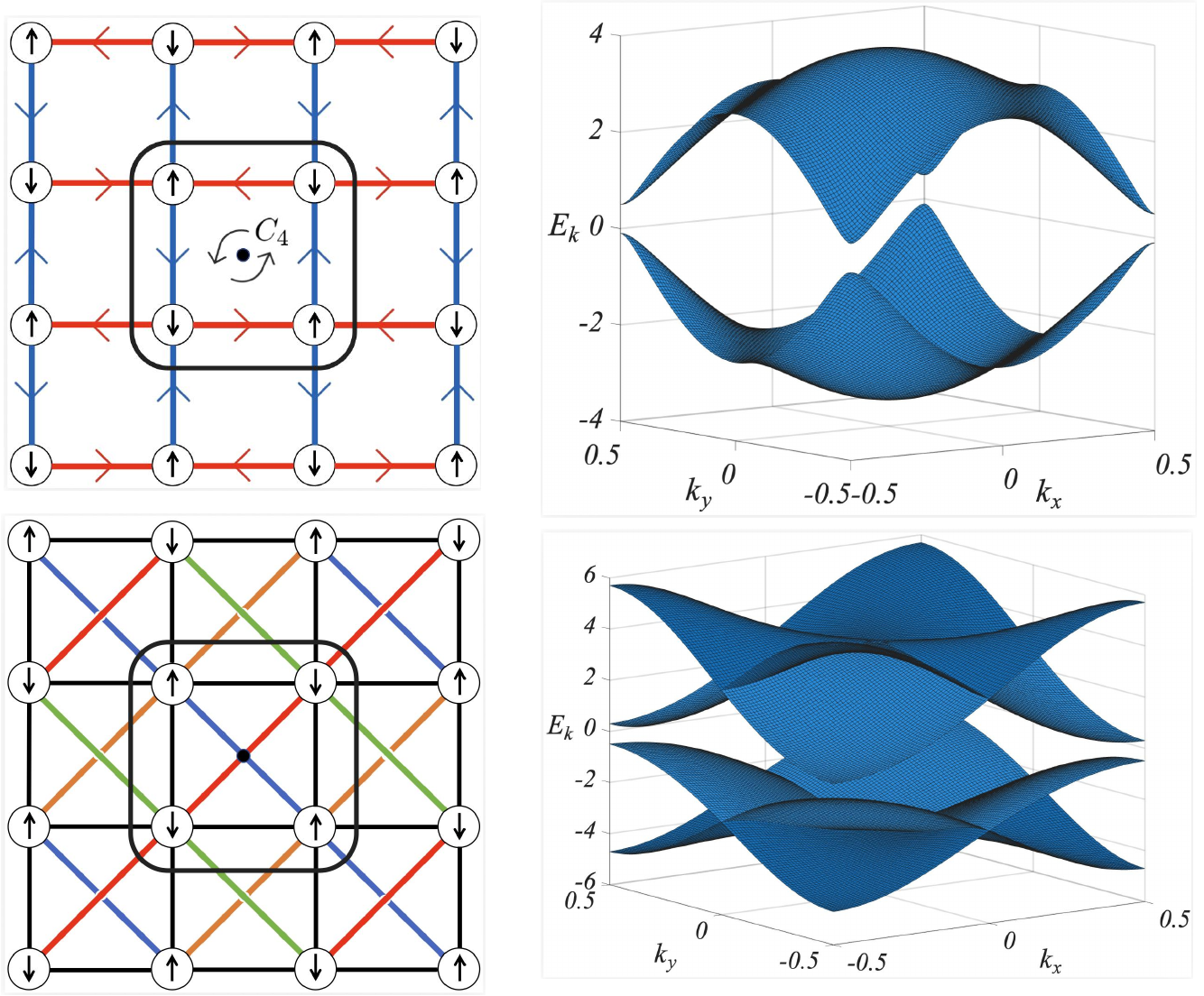}
\caption{Tight-binding model with spin twist. The black squares indicate the enlarged unit cell. The left panels show nearest-neighbor and next-nearest-neighbor hoppings, respectively. The right panels show the band structure of model~(\ref{model-spintwist}) with $a=1$, $b=0.8$, $\lambda=0.3$, $J=0.2$ (upper) and that with diagonal terms included (lower with Chern number $2$).}
\label{square-lattice}
\end{figure}

We now illustrate the physical consequence of AFM spin flux via a square-lattice model with $G_b= p4mm1'$ and $O_f=U(1)_s\star U(1)_c$. For simplicity, we assume that the charge sector is trivial. 
With basis $C ^\dag = ( c_\uparrow^\dag, c_\downarrow^\dag )$, we consider the tight-binding Hamiltonian
\beq\label{model-spintwist}
H_0 \!=\! \sum_{\langle ij\rangle} (C _i^\dagger \Lambda_{ij} C _j \!+\! \text{h.c.}) \! +\!  \sum_i \big[\lambda \hat N_i \!+\! 
(-1)^iM \hat S^z_i \big],
\eeq
where $\hat N_i=C _i^\dagger C _i$, $\hat S^z_i = C _i^\dagger {\sigma_z\over2} C _i$, $\langle ij\rangle$ denotes nearest-neighbor sites, $\lambda$ is the chemical potential, and $(-1)^iM$ stands for effective Zeeman field due to the onsite AFM moments. The hopping matrix reads $\Lambda_{ij} = a +\mi b\sigma_z$ for the horizental arrows {$i\to j$} and $\Lambda_{kl} = b + \mi a\sigma_z$ for the vertical arrows {$k \to l$} (see Fig.~\ref{square-lattice}), with $a,b$ real numbers. 

Due to the AFM order, the elementary translation $T_x$ is associated with spin flipping to form a symmetry $g_{T_x}=(e^{-\mi{\sigma_x\over2}\pi} \mid {T}_x)$. 
Hence the translation symmetry $g_{T_x}$ has nontrivial action on $U(1)_s$ via $ (U_\theta)^{T_x} = \phi_{T_x}U_\theta \phi_{T_x}^{-1}=U_{-\theta}$ with $\phi_{T_x}=e^{-\mi{\sigma_x\over2}\pi}$ and $U_\theta=e^{-\mi{\sigma_z\over2}\theta}$. Similarly, the other generators of $p4mm1'$ act on $U(1)_s$ via $\phi_{T_y}=\phi_{C_4}=\phi_{M_x}=\phi_{M_y} = e^{-\mi{\sigma_x\over2}\pi}$ and $\phi_{T} = e^{-\mi({\sigma_x\over2}+{\tau_y\over2})\pi}$, with $C_4$ the 4-fold lattice rotation, $M_{x,y}$ the mirror reflections, and $T=e^{-\mi({\sigma_y\over2}+{\tau_y\over2})\pi}K$ the time-reversal operation. 
The relation $g_{T_x}g_{T_{\pm y}}=P_f T_{x\pm y}$ shows that $T_{x\pm y}$ are symmetry operations; likewise, $T_{2x}, T_{2y}$ are also symmetry operations.
Furthermore, $U(1)_s$ gauge transformations $\gamma_g(r)$ should be introduced to make the Hamiltonian (\ref{model-spintwist}) invariant. It turns out that $\gamma_g(r)$ have a 4-sublattice structure (see the SM).  Choosing the plaquette center as the origin, then the coordinates of lattice sites are given by $r=({1\over2},{1\over2})+(x,y)$ with $x,y\in\mathbb{Z}$. The  gauge transformations of $C_4$ and $T$ are given as the following: 
if $x,y=$even, then $\gamma_{C_4}(r)=e^{-\mi\sigma_z\pi/4}$, $\gamma_T(r)=1$; 
if $x=$even, $y=$odd, then $\gamma_{C_4}(r)=e^{\mi\sigma_z\pi/4}$, $\gamma_T(r)=-e^{-2\theta_0\mi \sigma_z}$ with $\theta_0=\arctan{a\over b}$; 
if $x=$odd, $y=$even, then $\gamma_{C_4}(r)=e^{-\mi\sigma_z3\pi/4}$, $\gamma_T(r)=e^{-2\theta_0\mi \sigma_z}$; 
if $x=$odd, $y=$odd, then $\gamma_{C_4}(r)=e^{\mi\sigma_z3\pi/4}$, $\gamma_T(r)=-1$.
The final symmetry operations read $g_{C_4} = (\gamma_{C_4}(r) e^{-\mi{\sigma_x\over2}\pi} \mid C_4)$ and $g_{T} = (\gamma_{T}(r) e^{-\mi{\sigma_x\over2}\pi} e^{-\mi{\sigma_y\over2}\pi}\mid e) K$, $g_{M_{x,y}} = (e^{-\mi{\sigma_x\over2}\pi} \mid M_{x,y})$. 

Noticing that $g_T$ commutes with $T_{2x}$ and $T_{2y}$ but anti-commutes with $T_{x\pm y}$, we set $T_{2x},T_{2y}$ as unit translations (if one alternatively adopts $T_{x\pm y}$ as unit translations then $g_T$ is non-symmorphic in momentum space), then the band structure is 4-fold degenerate everywhere, as shown in Fig.~\ref{square-lattice}. 
The 4-fold degeneracy is owing to the spin twist classified by $H^2_{\varphi }(G_b, U(1)_s)$ with invariants $(g_{M_x})^2 = (g_{M_y})^2 =-1, (g_T)^2=1$ and $(g_{C_4})^4=-1$. Specially, since $\phi_{C_4} = e^{-\mi{\sigma_x\over2}\pi}$, $g_{C_4}$ acts nontrivially on $U(1)_s$ and behaves like an anti-unitary operator with $U(1)_c$ factor system. As a result, the AFM spin flux is characterized by the quantized invariant $(g_{C_4})^4=-1$, unlike the continuous invariant $g_{T_x}g_{T_y}g_{T_x}^{-1}g_{T_y}^{-1}$ of FM spin flux. The noncommutability of $U(1)_s$ and $\phi_{T_x} = e^{-\mi{\sigma_x\over2}\pi}$ yields the 2-fold degeneracy in the spin sector, while the combined invariant $(g_{C_4}^2 g_T)^2=-1$ of the symmetry $g_{C_4}^2 g_T$ (keeping each momentum $\pmb k$ invariant) interprets the rest 2-fold degeneracy (see End Matter for the 4D irrep).

Interestingly, upon including next-nearest-neighbor hopping terms (explicit form given in the SM), the $g_T$ and $g_{M_{x,y}}$ symmetries are explicitly broken but their combination $g_{TM_{x,y}}$ are unbroken.
As a result, the generic 4-fold degeneracies are lifted, except along the high-symmetry lines $k_x = 0$ and $k_y = 0$ (see Fig.~\ref{square-lattice}). The remaining fourfold degeneracies are owing to the survived invariants 
$(g_{T_y}g_{TM_x})^2=-T_{2y}$ and $(g_{T_x}g_{TM_y})^2 =-T_{2x}$. Moreover, due to the broken $g_T$  and the presence of AFM spin-flux, the resulting band structure can carry a nonzero Chern number $C = 2$ at half filling, thereby realizing quantum anomalous Hall effect.

\paragraph*{SCs with Coexisting Magnetic Order.—}
The symmetries for electrons in SCs and magnets are usually well described by known groups.
For instance, on square lattice a SC with $p+ip$ triplet pairing has a PSG symmetry $(e^{i{\tau_z\over2}{\pi\over2}}| C_4)$, and electrons in a N\'eel ordered magnets have a spin point symmetry $(e^{-i{\sigma_x\over2}{\pi}}| C_4)$. However, for a SC coexisting with Neel order, the above symmetries break down and a new operation $(e^{-i ({\sigma_x\over2}+{\tau_z\over4}){\pi}} 
| C_4)$ -- a SCPG element, becomes a symmetry. 
If the SC further contains AFM spin-flux, then spin-singlet pairing and the $S_z = 0$ triplet pairing will be symmetry-related. 
In this case, the SC contains both singlet and triplet pairing components. 

\begin{figure}[b]
  \centering
\includegraphics[scale=0.23]{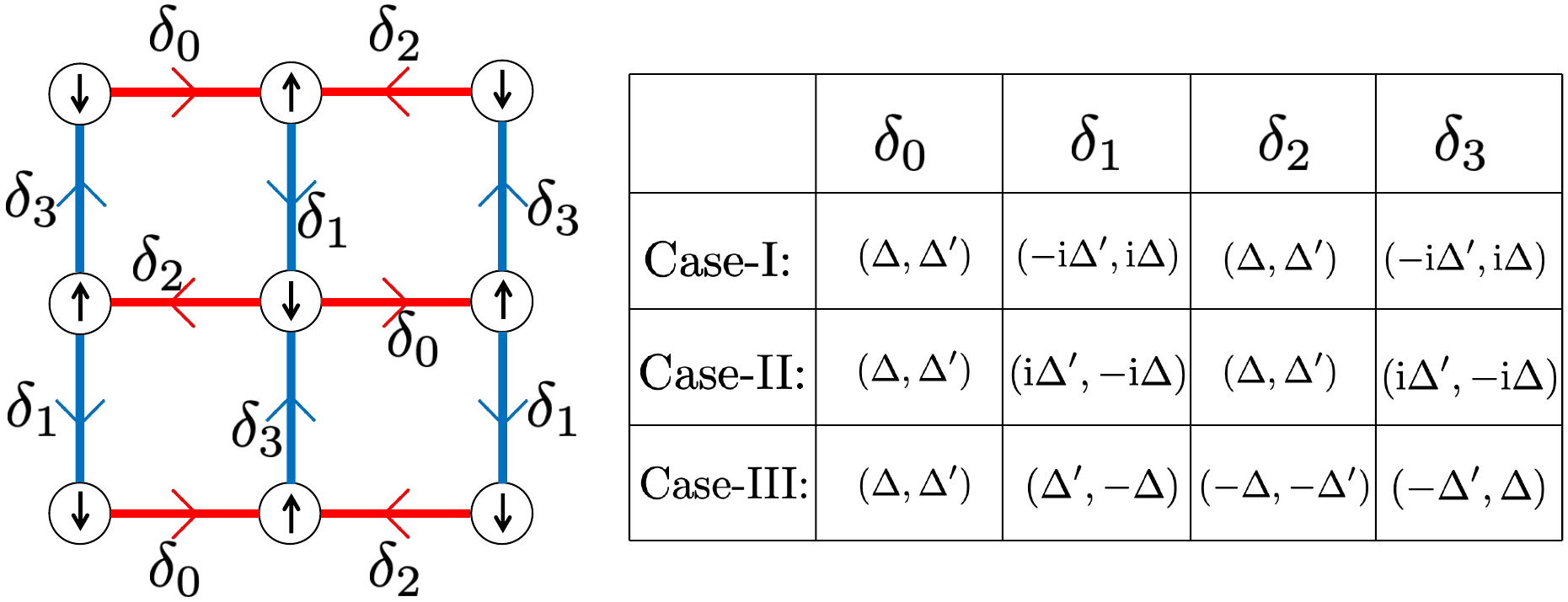}
\caption{Configurations for the three models exhibiting coexistence of AFM spin-flux and $s$, $d$, and $p_x-\mathrm{i}p_y$ type superconductivity. Here, \( \Delta \) and \( \Delta' \) are fixed complex numbers. Each directed bond carries \( \delta_{ij}=(\Delta_{ij}^{(s)},\Delta_{ij}^{(t)})\in\mathbb{C}^{2} \).
}\label{spintwist-pairing}
\end{figure}

Introducing fermion paring terms to the spin-twist model (\ref{model-spintwist}), one obtains the SC Hamiltonian, 
\beq\label{spintwist-SCs} 
H_1 &=& H_0 + \sum_{\langle i j\rangle} \big[ \Delta^{(s)}_{ij} (c_{i\uparrow}^\dag c_{j\downarrow}^\dag  - c_{i\downarrow}^\dag c_{j\uparrow}^\dag) + \notag \\
&&\ \ \ \ \ \ \ \ \ \ \ \ \ \ \!\Delta^{(t)}_{ij} (c_{i\uparrow}^\dag c_{j\downarrow}^\dag + c_{i\downarrow}^\dag c_{j\uparrow}^\dag ) +{\rm h.c.} \big],
\eeq
where $\Delta^{(s)}_{ij}, \Delta^{(t)}_{ij}$ stand for the singlet- and triplet-pairing terms respectively. The SCSG symmetry operations for the above model are given by $g_{C_4}=(\gamma_{C_4}({r})e^{-\mi{\pi\over 2}\sigma_x-\mi{\theta\over 2}\tau_z}\mid C_4)$, $g_{T_{x,y}}=(e^{-\mi{\pi\over 2}\sigma_x-\mi{\phi\over 2}\tau_z} \mid T_{x,y})$
where the spin gauge transformation $\gamma_{C_4}({r})$ is the same as in model (\ref{model-spintwist}) and the parameters $(\theta, \phi)$ have three possibilities $(\theta, \phi) = (0,0), (\pi,0), ({\pi\over 2},\pi)$ which are labeled as case I, II, III, respectively. Correspondingly, the values of $(\Delta^{(s)}_{ij}, \Delta^{(t)}_{ij})$ on each bond are shown in Fig.~\ref{spintwist-pairing} for the three cases.  The operations $g_{M_x}=\left(e^{-\mi\frac{\pi}{2}\sigma_x} \mid {M}_x\right)$ remain as symmetry in case I, II, but break down for case III. The effective time-reversal symmetry $g_T$ is broken in all three cases.

Because the effective mirror $g_{M_x}$ reverses the Berry curvature, cases I and II correspond to topologically trivial SCs. By contrast, case III realizes a topological SC with Chern number $\nu=\pm4$. Upon introducing next-nearest-neighbor hopping, the symmetry $g_{M_x}$ is broken in cases I and II, and nonzero Chern numbers $\nu=\pm2,\pm4,\pm8$ emerge. Incorporating spin-fluxes therefore provides an effective route to realizing topological SCs. Meanwhile, due to the presence of $g_{C_4}$, all these bands exhibit fourfold degeneracy at the $\Gamma$ point, as shown in the End Matter.

The representations of SCGs constrain the physical responses of fermionic systems through Neumann's principle. For example, in the \(A_1\) phase of \(^{3}\)He, all symmetry operations commute with \(S_z\). As a consequence, a nonuniform Zeeman field \(B_z\) can induce a pure mass superflow, known as the magnetic fountain effect \cite{PhysRevB.28.6582}, which is absent in the \(A\) and \(B\) phases. In the End Matter, we provide another example of an SCG-induced cross response, namely the {\it super-spin-Hall effect}.

\paragraph*{Conclusion.—} We introduced SCGs as a general symmetry framework for fermions in fluids and crystals. After deriving their group structure, we discussed several representative systems described by SCGs, including superfluids, charge-\(4e\) superconductors, antiferromagnets with spin fluxes, and superconductors coexisting with antiferromagnetic order. We showed that spin fluxes can enforce additional band degeneracies and nontrivial topological structures. In particular, SCGs can support cross spin-charge responses. Hence SCGs provide a symmetry-based route toward exploring new phases of matter, including strongly interacting symmetry protected topological phases, together with novel physical responses. 

\paragraph*{Acknowledgments.—} This work is supported by NSFC (Grants No. 12374166 and No. 12134020) and National Key Research and Development Program of China (Grants No. 2023YFA1406500 and No. 2022YFA1405300).

\bibliography{Ref}

\section{\bf End Matter}

\paragraph*{The four-dimensional irreps appearing in the AFM-flux model.--} Without including next-nearest-neighbor hopping terms, the band structure is fourfold degenerate everywhere. 
The reason is that the representation of the little co-group at a general point $\pmb k$ can be reduced into a direct sum of two 4-dimensional irreps,
\begin{small}
\begin{align}
 U(1) &= I_2\otimes e^{i\theta\sigma_z} \oplus I_2 \otimes e^{-i\theta\sigma_z}
,\notag\\ 
g_{T_x} &=
i\begin{pmatrix}
0& e^{ik_1} & 0 & 0 \\
1 & 0 & 0 & 0 \\
0 & 0 & 0 & 1 \\
0 & 0 & e^{ik_1} & 0
\end{pmatrix}\oplus i\begin{pmatrix}
0& e^{ik_1} & 0 & 0 \\
1 & 0 & 0 & 0 \\
0 & 0 & 0 & 1 \\
0 & 0 & e^{ik_1} & 0
\end{pmatrix} \notag\\
 g_{T_y} &=
i\begin{pmatrix}
0 & 0 & 0 & e^{ik_2} \\
0 & 0 & e^{ik_2} & 0 \\
0 & 1 & 0 & 0 \\
1 & 0 & 0 & 0
\end{pmatrix} \oplus i\begin{pmatrix}
0 & 0 & 0 & e^{ik_2} \\
0 & 0 & e^{ik_2} & 0 \\
0 & 1 & 0 & 0 \\
1 & 0 & 0 & 0
\end{pmatrix} 
,\notag\\ 
 g_T g_{C_2} &=\!\!
\begin{pmatrix}
0 & 0 & -1 & 0 \\
0 & 0 & 0 & -e^{-2i\theta_0} \\
1 & 0 & 0 & 0 \\
0 & e^{-2i\theta_0} & 0 & 0
\end{pmatrix}\!\! \oplus\!\! \begin{pmatrix}
0 & 0 & -1 & 0 \\
0 & 0 & 0 & e^{-2i\theta_0} \\
1 & 0 & 0 & 0 \\
0 & -e^{-2i\theta_0} & 0 & 0
\end{pmatrix}.\notag
\end{align}
\end{small}

When next-nearest-neighbor hopping terms are added, the degeneracy is lifted everywhere except along the high-symmetry lines \( k_x = 0 \) and \( k_y = 0 \). As an example, along \( k_y = 0 \),  the remaining fourfold degeneracy is protected by the following irreps 
\begin{small}
\begin{align}
U(1) &= I_2\otimes e^{\pm i\theta\sigma_z}, 
\ \ g_{T_x} =
i\begin{pmatrix}
0& e^{ik_1} & 0 & 0 \\
1 & 0 & 0 & 0 \\
0 & 0 & 0 & 1 \\
0 & 0 & e^{ik_1} & 0
\end{pmatrix},\notag\\
\!\!\! g_{T_y} &=
i\begin{pmatrix}
0 & 0 & 0 & e^{ik_2} \\
0 & 0 & e^{ik_2} & 0 \\
0 & 1 & 0 & 0 \\
1 & 0 & 0 & 0
\end{pmatrix}\!, \ \! 
g_{T_M}=\begin{pmatrix}
0 & 0 & 0 & -1 \\
0 & 0 & -e^{-2i\theta_0} & 0 \\
0 & 1 & 0 & 0 \\
e^{-2i\theta_0} & 0 & 0 & 0
\end{pmatrix}. \notag
\end{align}
\end{small}

%
%

\paragraph*{Neumann-constrained cross response.--} SCPG symmetries strongly constrain the linear responses of the system. Here we consider the cross response $J_i^{z}=\rho^z_{ij} A_j$, namely a spin supercurrent induced by the vector potential \( \pmb A \) in \(S_z\)-conserved superconductors on the square lattice. According to Neumann's principle, the response tensor \( \rho^z_{ij} \) must remain invariant under SCPG operations. Under symmetry transformations, the index \(i\) transforms under lattice rotations, the spin index \(z\) under spin rotations, and the index \(j\) under both lattice and charge operations. 

We consider four representative situations. (I) If the superconductor coexists with a flux-free AFM order and has SCPG symmetry \(G_1\) generated by \((e^{-i(\sigma_x/2+\tau_x/2)\pi}\mid C_4)\), \(\left(e^{-i\frac{\pi}{2}\sigma_x}\mid M_x\right)\), the response takes the form of an antisymmetric Hall tensor parameterized by \(\beta\). (II) In the presence of spin flux, the symmetry group \(G_2\) is generated by \((\gamma_{C_4}(r)e^{-i(\sigma_x/2+\tau_x/2)\pi}\mid C_4)\), \(
\left(e^{-i\frac{\pi}{2}\sigma_x}\mid M_x\right)\), where \(\gamma_{C_4}\) is defined as before. In addition to the antisymmetric Hall component \(\beta\), the response allows a staggered traceless diagonal component \(\alpha\). Under a uniform vector potential \( \pmb A \), the response remains Hall-like, whereas nonuniform fields can induce a nematic response. (III) If the superconductor coexists with AFM order and has spin point-group (SPG) symmetry \(G_3\) generated by \((e^{-i\sigma_x\pi/2}\mid C_4)\), \((e^{-i\sigma_x\pi/2}\mid M_x)\), \((e\mid e)K\), the response tensor becomes symmetric and off-diagonal. (IV) If the superconductor has SPG symmetry \(G_4\) generated by \((e^{-i\sigma_x\pi/2}\mid C_4)\), \(M_x\), \((e\mid e)K\), the off-diagonal components are forbidden, leaving a traceless diagonal nematic response. These results are summarized in Table~\ref{tab:LR_summary}, and detailed derivations are provided in the SM.

\begin{table}[htbp]
\centering
\resizebox{0.45\textwidth}{!}{
\renewcommand{\arraystretch}{1.25}
\begin{tabular}{c| c| c}
\hline\hline
 Group& Physical setting & Allowed $\rho^z$ in $\,J_i^{z}=\rho^z_{ij}A_j\,$ \\
\hline
$G_1$&AFM without spin flux
   & $\displaystyle
      \rho^z=\begin{pmatrix}0&\beta\\ -\beta&0\end{pmatrix}
     $ \\[8pt]
$G_2$&AFM with spin flux
   & $\displaystyle
      \rho^z_{\text{U}}=\begin{pmatrix}0&\beta\\ -\beta& 0\end{pmatrix},\quad
      \rho^z_{\text{nU}}=\begin{pmatrix}\alpha&0\\ 0&-\alpha\end{pmatrix}
     $ \\[8pt]
$G_3$& SPG-I
   & $\displaystyle
      \rho^z=\begin{pmatrix}0&a\\a&0\end{pmatrix}$
      \\[8pt]
$G_4$& SPG-II 
   & $\rho^z=\begin{pmatrix}b&0\\0&-b\end{pmatrix}$\\
\hline\hline
\end{tabular}}
\caption{Spin supercurrent responses for four representative cases. Here, U and nU stand for uniform and non-uniform respectively, $\alpha,\beta\in\mathbb{C}$, and $a,b\in\mathbb{R}$. 
}\label{tab:LR_summary}
\end{table}

\end{document}


\title{Supplemental Material for ``Spin-Charge Groups for Fermions in Fluids and Crystals: General Structures and Physical Consequences''}

\author{Arist Zhenyuan Yang}
\affiliation{School of Physics and Beijing Key Laboratory of Opto-electronic Functional Materials and Micro-nano Devices, Renmin University of China, Beijing, 100872, China}
\affiliation{Key Laboratory of Quantum State Construction and Manipulation (Ministry of Education), Renmin University of China, Beijing, 100872, China}
\affiliation{Department of Physics and HK Institute of Quantum Science \& Technology, The University of Hong Kong, Pokfulam Road, Hong Kong, China}

\author{Zheng-Xin Liu}
\email{liuzxphys@ruc.edu.cn}
\affiliation{School of Physics and Beijing Key Laboratory of Opto-electronic Functional Materials and Micro-nano Devices, Renmin University of China, Beijing, 100872, China}
\affiliation{Key Laboratory of Quantum State Construction and Manipulation (Ministry of Education), Renmin University of China, Beijing, 100872, China}

\date{\today}

\maketitle
\tableofcontents

\section{Solving spin-charge coupling from Goursat's lemma}

We introduce the so-called Goursat's lemma, and use it to uncover the spin-charge coupling structure in the internal symmetry group $O_f$. 
\begin{lemma}[Goursat's lemma]
Let \( H \subset G_1 \times G_2 \) be a subgroup such that \( \mathrm{pr}_i(H) = G_i \) for \( i = 1, 2 \), where \( \mathrm{pr}_i \) denotes the natural projection onto \( G_i \).  
Then \( H \) uniquely determines two normal subgroups \( N_1 \trianglelefteq G_1 \), \( N_2 \trianglelefteq G_2 \), and a group isomorphism  
\[
\varphi \colon G_1 / N_1 \to G_2 / N_2.
\]  
Here, $N_1$ is defined as $H \cap G_1$, where $G_1$ is viewed as a subgroup of $G_1 \times G_2$, namely the kernel of $\mathrm{pr}_2$. Similarly, we have $N_2 = H \cap G_2$. Moreover, there are natural isomorphisms:
\[
G_1 / N_1 \simeq H / (N_1 \times N_2) \simeq G_2 / N_2.
\]
\end{lemma}
For a proof, see page 8 of Serre’s book~\cite{serre2016finite}. Now, we can obtain the following
\begin{lemma}
Let $\mathcal{S}_{O_f}$ and $\mathcal{C}_{O_f}$ denote the representations of $O_f$ acting on the spin and charge sectors, respectively. Then $\mathcal{S}_0\lhd \mathcal{S}_{O_f}, \mathcal{C}_0\lhd \mathcal{C}_{O_f}$, and
\begin{align}
{\mathcal{S}_{O_f} \over \mathcal{S}_0} \cong {\mathcal{C}_{O_f} \over \mathcal{C}_0} \cong {O_f \over (\mathcal{S}_0 \star \mathcal{C}_0)}.
\end{align}
\end{lemma}
\begin{proof}
Each element of \( O_f \) corresponds to a coset representative of \( (SU(2)_s \times SU(2)_c) /\mathbb{Z}_2\).  
Accordingly, we define the group  
\[
O_f' \equiv \left( O_f \cup O_f \mathbb{Z}_2 \right) \subset SU(2)_s \times SU(2)_c,
\]
which lifts \( O_f \) to a subgroup of \( SU(2)_s \times SU(2)_c \).
By definition, \( O_f' \subset \mathcal{S}_{O_f} \times \mathcal{C}_{O_f} \), and its projections onto the spin ($\rm{pr}_s$) and charge ($\rm{pr}_c$) sectors are surjective, 
with kernels \( \mathcal{C}_0 \) and \( \mathcal{S}_0 \), respectively. 
Hence we have $\mathcal{S}_0 = \mathcal{S}_{O_f} \cap O_f', \mathcal{C}_0 = \mathcal{C}_{O_f} \cap O_f'$, and 
group isomorphisms 
\[
\mathcal{S}_{O_f}/\mathcal{S}_0\cong \mathcal{C}_{O_f}/\mathcal{C}_0\cong O'_f/(\mathcal{S}_0\times\mathcal{C}_0).
\]
Meanwhile, since $\mathbb{Z}_2$ is normal in both $O'_f$ and $\mathcal{S}_0\times\mathcal{C}_0$, we have have the following isomorphism
\[\frac{O'_f}{(\mathcal{S}_0\times\mathcal{C}_0)}\cong\frac{O'_f/\mathbb{Z}_2}{(\mathcal{S}_0\times\mathcal{C}_0)/\mathbb{Z}_2}={O_f \over (\mathcal{S}_0 \star \mathcal{C}_0)} .\]
\end{proof}

We apply Goursat's lemma to 
$S_z$-conserved charge-\(2e\) and charge-\(4e\) superconductors as illustrative examples.

For $S_z$-conserved charge-\(2e\) superconductors, we have \( \mathcal{S}_0 = U(1)_s \) and \( \mathcal{C}_0 = \mathbb{Z}_2^f \). Goursat's lemma then implies that the quotient group
\( Q = \mathcal{S}_{O_f}/U(1)_s = \mathcal{C}_{O_f}/\mathbb{Z}_2^f\) can only take values in \( \{E, \mathbb{Z}_2\} \). For \( Q = E \), we obtain \( O_f = \mathcal{S}_0 \star \mathcal{C}_0 = U(1)_s\). For \( Q = \mathbb{Z}_2 \), we have
\[
\mathcal{S}_{O_f}/U(1)_s = \{ U(1)_s, e^{-i {\sigma_x\over2} \pi} U(1)_s \}, \quad
\mathcal{C}_{O_f}/\mathbb{Z}_2^f = \{ \mathbb{Z}_2^f, e^{-i {\tau_z\over2} \pi} \mathbb{Z}_2^f \},
\]
which implies that \( O_f=[E+e^{-i({\sigma_x\over2}+{\tau_z\over2})\pi}]U(1)_{s}\).

For $S_z$-conserved charge-\(4e\) superconductors, we have \( \mathcal{S}_0 = U(1)_s \) and \( \mathcal{C}_0 = \mathbb{Z}_{4c} \), generated by \( e^{-i {\tau_z\over2} \pi} \). The quotient group \( Q \) again admits two possibilities, \( \{E, \mathbb{Z}_2\} \). For \( Q = E \), we obtain \( O_f = U(1)_s \star \mathbb{Z}_{4c}\). For \( Q = \mathbb{Z}_2 \), in contrast to the charge-\(2e\) case, there exist two inequivalent solutions for \( \mathcal{S}_{O_f} \) and \( \mathcal{C}_{O_f} \):
\Beq
&&\text{(i)}\quad 
\mathcal{S}_{O_f}/U(1)_s = \{ U(1)_s, e^{-i {\sigma_x\over2} \pi} U(1)_s \}, \quad
\mathcal{C}_{O_f}/\mathbb{Z}_{4c} = \{ \mathbb{Z}_{4c}, e^{-i {\tau_z\over4} \pi} \mathbb{Z}_{4c} \},\\
&&\text{(ii)}\quad 
\mathcal{S}_{O_f}/U(1)_s = \{ U(1)_s, e^{-i {\sigma_x\over2} \pi} U(1)_s \}, \quad
\mathcal{C}_{O_f}/\mathbb{Z}_{4c} = \{ \mathbb{Z}_{4c}, e^{-i {\tau_x\over2} \pi} \mathbb{Z}_{4c} \}.
\Eeq
Correspondingly, there are two possible forms of \( O_f \). The first is \(O_f= [ E+ e^{-i ({\sigma_x\over2} + {\tau_z\over4})\pi}]U(1)_s \star \mathbb{Z}_{4c} \), while the second is  \( O_f= [E+ e^{-i ({\sigma_x\over2} + {\tau_x\over2})\pi} ]U(1)_s \star \mathbb{Z}_{4c}\).

More examples of $O_f$ for given $\mathcal S_0$ and $\mathcal C_0$ are provided in Tab.~I of the main text (see also Tab.\ref{tab:Of_Infor}).


\section{Theory of Obstruction-free group extensions}

\subsection{From coupling $\Phi: G_b\to {\rm Out}(O_f)$ to homomorphism $\tilde{\phi}:G_b\to\mathscr{N}(O_f)/ O_f$}

Recall that, given two groups \( O_f \) and \( G_b \), an extension of \( G_b \) by \( O_f \) is a group \( G_f \) such that \( O_f \) is a normal subgroup of \( G_f \) and \( G_b \) is the corresponding quotient. The solutions to the extension problem, with \( O_f \) and \( G_b \) fixed, describe all consistent ways of coupling \( O_f \) and \( G_b \), and can be further classified in terms of the so-called coupling homomorphisms between them. A coupling homomorphism is a group homomorphism \( \Phi \) from \( G_b \) to \(\operatorname{Out}(O_f) = \operatorname{Aut}(O_f) / \operatorname{Inn}(O_f)\). For a given extension \( G_f \), the associated coupling homomorphism is induced by the conjugation action of \( G_f \) on \( O_f \), modulo inner automorphisms of \( O_f \). More precisely, let
\[
\mathscr{E}:\quad 1 \rightarrow O_f \xrightarrow{i} G_f \xrightarrow{\pi} G_b \rightarrow 1
\]
be a group extension, and let \( z \) be a section of \( \pi \), that is, a set-theoretic map satisfying \( \pi \circ z = \mathrm{id}_{G_b} \). Then one obtains a set-theoretic map \( \Phi^{\rm set}: G_b \to \operatorname{Aut}(O_f) \) defined by
\begin{align}
    \Phi^{\rm set}_g\cdot o = z(g)\, o\, z(g)^{-1}, \quad g \in G_b,\; o \in O_f.
\end{align}
The mapping \( \Phi^{\rm set} \) satisfies the property \( \Phi^{\rm set}_{g_1} \,\Phi^{\rm set}_{g_2} \, (\Phi^{\rm set}_{g_1 g_2})^{-1} \in \operatorname{Inn}(O_f)\), or, equivalently, the section \( z \) satisfies
\begin{align}\label{Section-property}
z(g_1)\, z(g_2)\, z(g_1 g_2)^{-1} \in O_f.
\end{align}
Thus, upon passing to the quotient via \( p: \operatorname{Aut}(O_f) \to \operatorname{Out}(O_f) \), this construction induces a homomorphism
\[
\Phi \equiv p \circ \Phi^{\rm set} :\quad G_b \xrightarrow{\ \Phi^{\rm set}\ } \operatorname{Aut}(O_f) \xrightarrow{\ p\ } \operatorname{Out}(O_f),
\]
namely, the coupling homomorphism. It is straightforward to verify that the coupling homomorphism \( \Phi \) is independent of the choice of the section \( z \). An extension is called central if \( O_f \) is contained in the center of \( G_f \), which is possible only when \( O_f \) is abelian. In this case, one has \( \operatorname{Inn}(O_f)=\{E\} \), and hence \( \operatorname{Out}(O_f)=\operatorname{Aut}(O_f) \). Furthermore, central extensions correspond precisely to the trivial coupling homomorphism, and conversely every extension with trivial coupling homomorphism is central. 

In the context of spin charge groups, our goal is not to exhaust all coupling homomorphisms \( \Phi: G_b \to \operatorname{Aut}(O_f) \). Instead, we restrict attention to the subset of maps \( \Phi^{\rm set} \) satisfying the following condition: for each \( g \in G_b \), there exists at least one \( \phi_g \in SO(4) \) such that
\beq\label{phiSO4}
\Phi^{\rm set}_g\cdot o=z(g)\, o\, z(g)^{-1} \equiv \, \phi_{g}o\, \phi_{g}^{-1},
\eeq 
This 
condition requires that the conjugation acts within the spin and charge degrees of freedom, whose maximal connected symmetry group is \( SO(4) \). It follows that \( \phi_{g} \in \mathscr{N}_{SO(4)}(O_f) \), the normalizer of \( O_f \) in \( SO(4) \). Hence, the set-theoretic map \( \Phi^{\rm set} \) in (\ref{phiSO4}) induces a map
\[
\phi : G_b \to \mathscr{N}_{SO(4)}(O_f). 
\]
Furthermore, the condition in Eq.~(\ref{Section-property}) implies that \( \phi \) satisfies
\begin{align}\label{phis-property}
\phi_{g_1}\, \phi_{g_2}\, \phi_{g_1 g_2}^{-1} \in O_f.
\end{align}
Therefore, upon passing to the quotient, one obtains a {\bf group homomorphism}
\[
\tilde{\phi} : G_b \to \mathscr{N}_{SO(4)}(O_f)/O_f.
\]
In later discussion, we restrict our attention to group extensions whose coupling homomorphisms $\Phi$ arise from the group homomorphisms \( \tilde{\phi} : G_b \to \mathscr{N}_{SO(4)}(O_f)/O_f \).

\subsection{Obstruction Free of Extensions Induced by \( \tilde\phi: G_b \to \mathscr{N}(O_f)/O_f \)}

It is well known that not every homomorphism from \( G_b \) to \( \operatorname{Out}(O_f) \) can serve as a coupling homomorphism for a group extension due to the presence of cohomological obstructions, namely nontrivial 3-cocycles classified by $H^3_{\Phi}(G_b, \mathcal Z(O_f))$\cite{eilenberg1947cohomology, serre2016finite}. Here, we show that all coupling homomorphisms \( G_b \to \operatorname{Out}(O_f) \) arising from the group homomorphisms \( G_b \to \mathscr{N}_{SO(4)}(O_f)/O_f \) are obstruction-free.

\begin{proposition}
The coupling homomorphism \( \Phi: G_b \to \operatorname{Out}(O_f) \) induced by \( \tilde{\phi}: G_b \to \mathscr{N}_{SO(4)}(O_f)/O_f \) is obstruction-free. 
\end{proposition}

\begin{proof}
Choose a lift \( \phi: G_b \to \mathscr{N}_{SO(4)}(O_f) \) of \( \tilde{\phi} \), and define
\[
f(g_1, g_2) := \phi_{g_1} \phi_{g_2} \phi_{g_1 g_2}^{-1} \in O_f, \quad \forall\, g_1, g_2 \in G_b.
\]
Since each inner automorphism of \( \mathscr{N}_{SO(4)}(O_f) \) induces an automorphism of \( O_f \), the lift \( \phi \) defines a mapping \( \Phi^{\rm set}: G_b \to \operatorname{Aut}(O_f) \), and one has
\begin{align}
\Phi^{\rm set}_{g_1}\Phi^{\rm set}_{g_2}
= \operatorname{Inn}[f(g_1,g_2)] \, \Phi^{\rm set}_{g_1 g_2},
\end{align}
where \( \operatorname{Inn} \) denotes the inner automorphism
\[
1 \to \mathcal{Z}(O_f) \to O_f \stackrel{\operatorname{Inn}}{\longrightarrow} \operatorname{Aut}(O_f) \to \operatorname{Out}(O_f) \to 1,
\]
and \( \mathcal{Z}(O_f) \) is the center of \( O_f \). 

The obstruction cocycle \( \omega_3 \) is detected by the associativity of \( \Phi^{\rm set}_{g_1} \Phi^{\rm set}_{g_2} \Phi^{\rm set}_{g_3} \) for \( g_1, g_2, g_3 \in G_b \), namely
\begin{align}\label{SecIIIB1:obstruction}
f(g_1,g_2) f(g_1 g_2,g_3)
= \omega_3(g_1,g_2,g_3)\left[\Phi^{\rm set}_{g_1}\cdot f(g_2,g_3)\right] f(g_1,g_2 g_3),
\end{align}
with \( \omega_3(g_1, g_2, g_3) \in \mathcal{Z}(O_f) \). 

Hence, we have
\beq
f(g_1,g_2) f(g_1 g_2,g_3)&=&\phi_{g_1}\phi_{g_2}\phi_{g_1 g_2}^{-1} \phi_{g_1 g_2}\phi_{g_3}\phi_{g_1 g_2 g_3}^{-1}\notag\\
&=& \phi_{g_1}\phi_{g_2}\phi_{g_3}\phi_{g_1 g_2 g_3}^{-1}.
\eeq

On the other hand,
\beq
\left[\Phi^{\rm set}_{g_1}\cdot f(g_2,g_3)\right] f(g_1,g_2 g_3)
&=& \phi_{g_1} f(g_2,g_3) \phi_{g_1}^{-1} f(g_1,g_2 g_3)\notag\\
&=& \phi_{g_1}\phi_{g_2}\phi_{g_3}\phi_{g_2 g_3}^{-1} \phi_{g_1}^{-1} \phi_{g_1}\phi_{g_2 g_3}\phi_{g_1 g_2 g_3}^{-1}\notag\\
&=& \phi_{g_1}\phi_{g_2}\phi_{g_3}\phi_{g_1 g_2 g_3}^{-1}.
\eeq

Hence, the obstruction 3-cocycle \( \omega_3(g_1, g_2, g_3) \) is trivial for any \( g_1, g_2, g_3 \in G_b \). 
\end{proof}

\subsection{ Illustration: \( \tilde{\phi}: G_b \to \mathscr{N}(O_f)/O_f \) for \( s \)-, \( d \)-, \( (p_x \pm i p_y) \)-wave superconductors}\label{spdwave}

We use a representative example to illustrate the homomorphism \( \tilde{\phi} \) and the notion of conjugate equivalence of group extensions. Consider the single-orbital BCS Hamiltonian for \( s \)-wave, \( d \)-wave, and \( (p_x \pm i p_y) \)-wave superconductors on a square lattice with 
$O_f=Z_2^f$ and \( G_b=\mathscr{C}_4 \):
\begin{align}\label{dwave}
H = \sum_{\bm k}
\begin{pmatrix}
c_{{\bm k}\uparrow}^\dagger & c_{-{\bm k}\downarrow}
\end{pmatrix}
\begin{pmatrix}
\chi_{\bm k} & \Delta_{\bm k} \\
\Delta_{\bm k}^* & -\chi_{-{\bm k}}^{T}
\end{pmatrix}
\begin{pmatrix}
c_{{\bm k}\uparrow} \\
c_{-{\bm k}\downarrow}^\dagger
\end{pmatrix}.
\end{align}

Under the \( C_4 \) rotation, the kinetic term \( \chi_{\bm k} \) remains invariant, while the gap function \( \Delta_{\bm k} \) acquires a phase \( e^{i\theta} \):
\begin{align}
C_4 H C_4^{-1}
= \sum_{\bm k}
\begin{pmatrix}
c_{{\bm k}\uparrow}^\dagger & c_{-{\bm k}\downarrow}
\end{pmatrix}
\begin{pmatrix}
\chi_{\bm k} & e^{i\theta} \Delta_{\bm k} \\
e^{-i\theta}\Delta_{\bm k}^* & -\chi_{-{\bm k}}^{T}
\end{pmatrix}
\begin{pmatrix}
c_{{\bm k}\uparrow} \\
c_{-{\bm k}\downarrow}^\dagger
\end{pmatrix},
\end{align}
where \( \theta = 0, \pi, \mp \frac{\pi}{2} \) for the \( s \)-wave, \( d \)-wave, and \( (p_x \pm i p_y) \)-wave superconductors, respectively.

The phase factors \( e^{i\theta} \) can be removed by a uniform charge \( U(1) \) transformation \( e^{-i \tau_z \theta/2} \). Accordingly, we define the operator
\begin{equation}\label{barC4}
g_{C_4} \equiv \left( e^{-i \tau_z \theta/2} \,\middle|\, C_4 \right),
\end{equation}
where \( E \) denotes the identity operator acting on the spin degree of freedom. This operator satisfies
\begin{equation*}
g_{C_4} \, c_{i\sigma} \, g_{C_4}^{-1} = e^{i \theta/2} c_{i' \sigma}, \quad
g_{C_4} \, c_{i\sigma}^\dagger \, g_{C_4}^{-1} = e^{-i \theta/2} c_{i' \sigma}^\dagger,
\end{equation*}
where \( \sigma \) labels the spin degree of freedom, and \( i' \) is the image of site \( i \) under the \( C_4 \) rotation. Consequently, the Hamiltonian is invariant under \( g_{C_4} \):
\begin{align}
g_{C_4} H g_{C_4}^{-1} = H.
\end{align}

Note that the fermion-parity operator \( P_f \) is always an element of the symmetry group. We therefore summarize the full symmetry groups \( G_f \) for the three pairing symmetries as follows:
\begin{itemize}
\item[(1)] For \( s \)-wave superconductors, the full symmetry group is the direct product
\[
G_f = \mathscr{C}_4 \times \mathbb{Z}_2^f,
\]
where \( \mathscr{C}_4 = \{E, C_4, C_4^2, C_4^3\} \) and \( \mathbb{Z}_2^f = \{E, P_f\} \).

\item[(2)] For \( d \)-wave superconductors, the full symmetry group is
\[
G_f = \bar{\mathscr{C}}_4 \times \mathbb{Z}_2^f,
\]
where \( \bar{\mathscr{C}}_4 = \{E, g_{C_4}, g_{C_4}^2, g_{C_4}^3\} \) and \( (g_{C_4})^4 = E \).

\item[(3)] For \( (p_x \pm i p_y) \)-wave superconductors, the full symmetry group is the cyclic group
\[
G_f = \bar{\mathscr{C}}_8,
\]
where \( \bar{\mathscr{C}}_8 = \{E, g_{C_4}, g_{C_4}^2, g_{C_4}^3, P_f, P_f g_{C_4}, P_f g_{C_4}^2, P_f g_{C_4}^3\}\), and \( (g_{C_4})^4 = P_f \).
\end{itemize}

All three symmetry groups \( G_f \) discussed above are central extensions of \( \mathscr{C}_4 \) by \( \mathbb{Z}_2^f \), and hence their coupling homomorphisms are trivial. It is well known that the classification of central extensions of \( \mathscr{C}_4 \) by \( \mathbb{Z}_2^f \) is given by
\[
H^2(\mathscr{C}_4,  {Z}_2^f) = \mathbb{Z}_2.
\]
Accordingly, the symmetry groups for the \( s \)-wave and \( d \)-wave superconductors correspond to the trivial extension, namely the direct product, whereas the symmetry group for the \( (p_x \pm i p_y) \)-wave superconductors corresponds to the nontrivial extension.

Noticing that $\mathscr N_{SO(4)}(O_f)=\mathscr N_{SO(4)}(Z_2^f)=SO(4)$ and $\mathscr N_{SO(4)}(Z_2^f)/Z_2^f=SO(3)_s\times SO(3)_c$, we now consider the homomorphism
\begin{align}\label{SecII-tildemu} 
\tilde{\phi} : \mathscr{C}_4 \to SO(3)_s\times SO(3)_c. 
\end{align}
Obviously, $\tilde{\phi}$ can be decomposed into two group homomorphisms $\tilde\phi_c: \mathscr{C}_4 \to SO(3)_c$ and $\tilde\phi_s:\mathscr{C}_4 \to SO(3)_s$. Here we focus on the map $\tilde\phi_c$ and assume that $\tilde\phi_s$ is the trivial one. Let \( \mathbb{B}_0 \) denote the kernel of \( \tilde{\phi}_c \). Then \( \mathbb{B}_0 \) can take only one of three values: \( \mathscr{C}_4 \), \( \mathscr{C}_2 \), or \( E \). These correspond to rotations in \( SO(3)_c \) with angles \( \theta = 0 \), \( \pi \), or \( \mp \pi/2 \), respectively, associated with the generator \( C_4 \). Upon lifting from \( SO(3)_c \) to \( SU(2)_c \), the corresponding angles in \( SU(2)_c \) become \( 0 \), \( \pi/2 \), and \( \mp \pi/4 \), respectively, consistent with Eq.~(\ref{barC4}). 

The discussion for $s$-, $d$-, $(p_x\pm ip_y)$-wave superconductors implies that nonconjugate group homomorphisms \( \tilde{\phi}_c: \mathscr{C}_4 \to SO(3)_c \), for instance those with different kernels, can nevertheless give rise to the same trivial coupling homomorphism $\Phi$. In other words, the trivial coupling homomorphism can be realized in multiple inequivalent ways. Different realizations of the same coupling homomorphism \( \Phi \) may give rise to identical or distinct cohomology classes, as illustrated by the \( s \)-wave, \( (p_x \pm i p_y) \)-wave, and \( d \)-wave superconductors.

Detailed information of $\mathscr{N}_{SO(4)}(O_f)/O_f$ for given $O_f$ are showned in Tab.\ref{tab:Of_Infor}. Here we provide an example, the $^3$He $A_1$ phase with $O_f=U(1)_{sc}\times Z_2^f$. Then one has $\mathscr N_{SO(4)}(O_f) = \Big(U(1)_{sc}\times Z_2^f\times U(1)_{sc}'\Big)\rtimes Z_{2sc} = \big( O_f\times U(1)_{sc}'\big)\rtimes Z_{2sc}$, with $U(1)_{sc}=\{e^{-i({\sigma_z\over2} + {\tau_z\over2})\theta};0\leq\theta<2\pi\},\ U(1)_{sc}'=\{e^{-i({\sigma_z\over2} - {\tau_z\over2})\theta};0\leq\theta<2\pi\}$ and $Z_{2sc}=\{ E, e^{-i({\sigma_x\over2}+{\tau_x\over2})\pi}\}$. Consequently, one has
\Beq
\mathscr N(O_f)/O_f \cong U(1)_{sc}'\rtimes Z_{2sc}\cong O(2).
\Eeq


\begin{table*}[htbp]
\caption{\large The internal symmetry groups $ O_f $ and related information.}\label{tab:Of_Infor}
\footnotesize
\centering
\resizebox{.85\linewidth}{!}{
\renewcommand{\arraystretch}{2}
\begin{tabular}{|c|c|c|c|c|c|c|}
\hline\hline
$\mathcal{S}_0$ & $\mathcal{C}_0$ & ${O_f\over (\mathcal{S}_0\times \mathcal{C}_0)/Z_2^f}$ & $O_f$ & $\mathcal Z(O_f)$ & $\mathscr{N}_{SO(4)}(O_f)/O_f$ & Physical systems\\ 
\hline
\multirow{5}{*}{$Z_2^f$} & \multirow{2}{*}{$Z_2^f$} &  $\mathbb Z_1$ & $Z_2^f$ & $Z_2^d$ &$SO(3)_s\times SO(3)_c$ & $^3$He-$B$, 2e SC-I \\
\cline{3-7}
& & $U(1),\! ...$ & ${ U(1)_{sc}\times Z_2^f}, ...$ &$U(1)_{sc}\times Z_2^f$ & O(2) & $^3$He-$A_1$\!,\! 2e SC-II,\!\! ... \\
\cline{2-7}
& $Z_{4c}$  & $\mathbb Z_1,...$ &$Z_{4c},...$ &$Z_{4c}$ & $SO(3)_s\times O(2)_c$&  4e SC-I, ...\\ 
\cline{2-7}
& \multirow{2}{*}{$U(1)_c$}  & $\mathbb Z_1$ &$U(1)_c$ &$U(1)_c$& $SO(3)_s\times Z_{2c}$ &Insulators-I \\ 
\cline{3-7}
& & $\mathbb{Z}_2$ & $[ E+ e^{-\mi({\sigma_z\over2}+{\tau_x\over2})\pi}]U(1)_c$ &$Z_2^f$&$O(2)_s$ & Insulator-II\\
\cline{2-7}
& $SU(2)_c$  & $\mathbb Z_1$ & $SU(2)_c$  &$Z_2^f$ & $SO(3)_s$ & Insulator-III \\
\hline
\multirow{8}{*}{$U(1)_s$\!}& \multirow{2}{*}{$Z_2^f$} & $\mathbb Z_1$&$U(1)_s$ &$U(1)_s$&$Z_{2s}\times SO(3)_c$ & 2e SC-III \\
\cline{3-7}
& & $\mathbb{Z}_2$ & $[E+e^{-\mi({\sigma_x\over2}+{\tau_{z}\over2})\pi}]U(1)_s$  &$Z_2^f$&$O(2)_c$ & {$^3$He-$A$,} 2e SC-IV\\
\cline{2-7}
& \multirow{3}{*}{$Z_{4c}$}  & $\mathbb Z_1$&$U(1)_s\star Z_{4c}$ &$U(1)_s\star Z_{4c}$&$Z_{2s}\times O(2)_c$& 4e SC-II \\
\cline{3-7}
& & \multirow{2}{*}{$\mathbb{Z}_2$} &$[E+e^{-\mi({\sigma_x\over2}+{\tau_{z}\over4})\pi}]U(1)_s\!\star\! Z_{4c}$  &$Z_2^f$&$O(2)_c$&  4e SC-III \\
\cline{4-7}
&  & &$[E+e^{-\mi({\sigma_x\over2}+{\tau_{x}\over2})\pi}]U(1)_s\!\star\! Z_{4c}$ &$Z_2\times Z_2^f$&$
Z_{2s}\times Z_{2c}$&  4e SC-IV\\
\cline{2-7}
& \multirow{2}{*}{$U(1)_{c}$}  &$\mathbb Z_1$ & $U(1)_s\star U(1)_c$ &$U(1)_s\star U(1)_c$&$Z_{2s}\times Z_{2c}$& Insulators-IV\\
\cline{3-7}
&  & $\mathbb{Z}_2$ &$[E\!+\!e^{-\mi({\sigma_x\over2}\!+\!{\tau_{x}\over2})\pi}]U(1)_s\!\!\star\! U(1)_c$ &$Z_2\times Z_2^f$&$Z_{2s}$& Insulators-V \\
\cline{2-7}
& $SU(2)_c$  & $\mathbb Z_1$ &$U(1)_s\star SU(2)_c$ &$U(1)_s$&$Z_{2s}$& Insulators-VI\\
\hline
\multirow{4}{*}{$SU(2)_s$\!}& $Z_2^f$ & $\mathbb Z_1$& $SU(2)_s$ &$Z_2^f$ & $SO(3)_c$ &  2e SC-V \\
\cline{2-7}
& $Z_{4c}$  & $\mathbb Z_1$&$SU(2)_s\star Z_{4c}$ & $Z_{4c}$ & $O(2)_c$ & 4e SC-V \\
\cline{2-7}
& $U(1)_c$  &$\mathbb Z_1$&$SU(2)_s\star U(1)_c$ & $U(1)_c$ & $Z_{2c}$ & Insulator-VII\\
\cline{2-7}
& $SU(2)_c$  & $\mathbb Z_1$ & $SO(4)$  & $Z_2^f$ & $Z_1$ & Insulator-VIII\\
\hline
\end{tabular}}
\end{table*}

\subsection{Solutions to group extensions with arbitrary kernels}

Group extensions can be classified rigorously within the mathematical literature\cite{eilenberg1947cohomology1,eilenberg1947cohomology}. Here, we translate the abstract mathematical formulation of these solutions into the framework adopted in the main text. In the following, we continue to use the notation introduced in the previous section.

\subsubsection{When \( O_f \) is abelian.} 

In this case, \( \tilde{\phi} \) induces a homomorphism \( \Phi: G_b \to \operatorname{Aut}(O_f) \). The isomorphism classes of extensions of \( G_b \) by \( O_f \) are in one-to-one correspondence with the elements of the second cohomology group \( H^2_{\Phi}(G_b, O_f) \). In particular, under this correspondence, the split extension, where \( G_f \) is the semidirect product of \( G_b \) and \( O_f \), corresponds to the trivial element \( 0 \in H^2_{\Phi}(G_b, O_f) \).

Different abstract realizations of \( G_f \) can be obtained by specifying the multiplication law on the set \( O_f \times G_b \), as follows:
\[
\left(o_1, g_1\right)\left(o_2 , g_2\right)
= \left( o_1\, o_2^{g_1}\, \omega_2(g_1, g_2),\; g_1 g_2 \right),
\]
where $o_2^{g_1} \equiv (\Phi^{\rm set}_{g_1}\cdot o_2) =\phi_{g_1}o_2\phi_{g_1}^{-1}$ and \( \omega_2(g_1, g_2) \in Z^2_{\Phi}(G_b, O_f) \) is a \(2\)-cocycle. Using \( \phi \) and the relation \( \phi_{g_1}\phi_{g_2}\phi_{g_1 g_2}^{-1} \in O_f \), these solutions can be reformulated in an alternative form as follows: 
\beq\label{Sec:transMultiplication}
\left(o_1, g_1\right)\left(o_2 , g_2\right)&=& \left( o_1 o_2^{g_1}\omega_2(g_1, g_2),\; g_1 g_2 \right)\notag\\
&=& \left( o_1 (\phi_{g_1} o_2\phi_{g_1}^{-1}) \omega_2(g_1, g_2),\; g_1 g_2 \right)\notag\\
&=& \left((o_1 \phi_{g_1})(o_2 \phi_{g_2}) \phi_{g_2}^{-1} \phi_{g_1}^{-1} \omega_2(g_1, g_2),\; g_1 g_2 \right)\notag\\
&=& \left((o_1\phi_{g_1})(o_2 \phi_{g_2}) \phi_{g_1 g_2}^{-1} \omega'_2(g_1, g_2),\; g_1 g_2 \right).
\eeq
where 
\[
\omega_2(g_1,g_2)
= \left[\phi_{g_1}\phi_{g_2}\phi_{g_1 g_2}^{-1}\right]\omega'_2(g_1,g_2).
\]
Here, the function \( \omega'_2(g_1, g_2) \in O_f \) is also a representative element of \( H^2_{\Phi}(G_b, O_f) \). {\it Note that \( \omega_2 \) and \( \omega'_2 \) may not be cohomologous.} 
This form enables us to translate the abstract multiplication law into a more conventional form. We first note that
\[
(o_1 \phi_{g_1})(o_2 \phi_{g_2}) \phi_{g_1 g_2}^{-1} \in O_f.
\]
Then, if we define
\begin{align}\label{Translate}
(o, g) \equiv (o \phi_{g} \mid g),
\end{align}
Eq.~(\ref{Sec:transMultiplication}) becomes
\beq\label{Sec_multi_trivial}
\left( o_1 \phi_{g_1} \mid g_1 \right) \left( o_2 \phi_{g_2} \mid g_2 \right) &=& 
(\omega_2(g_1, g_2)|e)\left( o_1 (\phi_{g_1} o_2\phi_{g_1}^{-1}) \phi_{g_1g_2}| g_1 g_2 \right)\notag\\
&=&
\left( [o_1 \phi_{g_1}][o_2 \phi_{g_2}] \phi_{g_1 g_2}^{-1} [\omega'_2(g_1, g_2)] \phi_{g_1 g_2} \mid g_1 g_2 \right) \notag\\
&=& \left( \omega'_2(g_1, g_2) [o_1 \phi_{g_1}][o_2 \phi_{g_2}] \phi_{g_1 g_2}^{-1} \phi_{g_1 g_2} \mid g_1 g_2 \right)\notag\\
&=& \left( \omega'_2(g_1, g_2) [o_1 \phi_{g_1}][o_2\phi_{g_2}] \mid g_1 g_2 \right) \notag \\
&=& \left( \omega'_2(g_1, g_2) \mid e \right) \left( [o_1 \phi_{g_1}][o_2 \phi_{g_2}] \mid g_1 g_2 \right).
\eeq
Remarkably, even when \( \omega'_2 \) is trivial, the cohomology class \( \omega_2 \) is not necessarily trivial, which can lead to nontrivial group extensions. The group structure for \( (p_x + i p_y) \)-wave superconductors discussed above is such an example.

\subsubsection{When \( O_f \) is non-abelian.} 

In this case, \( \tilde{\phi} \) induces a homomorphism \( \Phi: G_b \to \operatorname{Out}(O_f) \) without any obstruction, and the isomorphism classes of extensions are classified by \( H^2_{\varphi}(G_b, \mathcal{Z}(O_f)) \), where \( \mathcal{Z}(O_f) \) denotes the center of \( O_f \) and $\varphi$ the induced action of $G_b$ on $\mathcal{Z}(O_f)$. However, unlike the case where \( O_f \) is abelian, there is no natural correspondence between isomorphism classes and elements of \( H^2_{\varphi}(G_b, \mathcal{Z}(O_f)) \). Instead, the distinctions between isomorphism classes are determined by elements of \( H^2_{\varphi}(G_b, \mathcal{Z}(O_f)) \). Thus, to obtain all solutions for \( G_f \), one initial solution is required, and all other solutions can be derived sequentially by inserting elements from \( H^2_{\varphi}(G_b, \mathcal{Z}(O_f)) \) into this initial solution. With the aid of 
\[
\alpha(g_1, g_2) := \phi_{g_1} \phi_{g_2} \phi_{g_1 g_2}^{-1} \in O_f, \quad \forall\, g_1, g_2 \in G_b,
\]
the abstract form of the initial solution can also be defined on the set \( O_f \times G_b \), as demonstrated below\cite{eilenberg1947cohomology}:
\[
\left(o_1, g_1\right)\left(o_2 , g_2\right)
= \left( o_1 o_2^{g_1} \alpha(g_1, g_2),\; g_1 g_2 \right).
\]
Following (\ref{Sec:transMultiplication})--(\ref{Sec_multi_trivial}), the initial solution form is equivalent to 
\begin{align}\label{initial_form}
\left( o_1 \phi_{g_1}\! \mid g_1 \right) \left( o_2 \phi_{g_2}\! \mid g_2 \right)
= \left( (o_1\phi_{g_1})(o_2 \phi_{g_2}) \! \mid g_1 g_2 \right).
\end{align}
On the other hand, all solutions of extensions of \( G_b \) with non-abelian kernel \( O_f \) are given by
\begin{align}\label{all_solution_with_initial_form}
\left(o_1, g_1\right)\left(o_2 , g_2\right)
= \left( o_1 o_2^{g_1} \alpha(g_1, g_2) \omega'_2(g_1,g_2),\; g_1 g_2 \right).
\end{align}
where \( [\omega'_2] \in H^2_{\varphi}(G_b, \mathcal{Z}(O_f)) \). 
Hence, we have 
\beq\label{Sec_multi_trivial_nonabel}
\left( o_1 \phi_{g_1} \mid g_1 \right) \left( o_2\phi_{g_2} \mid g_2 \right)
&=& \left( \alpha(g_1, g_2) \omega'_2(g_1,g_2) (o_1 o_2^{g_1}) \phi_{g_1g_2}\mid g_1 g_2 \right)\notag \\
&=& \left( \omega'_2(g_1,g_2) (o_1 \phi_{g_1})(o_2 \phi_{g_2}) \mid g_1 g_2 \right)\notag \\
&=& \left( \omega'_2(g_1, g_2) \mid e \right) \left( (o_1 \phi_{g_1})(o_2 \phi_{g_2}) \mid g_1 g_2 \right).
\eeq

Notice that (\ref{Sec_multi_trivial_nonabel}) {\it formally} coincides with (\ref{Sec_multi_trivial}). It is therefore convenient to use the more physical notation \( (o \phi(g)\!\mid\! g) \) to represent group elements (as in the main text), rather than the abstract form \( (o, g) \).

\subsection{Solutions realized by introducing site-dependent \( SO(4) \) operations}

After providing the general form of the group multiplication law (\ref{Sec_multi_trivial_nonabel}), a natural question arises: how can it be realized? Here, we show that (\ref{Sec_multi_trivial_nonabel}) can be realized by introducing a background \( SO(4) \) gauge field. Specifically, introduce site-dependent \( SO(4) \) operations \( \gamma_{l}({r}) \) for each \( l \in G_b \), and require that \( \gamma_{l}({r}) \) commutes with \( O_f \). Then the group \( G_f=\left\{ \big(\gamma_{l}({r}) O_f \phi_{l}\mid l\big);\; l\in G_b\right\}\) has the following multiplication rule
\beq\label{multiply}
( \gamma_{l_1}({r}) x \phi_{l_1} \mid l_1 ) ( \gamma_{l_2}({r}) y \phi_{l_2} \mid l_2 )
= \omega_2(l_1, l_2)\Big(\gamma_{l_1 l_2}({r}) x y^{l_1} \phi_{l_1}\phi_{l_2} \mid l_1 l_2\Big),
\eeq
where
\[
x,y\in O_f,\quad \omega_2(l_1, l_2) = \gamma_{l_1}({r}) \gamma'_{l_2}({r}) \gamma_{l_1 l_2}^{-1}({r}),
\quad
\gamma'_{l_2}({r}) = \phi_{l_1}\gamma_{l_2}(l_1^{-1}{r})\phi_{l_1}^{-1}.
\]
The result follows from a direct calculation:
\beq
\left(\gamma_{l_{1}}(r)\,x\phi_{l_1}\mid l_{1}\right) \left(\gamma_{l_{2}}(r)\,y\phi_{l_{2}}\mid l_{2}\right)
&=&\left( \gamma_{l_{1}}(r)\,x\phi_{l_1} l_{1} \gamma_{l_{2}}(r)\,y\phi_{l_{2}}\mid  l_{2} \right)\notag\\
&=& \left( [\gamma_{l_{1}}(r)\,x\phi_{l_1}] l_{1} [\gamma_{l_{2}}(r)\,y\phi_{l_{2}}] l_{1}^{-1} \mid l_{1}l_{2} \right)\notag\\
&=& \left([\gamma_{l_{1}}(r)\,x\phi_{l_1}][ \gamma_{l_{2}}(l_{1}^{-1}r)\,y\phi_{l_2}] \mid l_{1}l_{2} \right)\notag\\
&=&\left(    [\gamma_{l_{1}}(r)\,x][\phi_{l_1}\gamma_{l_{2}}(l_{1}^{-1}r)\,\phi_{l_1}^{-1}][\phi_{l_1}y\phi_{l_1}^{-1}]\phi(l_1)\phi_{l_2}  \mid l_{1}l_{2}     \right)\notag\\
&=&\left(     [\gamma_{l_{1}}(r)][\phi_{l_1}\gamma_{l_{2}}(l_{1}^{-1}r)\,\phi_{l_1}^{-1}]x[\phi_{l_1}y\phi_{l_1}^{-1}]\phi_{l_1}\phi_{l_2} \mid l_{1}l_{2}       \right)\notag\\
&=&\left(  \gamma_{l_{1}}(r)\gamma'_{l_2}(r)x y^{l_1}   \phi_{l_1}\phi_{l_2} \mid l_{1}l_{2}       \right)\notag\\
&=& \left(  [\gamma_{l_{1}}(r)\gamma'_{l_2}(r)\gamma_{l_1l_2}^{-1}(r)]\gamma_{l_1l_2}(r)x y^{l_1}   \phi_{l_1}\phi_{l_2} \mid l_{1}l_{2}        \right)\notag\\
&=&\omega_2(l_1,l_2)\left( \gamma_{l_1l_2}(r)x y^{l_1}   \phi_{l_1}\phi_{l_2} \mid l_{1}l_{2}   \right)     \notag\\
&=&\alpha_2(l_1,l_2)\omega_2(l_1,l_2)\left( \gamma_{l_1l_2}(r)x y^{l_1}   \phi_{l_1l_2}\mid l_{1}l_{2}   \right)
\eeq
where the sixth equality follows from the fact that the centralizer group is a normal subgroup of the normalizer group. The fact that the factor $\omega_2(l_{1},l_{2})\in O_f$ is a product of centralizer elements implies $\omega(l_1,l_2)\in \mathcal{Z}(O_f)$.

\section{Enumeration of $\tilde\phi$ for SCPGs in 2D $S_z$-conserved charge $2e/4e$ superconductors}\label{sec: SCPG2D}

In this section we enumerate all possible group homomorphisms $\tilde\phi$ from 2D magnetic point groups to $\mathscr N(O_f)/O_f$, where $O_f$ varies in all possible internal groups for $\mathscr S_0=U(1)_s$, $\mathscr C_0=Z_2^f$ or $Z_{4c}$. 

\subsection{2e SC-III: $O_f=U(1)_s$.}

\begingroup\scriptsize
\begin{longtable}{@{}>{\centering\arraybackslash}p{0.08\textwidth} >{\centering\arraybackslash}p{0.15\textwidth} >{\centering\arraybackslash}p{0.055\textwidth} >{\RaggedRight\arraybackslash}p{0.69\textwidth}@{}}
\hline\hline
MPG & Generators & \makecell{\#\\SCPGs} & Representatives and images on generators\\
\hline
\endfirsthead

\hline\hline
MPG & Generators & \makecell{\#\\SCPGs} & Representatives and images on generators\\
\hline
\endhead

\hline\hline
\endfoot
$1$ & $-$ & 1 & $\tilde{\phi}_{1}:\; 1$ \\
\hline
$1^{\prime}$ & $T$ & 4 & $\tilde{\phi}_{1}:\; \left(e^{-i\left(\frac{\pi}{2}\tx + \frac{\pi}{2}\sy\right)}\mid T\right)$ \\
$1^{\prime}$ & $T$ & 4 & $\tilde{\phi}_{2}:\; \left(e^{-i\left(\frac{\pi}{2}\tx\right)}\mid T\right)$ \\
$1^{\prime}$ & $T$ & 4 & $\tilde{\phi}_{3}:\; \left(e^{-i\left(\frac{\pi}{2}\ty + \frac{\pi}{2}\sy\right)}\mid T\right)$ \\
$1^{\prime}$ & $T$ & 4 & $\tilde{\phi}_{4}:\; \left(e^{-i\left(\frac{\pi}{2}\ty\right)}\mid T\right)$ \\
\hline
$2$ & $C_{2,y}$ & 4 & $\tilde{\phi}_{1}:\; \left(e^{-i\left(\frac{\pi}{2}\tz\right)}\mid C_{2,y}\right)$ \\
$2$ & $C_{2,y}$ & 4 & $\tilde{\phi}_{2}:\; \left(e^{-i\left(\frac{\pi}{2}\tz + \frac{\pi}{2}\sy\right)}\mid C_{2,y}\right)$ \\
$2$ & $C_{2,y}$ & 4 & $\tilde{\phi}_{3}:\; \left(1\mid C_{2,y}\right)$ \\
\end{longtable}
\endgroup
\subsection{2e SC-IV: $O_f=\langle U(1)_s,\ e^{-i(\sigma_x+\tau_z)\frac{\pi}{2}}\rangle$}

\begingroup\scriptsize
\begin{longtable}{@{}>{\centering\arraybackslash}p{0.08\textwidth} >{\centering\arraybackslash}p{0.15\textwidth} >{\centering\arraybackslash}p{0.055\textwidth} >{\RaggedRight\arraybackslash}p{0.69\textwidth}@{}}
\hline\hline
MPG & Generators & \makecell{\#\\SCPGs} & Representatives and images on generators\\
\hline
\endfirsthead

\hline\hline
MPG & Generators & \makecell{\#\\SCPGs} & Representatives and images on generators\\
\hline
\endhead

\hline\hline
\endfoot
$1$ & $E$ & 1 & $\tilde{\phi}_{1}:$ $(E\mid E)$ \\
\hline
$1'$ & $T$ & 3 & $\tilde{\phi}_{1}:$ $(e^{-i\sigma_y\frac{\pi}{2}-i\tau_y\frac{\pi}{2}}\mid T)$ \\
$1'$ & $T$ & 3 & $\tilde{\phi}_{2}:$ $(e^{-i\tau_z\frac{\pi}{2}} e^{-i\sigma_y\frac{\pi}{2}-i\tau_y\frac{\pi}{2}}\mid T)$ \\
$1'$ & $T$ & 3 & $\tilde{\phi}_{3}:$ $(e^{-i\tau_x\frac{\pi}{2}} e^{-i\sigma_y\frac{\pi}{2}-i\tau_y\frac{\pi}{2}}\mid T)$ \\
\hline
$2$ & $C_{2y}$ & 3 & $\tilde{\phi}_{1}:$ $(E\mid C_{2y})$ \\
$2$ & $C_{2y}$ & 3 & $\tilde{\phi}_{2}:$ $(e^{-i\tau_z\frac{\pi}{2}}\mid C_{2y})$ \\
$2$ & $C_{2y}$ & 3 & $\tilde{\phi}_{3}:$ $(e^{-i\tau_x\frac{\pi}{2}}\mid C_{2y})$ \\
\hline
$2'$ & $C_{2y}T$ & 3 & $\tilde{\phi}_{1}:$ $(e^{-i\sigma_y\frac{\pi}{2}-i\tau_y\frac{\pi}{2}}\mid C_{2y}T)$ \\
\end{longtable}
\endgroup


\subsection{4e SC-II: $O_f=\langle U(1)_s,\;e^{-i\tau_z{\pi\over2}}\rangle$.}
\begingroup\scriptsize
\begin{longtable}{@{}>{\centering\arraybackslash}p{0.08\textwidth} >{\centering\arraybackslash}p{0.15\textwidth} >{\centering\arraybackslash}p{0.055\textwidth} >{\RaggedRight\arraybackslash}p{0.69\textwidth}@{}}

\hline\hline
MPG & Generators & \makecell{\#\\SCPGs} & Representatives and images on generators\\
\hline
\endfirsthead

\hline\hline
MPG & Generators & \makecell{\#\\SCPGs} & Representatives and images on generators\\
\hline
\endhead

\hline\hline
\endfoot
$1$ & $\emptyset$ & 1 & $\tilde{\phi}_{1}:$ $\text{trivial}$ \\
\hline
$1^{\prime}$ & $T$ & 6 & $\tilde{\phi}_{1}:$ $\left(e^{-i\sigma_y\frac{\pi}{2}} e^{-i\tau_y\frac{\pi}{2}}\mid T\right)$ \\
$1^{\prime}$ & $T$ & 6 & $\tilde{\phi}_{2}:$ $\left(e^{-i\sigma_y\frac{\pi}{2}} e^{-i\tau_z \frac{\pi}{4}} e^{-i\tau_y\frac{\pi}{2}}\mid T\right)$ \\
$1^{\prime}$ & $T$ & 6 & $\tilde{\phi}_{3}:$ $\left(e^{-i\sigma_y\frac{\pi}{2}} e^{-i\tau_x\frac{\pi}{2}} e^{-i\tau_y\frac{\pi}{2}}\mid T\right)$ \\
$1^{\prime}$ & $T$ & 6 & $\tilde{\phi}_{4}:$ $\left(e^{-i\sigma_z\frac{\pi}{2}} e^{-i\tau_y\frac{\pi}{2}}\mid T\right)$ \\
$1^{\prime}$ & $T$ & 6 & $\tilde{\phi}_{5}:$ $\left(e^{-i\sigma_z\frac{\pi}{2}} e^{-i\tau_z \frac{\pi}{4}} e^{-i\tau_y\frac{\pi}{2}}\mid T\right)$ \\
$1^{\prime}$ & $T$ & 6 & $\tilde{\phi}_{6}:$ $\left(e^{-i\sigma_z\frac{\pi}{2}} e^{-i\tau_x\frac{\pi}{2}} e^{-i\tau_y\frac{\pi}{2}}\mid T\right)$ \\
\hline
$2$ & $C_{2,y}$ & 6 & $\tilde{\phi}_{1}:$ $\left(1\mid C_{2,y}\right)$ \\
\end{longtable}
\endgroup


\subsection{4e SC-III: $O_f=\big\langle U(1)_s, e^{-i\tau_z{\pi\over 2}}, e^{-i(\sigma_x+{\tau_z\over 2}){\pi\over 2}} \big\rangle$}
\begingroup\scriptsize
\begin{longtable}{@{}>{\centering\arraybackslash}p{0.08\textwidth} >{\centering\arraybackslash}p{0.15\textwidth} >{\centering\arraybackslash}p{0.055\textwidth} >{\RaggedRight\arraybackslash}p{0.69\textwidth}@{}}

\hline\hline
MPG & Generators & \makecell{\#\\SCPGs} & Representatives and images on generators\\
\hline
\endfirsthead

\hline\hline
MPG & Generators & \makecell{\#\\SCPGs} & Representatives and images on generators\\
\hline
\endhead

\hline\hline
\endfoot
$1$ & --- & 1 & $\tilde{\phi}_{1}:$ $\text{trivial}$ \\
\hline
$1'$ & $T$ & 3 & $\tilde{\phi}_{1}:$ $(\J\mid T)$ \\
$1'$ & $T$ & 3 & $\tilde{\phi}_{2}:$ $(e^{-i\tau_z \frac{\pi}{4}}\,\J\mid T)$ \\
$1'$ & $T$ & 3 & $\tilde{\phi}_{3}:$ $(e^{-i\tau_x\frac{\pi}{2}}\,\J\mid T)$ \\
\hline
$2$ & $C_{y}$ & 3 & $\tilde{\phi}_{1}:$ $(1\mid C_{y})$ \\
$2$ & $C_{y}$ & 3 & $\tilde{\phi}_{2}:$ $(e^{-i\tau_z \frac{\pi}{4}}\mid C_{y})$ \\
$2$ & $C_{y}$ & 3 & $\tilde{\phi}_{3}:$ $(e^{-i\tau_x\frac{\pi}{2}}\mid C_{y})$ \\
\hline
$21'$ & $C_{y}$, $T$ & 10 & $\tilde{\phi}_{1}:$ $(1\mid C_{y})$, $\allowbreak\,(\J\mid T)$ \\
\end{longtable}
\endgroup


\subsection{4e SC-IV: $O_f=\big\langle U(1)_s,\ e^{-i\tau_z\frac{\pi}{2}},\ e^{-i(\sigma_x+\tau_x)\frac{\pi}{2}}\big\rangle$}

\begingroup\scriptsize
\begin{longtable}{@{}>{\centering\arraybackslash}p{0.08\textwidth} >{\centering\arraybackslash}p{0.15\textwidth} >{\centering\arraybackslash}p{0.055\textwidth} >{\RaggedRight\arraybackslash}p{0.69\textwidth}@{}}

\hline\hline
MPG & Generators & \makecell{\#\\SCPGs} & Representatives and images on generators\\
\hline
\endfirsthead

\hline\hline
MPG & Generators & \makecell{\#\\SCPGs} & Representatives and images on generators\\
\hline
\endhead

\hline\hline
\endfoot
$ 1 $ & $ \varnothing $ & 1 & $\tilde{\phi}_{1}:$ $\text{trivial}$ \\
\hline
$ 1' $ & $ T $ & 4 & $\tilde{\phi}_{1}:$ $(\U\mid T)$ \\
$ 1' $ & $ T $ & 4 & $\tilde{\phi}_{2}:$ $(\SU\mid T)$ \\
$ 1' $ & $ T $ & 4 & $\tilde{\phi}_{3}:$ $(\RU\mid T)$ \\
$ 1' $ & $ T $ & 4 & $\tilde{\phi}_{4}:$ $(\SRU\mid T)$ \\
\hline
$ 2 $ & $ C_{y} $ & 4 & $\tilde{\phi}_{1}:$ $(\Id\mid C_{y})$ \\
$ 2 $ & $ C_{y} $ & 4 & $\tilde{\phi}_{2}:$ $(\Sx\mid C_{y})$ \\
$ 2 $ & $ C_{y} $ & 4 & $\tilde{\phi}_{3}:$ $(\Rz\mid C_{y})$ \\
\end{longtable}
\endgroup

\section{Projective Twists $H^2_\varphi(G_b, \mathcal Z(O_f))$, invariants, and irreducible representations of $G_b$} 

\subsection{The Second Group Cohomology}

The extension include the projective twist whose factor systems are classified by the second group cohomological $H^2_\varphi(G_b, A)$, where $A$ is the center of $O_f$, namely $A=\mathcal Z(O_f)$, and $\varphi$ is the action on the coefficient group $A$ induced by the conjugation of $\phi(G_b)$ and will be discussed in section \ref{sec:action}. The group $O_f$ is the kernel of a spin-charge group SCG as an extensions of the `physical' group $G_b$.

\subsubsection{$A=\mathcal Z(O_f)$ as coefficient group}
We mainly consider the following ten groups as $O_f$: $Z_2^f,  U(1)_{sc}\times Z_2^f, Z_{4c}, U(1)_s, U(1)_c, [E+e^{-i({\sigma_x\over2} +{\tau_z\over2})\pi}]U(1)_s \cong Pin(2), U(1)_s\star Z_{4c}, [E+e^{-i({\sigma_x\over2} +{\tau_z\over4})\pi}]U(1)_s\star Z_{4c}, [E+e^{-i({\sigma_x\over2} +{\tau_x\over2})\pi}]U(1)_s\star Z_{4c}, U(1)_s\star U(1)_c,  [E+e^{-i({\sigma_x\over2} +{\tau_x\over2})\pi}]U(1)_s\star U(1)_{c}$. Their centers are given by:
\Beq
&&\mathcal Z(Z_2^f)\cong Z_2=\{1,-1\},\\ 
&&\mathcal Z(U(1)_{sc}\times Z_2^f)\cong U(1)\times Z_2,\\ 
&&\mathcal Z(Z_{4c})\cong Z_{4}=\{\pm1, e^{\pm i{\tau_z}{\pi\over2}}\},\\ 
&&\mathcal Z(U(1)_s)=\mathcal Z(U(1)_c)\cong Spin(2),\\ 
&&\mathcal Z\big([E+e^{-i({\sigma_x\over2} +{\tau_z\over2})\pi}]U(1)_s\big)\cong Z_2,\\ 
&&\mathcal Z\big(U(1)_s\star Z_{4c}\big)\cong U(1)\times Z_4, \\
&&\mathcal Z\big([E+e^{-i({\sigma_x\over2} +{\tau_z\over4})\pi}]U(1)_s\star Z_{4c}\big)\cong Z_4, \\
&&\mathcal Z\big([E+e^{-i({\sigma_x\over2} +{\tau_x\over2})\pi}]U(1)_s\star Z_{4c}\big)\cong Z_2\times Z_2, \\
&&\mathcal Z\big(U(1)_s\star U(1)_c\big)\cong U(1)\times Spin(2),  \\
&&\mathcal Z\big([E+e^{-i({\sigma_x\over2} +{\tau_x\over2})\pi}]U(1)_s\star U(1)_{c}\big)\cong Z_2\times Z_2.
\Eeq
More information can be found in Tab.~\ref{tab:Of_Infor}.

\subsubsection{Decoupling of cohomological twist}

When $\mathcal Z(O_f)$ is a product group $A=A_1\times A_2$ and if the action of $G_b$ on $A$ does not include exchanging the generators, then the group cohomology can be decoupled,
\Beq
&&H^2_\varphi(G_b, U(1)\times Z_4) = H^2_{\varphi_1}(G_b, U(1)) \times H^2_{\varphi_2}(G_b, Z_4),\\
&&H^2_\varphi(G_b, U(1)\times Spin(2)) = H^2_{\varphi_1}(G_b, U(1)) \times H^2_{\varphi_2}(G_b, U(1)),\\
&&H^2(G_b, Z_2\times Z_2) = H^2(G_b, Z_2) \times H^2(G_b, Z_2).
\Eeq
In the second equation we used the conclusion that $H^2_{\varphi}(G_b, Spin(2)) =H^2_{\varphi}(G_b, U(1)).$ Here $\varphi_1$ and $\varphi_2$ stand for the actions of $G_b$ on coefficient groups $A_1$ and $A_2$ respectively. In most cases the subscripts $1, 2$ correspond to the spin/charge sectors. If $\varphi_i$ (with $i=1,2$) is nontrivial, then half of the elements in $G_b$ reverse the coefficients (elements of groups $U(1)$ and $Z_4$) on which they act, namely, there exist certain $g\in G_b$ such that $\varphi_i(g)\cdot \omega =\omega^{-1}$ for any $\omega\in A_i$.

For the coefficient group $$Z_2\times Z_2= \{E, P_f, e^{-i({\sigma_z\over2}+{\tau_z\over2})\pi}, P_fe^{-i({\sigma_z\over2}+{\tau_z\over2})\pi}\},$$ $G_b$ can act nontrivially by exchanging the generators. 
If $G_b = H + uH$ and $\phi_u$ exchanges the generators via the relation $\phi_u e^{-i({\sigma_z\over2}+{\tau_z\over2})\pi}\phi_u^{-1} =P_fe^{-i({\sigma_z\over2}+{\tau_z\over2})\pi}$ while $\phi(H)$ has trivial action, then the cohomology  generally reduces to 
\Beq
H^2_\varphi(G_b, Z_2\times Z_2) \cong H^2(H, Z_2).
\Eeq
Therefore, the projective twists are composed of the factor systems with simple coefficient groups $A=U(1), Z_2, Z_4$.

\subsubsection{Action of $\varphi(G_b)$ on the coefficient group $A$}\label{sec:action}

The coefficient group $A$ is a $G_b$ module with action $\varphi$, where $\varphi$ is determined by the map $\tilde \phi: G_b\to N(O_f)/O_f$ or simply by the representative $\phi(G_b)\subset N(O_f)$. For unitary element $l\in G_b$, the action of $l$ on $A$ is defined by $\tilde\phi_l$ or $\phi_l\in N(O_f)$. Supposing $\omega_2(l_1,l_2)\in A$, then one has
\beq\label{unitaction}
\varphi(l)\cdot \omega_2(l_1,l_2) = \phi_l \omega_2(l_1,l_2) \phi_l^{-1}.
\eeq
Notice that there exist a more-to-one correspondence from $\tilde \phi$ to $\varphi$.

For anti-unitary groups, we adopt the definition of time reversal operation with $T = e^{-i(\sigma_y+\tau_y){\pi\over2}}K$ such that $T$ commutes with all of the spin-charge operations in $SO(4)$. In this convention, $\phi_T$ is defined on top of $e^{-i(\sigma_y+\tau_y){\pi\over2}}$ such that the action of $T$ on $A$ is fully determined by $\phi_T\in N(O_f)$,
\Beq
\varphi(T)\cdot \omega_2(l_1,l_2) &=& \phi_Te^{-i(\sigma_y+\tau_y){\pi\over2}}K \omega_2(l_1,l_2)K e^{i(\sigma_y+\tau_y){\pi\over2}}\phi_T^{-1}\\ 
&=& \phi_T \omega_2(l_1,l_2) \phi_T^{-1},
\Eeq
which agrees with the action of unitary elements in (\ref{unitaction}). Similarly, for any anti-unitary element $l\in G_b$, $\phi_l$ is defined on top of $e^{-i(\sigma_y+\tau_y){\pi\over2}}$ such that (\ref{unitaction}) applies for anti-unitary groups. 

{\it Sometimes it is more convenient to adopt another convention for the section of anti-unitary elements $l\in G_b$ with $\phi''_l=\phi_l e^{-i(\sigma_y+\tau_y){\pi\over2}}$. The advantage of this convention is that $\phi''_l$ is the complete spin-charge operation associated with $l$. For instance, when $\phi''_T=E$, the `pure time-reversal' operation reads $T''=K$. Notice that the time-reversal operation $T''$ in this new convention is not commuting with $SO(4)$ anymore.}

If an unitary group $G_b$ and an anti-unitary group $G_b'$ have the same multiplication table, they are said to be pseudo-isomorphic. If they have the same action on a coefficient group $A$, then one has
$$
H^2_\varphi(G_b',A) =H^2_\varphi(G_b,A).
$$
Namely, the classification only depends on the action $\varphi$, no matter if the group is unitary or anti-unitary. However, as discussed in section \ref{sec:rep}, the representation theory for anti-unitary groups are very different from unitary ones even if their invariants (i.e. projective classes) are the same.

\subsection{Invariants of $H^2_\varphi(G_b, A)$}

In the following, we will provide the invariants (namely the generators of the second group cohomology $H^2_\varphi(G_b, A)$ with $G_b$ abstract groups, $A$ abelian coefficient groups $A=U(1), Z_4, Z_2, Z_2\times Z_2$ and $\varphi$ the various actions of $G_b$ on $A$.  From these invariants, one can obtain the invariants of the original cohomology $H^2_\varphi\big(G_b, \mathcal Z(O_f))$.


When 
$\varphi$ is trivial action on the coefficient groups $A=U(1), Z_4, Z_2$, then the invariants for the projective classes are given in Tab.\ref{tab:unit_invar}. 

When the nontrivial action of $\varphi$ is 
taking inversion (or complex conjugation) for coefficient groups $A=U(1), Z_4$, then the invariants for the projective classes are given in Tab.\ref{tab:antiunit_invar}. 

When 
the nontrivial action of $\varphi$ is exchanging the generators of the coefficient group $A=Z_2\times Z_2$, then the invariants for the projective classes are given in Tab.\ref{tab:exchange_invar}.

\begin{table}[tbp]
\caption{Invariants of $H^2(G_b, A)$ with trivial action.}\label{tab:unit_invar}
\centering
\resizebox{0.9\textwidth}{!}{
\renewcommand{\arraystretch}{2}
\begin{tabular}{|c|c|c|c|c|}
\hline
$ {G}_b$ & generators & coefficient $A$ & $H^2(G_b,A)$ & invariants \\ 
\hline 
\multirow{3}{*}{$Z_2$} & \multirow{3}{*}{$a; a^2=E$}  & $U(1)$ & $\mathbb Z_1$ & $-$ \\
\cline{3-5}
 & &  $Z_2$  &$\mathbb Z_2$ &  $\hat{a}^2=\omega_2(a,a)\in\mathbb Z_2$ \\
\cline{3-5}
& & $Z_4$    &$\mathbb Z_2$ &  $\hat a^4 = \omega_2(a,a)^2\in\mathbb Z_2$ \\
\hline 
\multirow{3}{*}{$Z_2\times Z_2$} & \multirow{3}{*}{$a,b; a^2=b^2=E$}   
& $U(1)$  &$\mathbb Z_2$ &  $\hat a\hat b\widehat {a^{-1}}\widehat {b^{-1}}={\omega_2(a,b)\over\omega_2(b,a)}\in\mathbb Z_2$ \\
\cline{3-5} 
& &  $Z_2$ &$\mathbb Z_2^3$ &  ${\omega_2(a,b)\over\omega_2(b,a)},\ \omega_2(a,a),\ \omega_2(b,b)\in\mathbb Z_2$ \\
\cline{3-5}
& & $Z_4$  &$\mathbb Z_2^3$ &  ${\omega_2(a,b)\over\omega_2(b,a)},\ \omega_2(a,a)^2,\ \omega_2(b,b)^2\in\mathbb Z_2$ \\
\hline 
\multirow{3}{*}{$Z_3\rtimes Z_2$} & \multirow{3}{*}{$r, a; r^3=a^2=E,ra=ar^2$}  & $U(1)$ &$\mathbb Z_1$ &  $-$ \\
\cline{3-5}
& &  $Z_2$  &$\mathbb Z_2$ &  $\omega_2(a,a)\in \mathbb Z_2$ \\
\cline{3-5}
& & $Z_4$  & $\mathbb Z_2$ &  $\omega_2(a,a)^2\in \mathbb Z_2$ \\
\hline
\multirow{3}{*}{$Z_4$} & \multirow{3}{*}{$ r; r^4=E$}  & $U(1)$ &$\mathbb Z_1$ & $-$ \\
\cline{3-5}
& &  $Z_2$  &$\mathbb Z_2$ & $(\widehat{r^2})^2=\omega(r^2,r^2)\in\mathbb Z_2$ \\ 
\cline{3-5}
& & $Z_4$ & ${ \mathbb Z_4}$ & $\hat{r}^4=\omega(r,r)^2 \omega(r^2,r^2)\in{ \mathbb Z_4}$ \\
\hline
\multirow{3}{*}{$Z_4\rtimes Z_2$} & \multirow{3}{*}{$ r,a; r^4=a^2=E, ra=ar^3$}  & $U(1)$ & $\mathbb Z_2$ & ${\omega_2(r^2,a)\over\omega_2(a,r^2)} \in \mathbb Z_2$ \\
\cline{3-5}
& &  $Z_2$  &$\mathbb Z_2^3$ &  ${\omega_2(r^2,a)\over\omega_2(a,r^2)},\omega_2(a,a),\omega_2(ar,ar)\in\mathbb Z_2$ \\
\cline{3-5}
& & $Z_4$ & $\mathbb Z_2^3$ & ${\omega_2(r^2,a)\over\omega_2(a,r^2)},\omega_2(a,a)^2,\omega_2(ar,ar)^2\in\mathbb Z_2$ \\
\hline
\multirow{3}{*}{$Z_6$} & \multirow{3}{*}{$ r; r^6=E$}  & $U(1)$ &$\mathbb Z_1$ & $-$ \\
\cline{3-5}
& &  $Z_2$  &$\mathbb Z_2$ & $(\hat{r^3})^2=\omega_2(r^3,r^3)\in\mathbb Z_2$ \\ 
\cline{3-5}
& & $Z_4$ & $\mathbb Z_2$ & $(\hat{r^3})^4=\omega_2(r^3,r^3)^2\in\mathbb Z_2$ \\
\hline
\multirow{3}{*}{$Z_6\rtimes Z_2$} & \multirow{3}{*}{$ r,a; r^6=a^2=E, ra=ar^5$}  & $U(1)$ & $\mathbb Z_2$ & ${\omega_2(r^3,a)\over\omega_2(a,r^3)} \in \mathbb Z_2$ \\
\cline{3-5}
& &  $Z_2$  &$\mathbb Z_2^3$ &  $ {\omega_2(r^3,a)\over\omega_2(a,r^3)}, \omega_2(r^3,r^3),\omega_2(a,a)\in\mathbb Z_2$ \\
\cline{3-5}
& & $Z_4$ & $\mathbb Z_2^3$ & ${\omega_2(r^3,a)\over\omega_2(a,r^3)}, \omega_2(r^3,r^3)^2,\omega_2(a,a)^2 \in\mathbb Z_2$ \\
\hline
\hline
\multirow{3}{*}{$Z_2\times Z_2\times Z_2$} & \multirow{3}{*}{$ a,b,t; a^2=b^2=t^2=E$}  & $U(1)$ & $\mathbb Z_2^3$ & ${\omega_2(a,b)\over\omega_2(b,a)}, {\omega_2(a,t)\over\omega_2(t,a)},{\omega_2(b,t)\over\omega_2(t,b)} \in \mathbb Z_2$ \\
\cline{3-5}
& &  $Z_2$  &$\mathbb Z_2^6$ & ${\omega_2(a,b)\over\omega_2(b,a)}, {\omega_2(a,t)\over\omega_2(t,a)},{\omega_2(b,t)\over\omega_2(t,b)}$,  $\omega_2(a,a),\omega_2(b,b),\omega_2(t,t)\in \mathbb Z_2$ \\
\cline{3-5}
& & $Z_4$ & $\mathbb Z_2^6$ &${\omega_2(a,b)\over\omega_2(b,a)}, {\omega_2(a,t)\over\omega_2(t,a)},{\omega_2(b,t)\over\omega_2(t,b)}$, $\omega_2(a,a)^2,\omega_2(b,b)^2,\omega_2(t,t)^2\in \mathbb Z_2$ \\
\hline
\multirow{3}{*}{$Z_4\times Z_2$} & \multirow{3}{*}{$ r,t; r^4=t^2=E, rt=t r$}  & $U(1)$ & $\mathbb Z_2$ & ${\omega_2(r,t)\over\omega_2(t,r)} \in \mathbb Z_2$ \\
\cline{3-5}
& &  $Z_2$  &$\mathbb Z_2^3$ &  ${\omega_2(r,t)\over\omega_2(t,r)},\omega_2(t,t), \omega_2(r^2,r^2)\in\mathbb Z_2$ \\
\cline{3-5}
& & $Z_4$ & $\mathbb Z_2^2\times { \mathbb Z_4}$ & $ {\omega_2(r,t)\over\omega_2(t,r)},\omega_2(t,t)^2\in\mathbb Z_2,\ \hat r^4=\omega(r,r)^2 \omega_2(r^2,r^2)\in{ \mathbb Z_4}$ \\
\hline
\multirow{3}{*}{$(Z_4\rtimes Z_2)\times Z_2$} & \multirow{3}{*}{$ r,a,t; r^4=a^2=t^2=E, ra=ar^3$}  & $U(1)$ & $\mathbb Z_2^3$ & ${\omega_2(r^2,a)\over\omega_2(a,r^2)},{\omega_2(r,t)\over\omega_2(t,r)},  {\omega_2(a,t)\over\omega_2(t,a)} \in \mathbb Z_2$ \\
\cline{3-5}
& &  $Z_2$  &$\mathbb Z_2^6$ &  ${\omega_2(r^2,a)\over\omega_2(a,r^2)},{\omega_2(r,t)\over\omega_2(t,r)},  {\omega_2(a,t)\over\omega_2(t,a)},  \omega_2(a,a),\omega_2(ar,ar),\omega_2(t,t)\in\mathbb Z_2$ \\
\cline{3-5}
& & $Z_4$ & $\mathbb Z_2^6$ & ${\omega_2(r^2,a)\over\omega_2(a,r^2)},{\omega_2(r,t)\over\omega_2(t,r)},  {\omega_2(a,t)\over\omega_2(t,a)},  \omega_2(a,a)^2,\omega_2(ar,ar)^2,\omega_2(t,t)^2\in\mathbb Z_2$ \\
\hline 
\multirow{3}{*}{$Z_6\times Z_2$} & \multirow{3}{*}{$ r,t; r^6=t^2=E, rt=t r$}  & $U(1)$ & $\mathbb Z_2$ & ${\omega_2(r,t)\over\omega_2(t,r)} ={\omega_2(r^3,t)\over\omega_2(t,r^3)} \in \mathbb Z_2$ \\
\cline{3-5}
& &  $Z_2$  &$\mathbb Z_2^3$ &  ${\omega_2(r^3,t)\over\omega_2(t,r^3)},\omega_2(t,t), \omega_2(r^3,r^3)\in\mathbb Z_2$ \\
\cline{3-5}
& & $Z_4$ & $\mathbb Z_2^3$ & $ {\omega_2(r^3,t)\over\omega_2(t,r^3)}, \omega_2(t,t)^2, \omega_2(r^3,r^3)^2\in\mathbb Z_2$ \\
\hline
\multirow{3}{*}{$(Z_6\rtimes Z_2)\times Z_2$} & \multirow{3}{*}{$ r,a,t; r^6=a^2=t^2=E, ra=ar^5$}  & $U(1)$ & $\mathbb Z_2^3$ & ${\omega_2(r^3,a)\over\omega_2(a,r^3)},{\omega_2(r^3,t)\over\omega_2(t,r^3)},  {\omega_2(a,t)\over\omega_2(t,a)} \in \mathbb Z_2$ \\
\cline{3-5}
& &  $Z_2$  &$\mathbb Z_2^6$ &  ${\omega_2(r^3,a)\over\omega_2(a,r^3)},{\omega_2(r^3,t)\over\omega_2(t,r^3)},  {\omega_2(a,t)\over\omega_2(t,a)},  \omega_2(r^3,r^3),\omega_2(a,a),\omega_2(t,t)\in\mathbb Z_2$ \\
\cline{3-5}
& & $Z_4$ & $\mathbb Z_2^6$ & ${\omega_2(r^3,a)\over\omega_2(a,r^3)},{\omega_2(r^3,t)\over\omega_2(t,r^3)},  {\omega_2(a,t)\over\omega_2(t,a)},  \omega_2(r^3,r^3)^2,\omega_2(a,a)^2,\omega_2(t,t)^2\in\mathbb Z_2$ \\
\hline
\end{tabular}
}
\end{table}


\begin{table}[tbp]
\caption{Invariants of $H^2_\varphi(G_b, A)$ where $H$ acts trivially on $A$ and the elements in $G_b - H$ act on $A$ by taking inversion.} \label{tab:antiunit_invar}
\centering
\resizebox{1\textwidth}{!}{
\renewcommand{\arraystretch}{2}
\begin{tabular}{|c|c|c|c|c|}
\hline
$ {G}_b$ &action-free group $H$ and $G_b-H$ & $A$ & $H^2(G_b,A)$ & invariants \\ 
\hline 
\multirow{2}{*}{$Z_2$} & {$Z_1;$}  & $U(1)$ & $\mathbb Z_2$ & $\hat a^2 = \omega_2(a,a)$ \\
\cline{3-5}
& \multirow{1}{*}{$\{ a\}$} & $Z_4$    &$\mathbb Z_2$ &  $\hat a^2 = \omega_2(a,a) $ \\
\hline 
\multirow{2}{*}{$Z_2\times Z_2$} & {$Z_2=\{E, ab\};$}   
& $U(1)$  &$\mathbb Z_2^2$ &  $\hat {a}^2 = \omega_2(a,a) , \ \hat{b}^2=\omega_2(b,b) $ \\
\cline{3-5}
& \multirow{1}{*}{$\{a, b\}$} & $Z_4$  &$\mathbb Z_2^3$ &  $\hat {a}^2 =\omega_2(a,a), \ \hat {b}^2 =\omega_2(b,b),\  (\widehat{ab})^4=\omega_2(ab,ab)^2$ \\
\hline 
\multirow{2}{*}{$Z_3\rtimes Z_2$} & {$Z_3=\{E,r,r^2\}; $}  & $U(1)$ &$\mathbb Z_2$ &  $\hat a^2=\omega_2(a,a)$ \\
\cline{3-5}
& \multirow{1}{*}{$\{a, ar, ar^2\}$} & $Z_4$  & $\mathbb Z_2$ &  $\hat a^2=\omega_2(a,a)$ \\
\hline
\multirow{2}{*}{$Z_4$} & {$Z_2=\{E,r^2\};$}  & $U(1)$ &$\mathbb Z_2$ & $\hat r^4 = \omega_2(r,r)^2 \omega_2(r^2,r^2)$ \\
\cline{3-5}
& \multirow{1}{*}{$\{r, r^3\}$} & $Z_4$ & $ \mathbb Z_2$ & $\hat{r}^4=\omega_2(r,r)^2 \omega(r^2,r^2)$ \\
\hline
\multirow{4}{*}{$Z_4\rtimes Z_2$} & {$Z_2\times Z_2=\{E,r^2,a,ar^2\};$}  & $U(1)$ & $\mathbb Z_2^2$ & ${\omega_2(r^2,a)\over\omega_2(a,r^2)},\ \widehat {ar}^2 = \omega_2(ar,ar)$  \\
\cline{3-5}
& \multirow{1}{*}{$\{r, r^3, ar, ar^3\}$} & $Z_4$ & $\mathbb Z_2^3$ & ${\omega_2(r^2,a)\over\omega_2(a,r^2)},\ \widehat {ar}^2 =\omega_2(ar,ar), \  \omega_2(a,a)^2 $\\
\cline{2-5} 
& {$Z_4=\{E,r,r^2,r^3\};  $}  & $U(1)$ & $\mathbb Z_2^2$ & $\hat a^2 = \omega_2(a,a), \ \widehat{ar}^2 =\omega_2(ar, ar)$ \\
\cline{3-5}
& \multirow{1}{*}{$\{a, ar, ar^2, ar^3\}$} & $Z_4$ & $\mathbb Z_2^2\times { \mathbb Z_4}$ & $\hat a^2 = \omega_2(a,a), \ \widehat{ar}^2 =\omega_2(ar, ar),\  \hat r^4=\omega_2(r,r)^2\omega_2(r^2,r^2)\in{ \mathbb Z_4}$ \\
\hline
\multirow{2}{*}{$Z_6$} & {$ Z_3=\{E,r^2,r^4\};$}  & $U(1)$ &$\mathbb Z_2$ & $(\hat{r^3})^2=\omega_2(r^3,r^3)$ \\
\cline{3-5}
& \multirow{1}{*}{$\{r,r^3,r^5\}$} & $Z_4$ & $\mathbb Z_2$ & $(\hat{r^3})^2=\omega_2(r^3,r^3)$ \\
\hline
\multirow{4}{*}{$Z_6\rtimes Z_2$} & {$Z_3\rtimes Z_2=\{E,r^2,r^4,a,ar^2,ar^4\};$}  & $U(1)$ & $\mathbb Z_2^2$ & $(\widehat{r^3})^2 = {\omega_2(r^3,r^3)},\ \widehat {ar}^2 = \omega_2(ar,ar)$  \\
\cline{3-5}
& \multirow{1}{*}{$\{r, r^3,  r^5, ar, ar^3,a r^5\}$} & $Z_4$ & $\mathbb Z_2^3$ & $(\widehat{r^3})^2 = {\omega_2(r^3,r^3)},\ \widehat {ar}^2 =\omega_2(ar,ar), \  \hat a^4=\omega_2(a,a)^2 $\\
\cline{2-5} 
& {$Z_6=\{E,r,r^2,r^3,r^4,r^5\};$}  & $U(1)$ & $\mathbb Z_2^2$ & $\hat a^2 = \omega_2(a,a), \ (\widehat{ar})^2 =\omega_2(ar, ar)$ \\
\cline{3-5}
& \multirow{1}{*}{$\{a, ar, ar^2, ar^3,ar^4,ar^5\}$} & $Z_4$ & $\mathbb Z_2^3$ & $\hat a^2 = \omega_2(a,a), \ (\widehat{ar})^2 =\omega_2(ar, ar),\ {(\widehat {r^3})^4=\omega_2(r^3,r^3)^2}$ \\
\hline
\hline
\multirow{2}{*}{$(Z_2\times Z_2)\times Z_2$} & {$ Z_2\times Z_2=\{E,a,b,ab\};$}  & $U(1)$ & $\mathbb Z_2^4$ & ${\omega_2(a,b)\over\omega_2(b,a)}, \hat t^2 = \omega_2(t,t), (\widehat{ta})^2=\omega_2(ta,ta), (\widehat{tb})^2=\omega_2(tb,tb) $ \\
\cline{3-5}
& \multirow{1}{*}{$\{t,ta,tb,tab\}$} & $Z_4$ & $\mathbb Z_2^6$ &${\omega_2(a,b)\over\omega_2(b,a)}, \hat t^2 = \omega_2(t,t), (\widehat{ta})^2=\omega_2(ta,ta), (\widehat{tb})^2=\omega_2(tb,tb), \hat a^4=\omega_2(a,a)^2,\hat b^4=\omega_2(b,b)^2$ \\%
\hline
\multirow{4}{*}{$Z_4\times Z_2$} & {$Z_2\times Z_2=\{E,r^2,t,tr^2\};$}  & $U(1)$ & $\mathbb Z_2^2$ &  $ {\omega_2(t,t)\omega_2(r, t))\over\omega_2(t,r)},\ \hat {r}^4 = \omega_2(r,r)^2\omega_2(r^2,r^2)$  \\ 
\cline{3-5}
& \multirow{1}{*}{$\{r, r^3, tr, tr^3\}$} & $Z_4$ & $\mathbb Z_2^3$ & ${\omega_2(t,t)\omega_2(r,t))\over\omega_2(t,r)},\ \widehat {r}^4 = \omega_2(r,r)^2\omega_2(r^2,r^2),\ \hat{t}^4=\omega_2(t,t)^2$  \\ 
\cline{2-5} 
& {$Z_4=\{E,r,r^2,r^3\};$}  & $U(1)$ & $\mathbb Z_2^2$ & $\hat t^2 = \omega_2(t,t), \ (\widehat{tr^2})^2 =\omega_2(tr^2, tr^2)$ \\
\cline{3-5}
& \multirow{1}{*}{$\{t, tr, tr^2, tr^3\}$} & $Z_4$ & $\mathbb Z_2^3$ & $\hat t^2 = \omega_2(t,t), \ (\widehat{tr^2})^2 =\omega_2(tr^2, tr^2),\ \ \hat{r}^4=\omega_2(r,r)^2\omega_2(r^2,r^2) $ \\
\hline
\multirow{6}{*}{$(Z_4\rtimes Z_2)\times Z_2$} & {$Z_2\times Z_2\times Z_2=\{E,r^2\}\times\{E,a\}\times\{E,t\};$}  & $U(1)$ & $\mathbb Z_2^4$ & $ {\omega_2(r^2,a)\over\omega_2(a,r^2)},\ {\omega_2(t,a)\over\omega_2(a,t)},\ \widehat {ar}^2 = \omega_2(ar,ar),\  \widehat {tar}^2 = \omega_2(tar,tar)$  \\
\cline{3-5}
& \multirow{1}{*}{$\{r, r^3, ar, ar^3, tr, tr^3, tar, tar^3\}$} & $Z_4$ & $\mathbb Z_2^6$ & $ {\omega_2(r^2,a)\over\omega_2(a,r^2)},\ {\omega_2(t,a)\over\omega_2(a,t)},\ \widehat {ar}^2 = \omega_2(ar,ar),\ \widehat {tar}^2 = \omega_2(tar,tar),\ \omega_2(a,a)^2, \ \omega_2(ta,ta)^2$  \\
\cline{2-5} 
& {$Z_4\times Z_2=\{E,r,r^2,r^3\}\times\{E,t\};$}  & $U(1)$ & $\mathbb Z_2^4$ & ${{\omega_2(r,t)\over\omega_2(t,r)}}, \ \hat{a}^2 =\omega_2(a, a),\ \widehat{ar}^2=\omega_2(ar,ar), \ \widehat{ta}^2=\omega_2(ta,ta)$ \\
\cline{3-5}
& \multirow{1}{*}{$\{a, ar, ar^2, ar^3,ta, tar, tar^2, tar^3\}$} & $Z_4$ & $\mathbb Z_2^5\times{\mathbb Z_4}$ & ${{\omega_2(r,t)\over\omega_2(t,r)}}, \hat{a}^2\!\!=\!\omega_2(a, a), \widehat{ar}^2\!\!=\!\omega_2(ar,ar), \ \widehat{ta}^2\!\!=\!\omega_2(ta,ta), \hat t^4\!\!=\!\omega_2(t,t)^2, \hat r^4\!\!=\!\omega_2(r,r)^2\omega_2(r^2,r^2)\!\in\!{\mathbb Z_4}$ \\
\cline{2-5} 
& {$Z_4\rtimes Z_2=\{E,r,r^2,r^3\}\rtimes\{E,a\};$}  & $U(1)$ & $\mathbb Z_2^4$ & ${\omega_2(r^2,a)\over\omega_2(a,r^2)}, \hat t^2 = \omega_2(t,t), \ \widehat{ta}^2 =\omega_2(ta, ta), \ (\widehat{tar})^2 =\omega_2(tar, tar)$ \\ 
\cline{3-5}
& \multirow{1}{*}{$\{t, tr, tr^2, tr^3, ta,tar,tar^2,tar^3\}$} & $Z_4$ & $\mathbb Z_2^6$ & ${\omega_2(r^2,a)\over\omega_2(a,r^2)}, \hat t^2\! =\! \omega_2(t,t), \ \widehat{ta}^2 \!=\!\omega_2(ta, ta),\ (\widehat{tar})^2 \!=\!\omega_2(tar, tar),\ \hat a^4\!=\!\omega_2(a,a)^2, \ \widehat{ar}^4\!=\!\omega_2(ar, ar)^2$\\ 
\hline 
\multirow{2}{*}{$Z_6\times Z_2$} & {$Z_6=\{E,r,r^2,r^3,r^4,r^5\};$}  & $U(1)$ & $\mathbb Z_2^2$ & $  \hat t^2 = \omega_2(t,t),\ (\widehat{tr^3})^2= \omega_2(tr^3,tr^3) $ \\
\cline{3-5}
& \multirow{1}{*}{$\{t,tr,tr^2,tr^3,tr^4,tr^5\}$} & $Z_4$ & $\mathbb Z_2^3$ & $ \hat t^2 \!=\!\omega_2(t,t),\ (\widehat{tr^3})^2\!=\! \omega_2(tr^3,tr^3),\ (\widehat{r^3})^4\!=\!\omega_2(r^3,r^3)^2 $ \\
\hline
\multirow{4}{*}{$(Z_6\rtimes Z_2)\times Z_2$} & {$Z_6\times Z_2=\{E,r,r^2,r^3,r^4,r^5\}\times\{E,t\};$}  & $U(1)$ & $\mathbb Z_2^4$ & $ {\omega_2(r,t)\over\omega_2(t,r)}={\omega_2(r^3,t)\over\omega_2(t,r^3)},\ \hat {a}^2 = \omega_2(a,a),\ \widehat {ar}^2 = \omega_2(ar,ar),\ \widehat {ta}^2 = \omega_2(ta,ta)$  \\
\cline{3-5}
& \multirow{1}{*}{$\{a, ar, ar^2, ar^3, ar^4, ar^5, ta, tar, tar^2,tar^3, tar^4, tar^5\}$} & $Z_4$ & $\mathbb Z_2^6$ & $ {\omega_2(r^3,t)\over\omega_2(t,r^3)},\  \hat {a}^2\! =\! \omega_2(a,a),\ \widehat {ar}^2 \!=\! \omega_2(ar,ar),\ \widehat {ta}^2 \!=\! \omega_2(ta,ta),\ \hat t^4 \!=\! \omega_2(t,t)^2,\ (\widehat{r^3})^4 \!=\! \omega_2(r^3,r^3)^2$  \\
\cline{2-5} 
& {$Z_6\rtimes Z_2=\{E,r,r^2,r^3,r^4,r^5\}\rtimes\{E,a\};$}  & $U(1)$ & $\mathbb Z_2^4$ & ${\omega_2(r^3,a)\over\omega_2(a,r^3)}, \hat t^2 = \omega_2(t,t), \ (\widehat{tr^3})^2 =\omega_2(tr^3, tr^3),\ \widehat{ta}^2 =\omega_2(ta, ta)$ \\
\cline{3-5}
& \multirow{1}{*}{$\{t, tr, tr^2, tr^3, tr^4,tr^5,ta,tar,tar^2,tar^3, tar^4, tar^5\}$} & $Z_4$ & $\mathbb Z_2^6$ & ${\omega_2(r^3,a)\over\omega_2(a,r^3)}, \hat t^2 \!=\! \omega_2(t,t), \ (\widehat{tr^3})^2 \!=\!\omega_2(tr^3, tr^3),\ \widehat{ta}^2 \!=\!\omega_2(ta, ta),\ (\widehat{r^3})^4\!=\!\omega_2(r^3,r^3)^2,\ \hat a^4\!=\!\omega_2(a,a)^2$\\
\hline
\end{tabular}
}
\end{table}


\begin{table}[tbp]
\caption{Invariants of $H^2_\varphi(G_b, Z_2\times Z_2)$ where $H$ acts trivially and $G_b - H$ exchanges the generators of $Z_2\times Z_2$.}\label{tab:exchange_invar}
\centering
\resizebox{1\textwidth}{!}{
\renewcommand{\arraystretch}{2}
\begin{tabular}{|c|c|c|c|}
\hline
$ {G}_b$ &action-free group $H$ and $G_b-H$ & $H^2(G_b,Z_2^{(1)}\!\!\!\times\!\! Z_2^{(2)})$ & invariants \\ 
\hline 
\multirow{2}{*}{$Z_2$} & {$Z_1;$}  &  \multirow{2}{*}{$\mathbb Z_1$} & \multirow{2}{*}{$-$} \\
& \multirow{1}{*}{$\{ a\}$} & &\\ 

\hline 
\multirow{2}{*}{$Z_2\times Z_2$} & {$Z_2=\{E, ab\};$}   
& \multirow{2}{*}{$\mathbb Z_2$} &  \multirow{2}{*}{$\omega_2^{(1)}(ab,ab)$} \\
& \multirow{1}{*}{$\{a, b\}$} &   &   \\

\hline 
\multirow{2}{*}{$Z_3\rtimes Z_2$} & {$Z_3=\{E,r,r^2\}; $}  & \multirow{2}{*}{$\mathbb Z_1$} &  \multirow{2}{*}{$-$} \\
& \multirow{1}{*}{$\{a, ar, ar^2\}$} &    &    \\

\hline
\multirow{2}{*}{$Z_4$} & {$Z_2=\{E,r^2\};$}  & \multirow{2}{*}{$\mathbb Z_2$} & \multirow{2}{*}{${\omega^{(1)}_2(r^2,r^2)}$} \\
& \multirow{1}{*}{$\{r, r^3\}$} &    &  \\

\hline
\multirow{4}{*}{$Z_4\rtimes Z_2$} & {$Z_2\times Z_2=\{E,r^2,a,ar^2\};$}  &  \multirow{2}{*}{$\mathbb Z_2^3$} & \multirow{2}{*}{${\omega_2^{(1)}(a,ar^2)\over\omega_2^{(1)}(ar^2,a)},\ \omega_2^{(1)}(a,a),\ \omega_2^{(1)}(ar^2,ar^2)$}  \\
& \multirow{1}{*}{$\{r, r^3, ar, ar^3\}$} &   &  \\
\cline{2-4} 
& {$Z_4=\{E,r,r^2,r^3\};  $}  &  \multirow{2}{*}{$\mathbb Z_2$} & \multirow{2}{*}{${\omega_2^{(1)}(r^2,r^2)}$} \\
& \multirow{1}{*}{$\{a, ar, ar^2, ar^3\}$} &  &   \\

\hline
\multirow{2}{*}{$Z_6$} & {$ Z_3=\{E,r^2,r^4\};$}  & \multirow{2}{*}{$\mathbb Z_1$} & \multirow{2}{*}{$-$} \\
& \multirow{1}{*}{$\{r,r^3,r^5\}$} &    &   \\

\hline
\multirow{4}{*}{$Z_6\rtimes Z_2$} & {$Z_3\rtimes Z_2=\{E,r^2,r^4,a,ar^2,ar^4\};$}  &  \multirow{2}{*}{$\mathbb Z_2$} & \multirow{2}{*}{$\omega_2^{(1)}(a,a)$}  \\
& \multirow{1}{*}{$\{r, r^3,  r^5, ar, ar^3,a r^5\}$} &   &  \\
\cline{2-4} 
& {$Z_6=\{E,r,r^2,r^3,r^4,r^5\};$}  &  \multirow{2}{*}{$\mathbb Z_2$} & \multirow{2}{*}{${\omega_2^{(1)}(r^3,r^3)}$} \\
& \multirow{1}{*}{$\{a, ar, ar^2, ar^3, ar^4, ar^5\}$} &    &   \\

\hline
\hline
\multirow{2}{*}{$(Z_2\times Z_2)\times Z_2$} & {$ Z_2\times Z_2=\{E,a,b,ab\};$}  &  \multirow{2}{*}{$\mathbb Z_2^3$} & \multirow{2}{*}{${\omega_2^{(1)}(a,b)\over\omega_2^{(1)}(b,a)}, \ \omega_2^{(1)}(a,a),\ \omega_2^{(1)}(b,b)$} \\
& \multirow{1}{*}{$\{t,ta,tb,tab\}$} &   & \\%

\hline
\multirow{4}{*}{$Z_4\times Z_2$} & {$Z_2\times Z_2=\{E,r^2,t,tr^2\};$}  &  \multirow{2}{*}{$\mathbb Z_2^3$} & \multirow{2}{*}{${\omega_2^{(1)}(r^2,t)\over  \omega_2^{(1)}(t,r^2)},\ \omega_2^{(1)}(r^2,r^2),\ \omega_2^{(1)}(t,t)$}  \\
& \multirow{1}{*}{$\{r, r^3, tr, tr^3\}$} &   &   \\
\cline{2-4} 
& {$Z_4=\{E,r,r^2,r^3\};$}  &  \multirow{2}{*}{$\mathbb Z_2$} & \multirow{2}{*}{$\omega_2^{(1)}(r^2,r^2)$} \\
& \multirow{1}{*}{$\{t, tr, tr^2, tr^3\}$} &   &  \\

\hline
\multirow{6}{*}{$(Z_4\rtimes Z_2)\times Z_2$} & {$Z_2\times Z_2\times Z_2=\{E,r^2\}\times\{E,a\}\times\{E,t\};$}  &  \multirow{2}{*}{$\mathbb Z_2^6$} & \multirow{2}{*}{${\omega_2^{(1)}(r^2,t)\over  \omega_2^{(1)}(t,r^2)},\ { \omega_2^{(1)}(r^2,a)\over  \omega_2^{(1)}(a,r^2)},\ {\omega_2^{(1)}(a,t)\over  \omega_2^{(1)}(t,a)},\  \omega_2^{(1)}(r^2,r^2),\ \omega_2^{(1)}(t,t),\ \omega_2^{(1)}(a,a)$}  \\
& \multirow{1}{*}{$\{r, r^3, ar, ar^3, tr, tr^3, tar, tar^3\}$} &   &  \\
\cline{2-4} 
& {$Z_4\times Z_2=\{E,r,r^2,r^3\}\times\{E,t\};$}  &  \multirow{2}{*}{$\mathbb Z_2^3$} & \multirow{2}{*}{${\omega_2^{(1)}(r,t)\over  \omega_2^{(1)}(t,r)},\  {\omega_2^{(1)}( r^2, r^2)},\  {\omega_2^{(1)}(t,t)}$} \\
& \multirow{1}{*}{$\{a, ar, ar^2, ar^3,ta, tar, tar^2, tar^3\}$} &   &   \\
\cline{2-4} 
& {$Z_4\rtimes Z_2=\{E,r,r^2,r^3\}\rtimes\{E,a\};$}  &  \multirow{2}{*}{$\mathbb Z_2^3$} & \multirow{2}{*}{${\omega_2^{(1)}(r^2,a)\over  \omega_2^{(1)}(a,r^2)},\ {\omega_2^{(1)}( a, a)},\  {\omega_2^{(1)}(ar,ar)}$} \\
& \multirow{1}{*}{$\{t, tr, tr^2, tr^3, ta,tar,tar^2,tar^3\}$} &  & \\

\hline 
\multirow{2}{*}{$Z_6\times Z_2$} & {$Z_6=\{E,r,r^2,r^3,r^4,r^5\};$}  &  \multirow{2}{*}{$\mathbb Z_2$} & \multirow{2}{*}{${\omega_2^{(1)}(r^3,r^3)}$} \\
& \multirow{1}{*}{$\{t,tr,tr^2,tr^3,tr^4,tr^5\}$} &    &   \\

\hline 
\multirow{4}{*}{$(Z_6\rtimes Z_2)\times Z_2$} & {$Z_6\times Z_2=\{E,r,r^2,r^3,r^4,r^5\}\times\{E,t\};$}  &  \multirow{2}{*}{$\mathbb Z_2^3$} & \multirow{2}{*}{${\omega_2^{(1)}(r^3,t)\over  \omega_2^{(1)}(t,r^3)},\ \omega_2^{(1)}(r^3,r^3),\ \omega_2^{(1)}(t,t)$}  \\
& \multirow{1}{*}{$\{a, ar, ar^2, ar^3, ar^4, ar^5, ta, tar, tar^2,tar^3, tar^4, tar^5\}$} &  &  \\
\cline{2-4} 
& {$Z_6\rtimes Z_2=\{E,r,r^2,r^3,r^4,r^5\}\rtimes\{E,a\};$}  &  \multirow{2}{*}{$\mathbb Z_2^3$} & \multirow{2}{*}{${ \omega_2^{(1)}(r^3,a)\over  \omega_2^{(1)}(a,r^3)},\ \omega_2^{(1)}(r^3,r^3),\ \omega_2^{(1)}(a,a)$} \\
& \multirow{1}{*}{$\{t, tr, tr^2, tr^3, tr^4,tr^5,ta,tar,tar^2,tar^3, tar^4, tar^5\}$} &  & \\
\hline

\end{tabular}
}
\end{table}

\subsection{Projective Representations of $G_b$: a few Examples}\label{sec:rep}
In this section, we provide concrete examples to illustrate the invariants and the generalized projective representations in the corresponding class. 


We focus on the cases in which $G_b$ is a magnetic point group. According to the unitarity/anti-unitarity and the action $\varphi$ on the spin and charge coefficients, we label the elements of $G_b$ in the following way: \\
\indent (i) an anti-unitary element $g$ is labeled by $g'$ as the usual notation of magnetic groups;\\
\indent (ii) an element $g$ which flips the spin $\sigma_z$ is labelled by adding a superscript $\tilde g$;\\
\indent (iii) an element $g$ which reverses the charge $\tau_z$ is labelled by adding a superscript $\dot g $.\\

For instance, for the magnetic point group 
$$
2'mm'=\{E, C_{2z}T, M_{x}, M_{y}T\},
$$
if $C_{2z}$ and $M_{x}$ reverse the spin $\sigma_z$, while $M_{x}$ and $M_{y}$ reverse the charge $\tau_z$, then the corresponding group can be denoted as 
$$
\tilde 2'\tilde{\dot m}\dot m'=\{E, \tilde C_{2z}T, \tilde {\dot M}_{x}, {\dot M}_{y}T.\}
$$

It should be mentioned that the above notation only distinguishes different action $\varphi$ on the coefficient group $\mathcal Z(O_f)$. Different maps $\tilde \phi$ may corresponds to the same $\varphi$, then they will be labeled by the same notation with $\tilde\ $ and $\dot\ $.


\subsubsection{Example-1: $G_b=\mathscr C_2=2=\{E,C_{2}\}\cong Z_2$} 

{\bf Trivial action $\varphi=E$}

Tab.\ref{tab:unit_invar} list the invariants for trivial action $\varphi=E$. The trivial action include two different cases:

\indent (I) unitary group $G_b=2$ with $\phi(G_b)=E$. There is no need to distinguish the spin coefficient group and the charge coefficient group. The classification reads
\beq\label{Z2unit}
H^2(2,U(1))=\mathbb Z_1,\ \ H^2(2, Z_2)=H^2(2, Z_4)=\mathbb Z_2,
\eeq
where nontrivial twist exists for the coefficient groups $Z_2$ and $Z_4$. 

The trivial class for $A=U(1), Z_2, Z_4$ corresponds to 1-dimensional linear reps of $G_b=2$, namely, $D(C_2)=\pm 1$.

For the $Z_2$ coefficient group, the nontrivial class is characterized by the invariant $\hat C_2^2= \omega_2(C_2, C_2)=-1$, and the corresponding nontrivial projective rep reads
$$
D(C_2) = \pm i,
$$
which can be realized by the representative $\phi_{C_{2}} =e^{-i\sigma_z{\pi\over2}}$ or $\phi_{C_{2}} =e^{-i\tau_z{\pi\over2}}$.

For $Z_{4c}$ coefficient group, the nontrivial class is characterized by the invariant $\hat C_2^4 = \omega_2(C_2, C_2)^2=-1$ (namely $\omega_2(C_2,C_2)=\pm i$), and the corresponding nontrivial projective rep is given by $$D(C_2)=e^{\pm i{\pi\over4}},$$
which can be realized by the representative $\phi_{C_{2}} =e^{-i\sigma_z{\pi\over4}}$ or $\phi_{C_{2}} =e^{-i\tau_z{\pi\over4}}$.\\

\indent (II) anti-unitary group $G_b=2'=\{E, C_2T\}$, where $\phi(G_b)$ is not necessarily trivial. Since $T = e^{-i(\sigma_y+\tau_y){\pi\over2}}K$ commutes with both $U(1)_s$ and $U(1)_c$, the classification for anti-unitary group $2'$ is the same as that of the unitary group $2$ in (\ref{Z2unit}). However, the representations are different. 

For $U(1)$-coefficient, there is only one trivial class
$$
H^2(2',U(1)_s)=H^2(2',U(1)_c)=\mathbb Z_1.
$$
However, due to the definition of $T = e^{-i(\sigma_y+\tau_y){\pi\over2}}K$, the anti-unitary element $2'$ is associated with nontrivial spin-charge operation $e^{-i(\sigma_y+\tau_y){\pi\over2}}$, which results in irreducible representation with dimension higher than 1. For instance, if one considers the spin sector $A=U(1)_s$, an irreducible representation of $2'$ reads
\beq\label{linearrepZ2T}
D(C_2T)K = e^{-i \sigma_x{\pi\over2}}K
\eeq
with $[D(C_2T)K]^2=1$. To ensure the trivial action on $U(1)_s$, the dimension of the irreducible rep of $C_2T$ is at least 2. This 2-dimensional irrep of $C_2T$ is owing to the nontrivial action instead of nontrivial twist. The above representation corresponds to the representative $\phi_{C_2T} = e^{-i(\sigma_z+\tau_y){\pi\over2}}$. 

Similarly, for $A=U(1)_c$, the corresponding irrep reads
$$
D(C_2T)K = e^{-i \tau_x{\pi\over2}}K,
$$
which can be realized by the representative $\phi_{C_2T} = e^{-i(\sigma_y + \tau_z){\pi\over2}}$. 

When $A=Z_2=\{1,-1\}$, the elements in $A$ are real, hence the action $\varphi$ is always trivial and one has $H^2(2',Z_2)=H^2(2,Z_2)=\mathbb Z_2$. In the trivial class with $\omega_2(C_2T,C_2T)=1$, the irrep can be chosen as 
$$
D(C_2T)=1K,
$$
which can be realized by the representative $\phi_{C_2T} = e^{-i(\sigma_y+\tau_y){\pi\over2}}$. 
In the nontrivial class with $\omega_2(C_2T,C_2T)=-1$, the irrep can be chosen as 
$$
D(C_2T)K=e^{-i\sigma_y{\pi\over2}}K,
$$
which can be realized by the representative $\phi_{C_2T} = e^{-i \tau_y{\pi\over2}}$.

When $A=Z_{4c}=\{\pm1, e^{\pm i{\tau_z}{\pi\over2}}\}$,one has $H^2(2',Z_{4c})=\mathbb Z_2$. The irreducible rep in the trivial class with $\omega_2(C_2T,C_2T)^2 = 1$ can be chosen as $D(C_2T)K = e^{-i\tau_x{\pi\over2}}K$ or $D(C_2T)K = e^{-i\tau_y{\pi\over2}}K$. The irrep in the nontrivial class with $\omega_2(C_2T,C_2T)^2 = -1$ (namely $\omega_2(C_2,C_2) = e^{\pm i\tau_z{\pi\over2}}$) can be chosen as 
$$
D(C_2T)K = e^{-i {1\over\sqrt2}(\tau_y\pm \tau_x){\pi\over2}}K,
$$
which can be realized by the representative $\phi_{C_2T} = e^{-i (\sigma_y{\pi\over2}  \pm\tau_z{\pi\over4})}$.

{\bf Charge coefficient groups with nontrivial action $\varphi$}

Now we consider the case where $\phi(G_b)$ is nontrivial. When acting on the charge coefficient groups $U(1)_c=\{e^{-i{\tau_z\over2}\theta};\theta\in[0,4\pi)\}$ and $Z_{4c}=\{\pm 1,  e^{\pm i{\tau_z}{\pi\over2}}\}$, the nontrivial action is taking inversion or complex conjugation. In the following we consider two different cases of $\phi(G_b)$:\\
\indent (I) unitary group $G_b=\dot2 = \{E,\dot C_2\}$, where the generator $C_2$ is associated with nontrivial charge operation such as $\phi_{C_2}=e^{\pm i{\tau_x}{\pi\over2}}$. Then the cohomology group
\Beq
H^2(\dot 2, U(1)_c) = H^2(\dot 2, Z_{4c}) =\mathbb Z_2,
\Eeq
is generated by the invariant $\widehat{\dot C_2}^2 = \omega_2(\dot C_2, \dot C_2)=\pm1$. 

The charge operation $\phi_{C_2}=e^{\pm i{\tau_x}{\pi\over2}}$ associated to $C_2$ directly gives rise to a representation
$$
D(\dot C_2) = e^{\pm i{\tau_x}{\pi\over2}}= \pm i\tau_x.
$$ 
Since $D(\dot C_2)$ does not commute with $U(1)_c$, this 2-dimensional rep is irreducible. Because $\omega_2(\dot C_2, \dot C_2) =D(\dot C_2)^2 = -1$, this rep belongs to the nontrivial class. 


For the trivial class with $\omega_2(\dot C_2, \dot C_2)=1$, the representation is still 2-dimensional, with 
$$
D(\dot C_2) =  i e^{\pm i{\tau_x}{\pi\over2}} = \mp\tau_x
$$ 
and $D(\dot C_2)^2 =1$. Notice that the factor $i$ here cannot be gauged away since the coefficient group $U(1)_c$ or $Z_{4c}$ is generated by ${\tau_z\over2}$. It should be considered as a spin rotation $e^{is_z{\pi}}$ with $s_z={1\over2}$. In this case, the action of $G_b$ on charge coefficient groups is realized by associating $e^{-i{(\tau_x\pm \sigma_z)}{\pi\over2}}$ to $C_2$, namely $\phi_{C_2} = e^{-i{(\tau_x\pm \sigma_z)}{\pi\over2}}$. \\

\indent (II) anti-unitary group $G_b=\dot 2'=\{E, \dot C_2T\}$ where the generator $C_2T$ is associated with nontrivial charge operation such as $e^{\pm i{\tau_x}{\pi\over2}}$ or $e^{\pm i{\tau_y}{\pi\over2}}$.

For both $Z_{4c}$ and $U(1)_c$ coefficient groups, the cohomology group
\Beq
H^2(\dot 2', U(1)_c) =H^2(\dot 2', Z_{4c}) =\mathbb Z_2
\Eeq
is generated by the invariant $\widehat{C_2T}^2 = \omega_2(\dot C_2T, \dot C_2T)=\pm1$.

For the trivial class with $\widehat{\dot C_2T}^2 = \omega_2(\dot C_2T, \dot C_2T)=1$, the 2-dimensional rep $D(\dot C_2T)K = e^{-i{\tau_z}{\pi\over2}}K$ can be reduced to direct sum of 1-dimensional irreps, 
$$
D(\dot C_2T)K = \pm1 K.
$$ 
This irrep can be realized by the representative $\phi_{C_2T} = e^{\pm i{(\sigma_y + \tau_x)}{\pi\over2}}$.

The nontrivial class is of Kramers type with $\widehat{\dot C_2T}^2 = \omega_2(\dot C_2T,\dot C_2T)=-1$. The corresponding 2-dimensional projective rep of $C_2T$ reads 
$$
D(\dot C_2T)K = e^{\pm i{\sigma_y}{\pi\over2}}K = \pm i\sigma_y K,
$$
which has two spin components and a single charge component, and can be realized via the representative $\phi_{C_2T} = \pm e^{-i{\tau_y}{\pi\over2}}$. 

%
%
%


{\bf Spin coefficient groups with nontrivial action $\varphi$}

Now we consider the case where $G_b$ has nontrivial action (taking inversion or complex conjugation) on the spin coefficient groups $U(1)_s=\{e^{-i{\sigma_z\over2}\theta};\theta\in[0,4\pi)\}$. 
Similar to the charge coefficient groups, there are also two possibilities to realize this nontrivial action:\\
\indent (I) unitary group $G_b=\tilde 2=\{E, \tilde C_2\}$ with the generator $C_2$ being associated with nontrivial charge operation such as $e^{-i{\sigma_x}{\pi\over2}}$. Then 
\Beq
H^2(\tilde 2, U(1)_s) =\mathbb Z_2,
\Eeq
which are generated by the invariant $\widehat{\tilde C_2}^2 = \omega_2(\tilde C_2, \tilde C_2)=\pm1$. 

For the nontrivial class $\omega_2(\tilde C_2, \tilde C_2)=-1$, the 2-dimensional representation of $\tilde C_2$ reads 
$$
D(\tilde C_2) = e^{\pm i{\sigma_x}{\pi\over2}}= \pm i\sigma_x
$$ with $D(\tilde C_2)^2 =-1$.  

For the trivial class $\omega_2(\tilde C_2, \tilde C_2)=1$, the representation is still 2-dimensional, with 
$$
D(\tilde C_2) = i e^{\pm i{\sigma_x}{\pi\over2}} = \mp\sigma_x
$$ 
and $D(\tilde C_2)^2 =1$. 
The factor $i$ comes from charge rotation $e^{i{\tau_z\over2}{\pi}}$. Hence the above rep can be realized by $\phi_{C_2} = e^{-i{(\sigma_x\pm\tau_z)}{\pi\over2}}$.\\ 

\indent (II) $G_b=\tilde 2'=\{E, \tilde C_2T\}$ with the generator $C_2T$ being associated with nontrival spin operation such as $e^{-i \sigma_x {\pi\over2}}$. 

Then we have
\Beq
H^2(\tilde 2', U(1)_s) =\mathbb Z_2,
\Eeq
with invariant $\widehat{\tilde C_2T}^2 = \omega_2(\tilde C_2T, \tilde C_2T)=\pm 1$.

For the trivial class with $\widehat{\tilde C_2T}^2 = \omega_2(\tilde C_2T,\tilde C_2T)=1$, the representation is 1-dimensional, like $D(\tilde C_2T)K = \pm1K$ (which is reduced from $D(\tilde C_2T)K = e^{-i\sigma_z{\pi\over2}}K$). This rep can be realized by the representative $\phi_{C_2T}=e^{-i(\sigma_x +\tau_y){\pi\over2}}$.

The nontrivial class is of Kramers type with $\widehat{\tilde C_2T}^2 = \omega_2(\tilde C_2T,\tilde C_2T)=-1$. The corresponding projective rep is 2-dimensional, namely,
$$
D(\tilde C_2T)K = e^{\pm i{\tau_y\over2}\pi} = \pm i\tau_y K.
$$
This rep has two charge components and a single spin component, and can be realized by the representative $\phi_{C_2T} =e^{-i{\sigma_y\over2}\pi}$.

%
%
%

\subsubsection{Example-2: $G_b = \mathscr C_{2v}=2mm=\{E, C_{2z}, M_{x}, M_{y}\}\cong Z_2\times Z_2$} 
Since the projective reps with charge coefficient groups and spin coefficient groups are very similar, in later discussion, we only consider the {\it charge sector}, and skip the discussion for the spin sector. 

{\bf Trivial action $\varphi(G_b)=E$}

The trivial action include two different cases:

\indent (I) unitary group $G_b=2mm$. 
For the $U(1)_c$ coefficient, one has 
$$
H^2(2mm, U(1)_c) =\mathbb Z_2
$$
generated by the invariant $\eta_{x,y}={\omega_2(M_{x}, M_{y})\over \omega_2(M_{y}, M_{x})}=\pm 1$. The trivial class with $\eta_{x,y}=1$ contains 4 nonequivalent 1-dimensional reps, and the nontrivial class with $\eta_{x,y}=-1$ contains a 2-dimensional projective rep 
\beq\label{eta-1}
\!D(M_{x}) \!=\!e^{\pm i\sigma_x{\pi\over2}},\  D(M_{y}) \!=\! e^{\pm i\sigma_y{\pi\over2}}. 
\eeq
This rep can be realized by associating spin operations to $G_b$, namely, $\phi_{M_{x}}=e^{\pm i\sigma_x{\pi\over2}}, \phi_{M_{y}}=e^{\pm i\sigma_y{\pi\over2}}$.

For the $Z_2$ coefficient, one has
$$
H^2(2mm, Z_2)=\mathbb Z_2^3,
$$
which is generated by the invariants $\eta_{x,y}={\omega_2(M_{x}, M_{y})\over \omega_2(M_{y}, M_{x})}=\pm 1, \ \eta_x =\omega_2(M_{x}, M_{x})=\pm1,\ \eta_y =\omega_2(M_{y}, M_{y})=\pm1$. 
The projective rep with $(\eta_{x,y},\eta_x,\eta_y)=(-1,1,1)$ can be either in the spin sector shown in (\ref{eta-1}) or in the charge sector $D(M_{x}) =e^{\pm i\tau_x{\pi\over2}},\  D(M_{y}) = e^{\pm i\tau_y{\pi\over2}}$. The latter rep can be realized by the representative $\phi_{M_{x}}=e^{\pm i\tau_x{\pi\over2}}, \phi_{M_{y}}=e^{\pm i\tau_y{\pi\over2}}$.

The projective rep with $(\eta_{x,y},\eta_x,\eta_y)=(1,-1,1)$ is given by 
\Beq
D(M_{x}) = \pm i,\ D(M_{y}) = \pm 1,
\Eeq
which can be realized by associating spin or charge operation to $M_{x}$, namely by the representative $\phi_{M_{x}} =e^{-i\sigma_z{\pi\over2}}$ 
or $\phi_{M_{x}} =e^{-i\tau_z{\pi\over2}}$. 
The projective rep with $(\eta_{x,y},\eta_x,\eta_y)=(1, 1, -1)$ is given by 
\Beq
D(M_{x}) = \pm 1, D(M_{y}) = \pm i,
\Eeq
which can be realized by the representative $\phi_{M_{y}} =e^{-i\sigma_z{\pi\over2}}$ 
or $\phi_{M_{y}} =e^{-i\tau_z{\pi\over2}}$. 
The projective reps of the rest ones can be obtained by the direct product of the above ones.

For the $Z_{4c}$ coefficient, one has
$$
H^2(2mm, Z_{4c})=\mathbb Z_2^3,
$$
which is generated by the invariant $\eta_{x,y}={\omega_2(M_{x}, M_{y})\over \omega_2(M_{y}, M_{x})}=\pm 1, \ \eta_x =\omega_2(M_{x}, M_{x})^2=\pm1,\ \eta_y =\omega_2(M_{y}, M_{y})^2=\pm1$. 
The projective rep with $(\eta_{x,y},\eta_x,\eta_y)=(-1,1,1)$ is given in (\ref{eta-1}). The projective rep in the class $(\eta_{x,y},\eta_x,\eta_y)=(1,-1,1)$ is given by 
\Beq
D(M_{x}) = e^{\pm i{\pi\over4}},\ D(M_{y}) = \pm 1,
\Eeq
which can be realized by the representative $\phi_{M_{x}} =e^{\pm i\tau_z{\pi\over4}}$ or $\phi_{M_{x}} =e^{\pm i\sigma_z{\pi\over4}}$.
The projective rep in the class $(\eta_{x,y},\eta_x,\eta_y)=(1, 1, -1)$ is given by 
\Beq
D(M_{x}) = \pm 1,\ D(M_{y}) = e^{\pm i{\pi\over4}},
\Eeq
which can be realized by the representative $\phi_{M_{y}} =e^{\pm i\tau_z{\pi\over4}}$ or $\phi_{M_{y}} =e^{\pm i\sigma_z{\pi\over4}}$.
The projective reps of the rest classes can be obtained by the direct product of the above ones.\\

\indent (II) antiunitary group $G_b=2'mm'=\{E, C_{2z}T, M_{x}, M_{y}T\}$. \\
For $U(1)_c$ coefficient, since $T = e^{-i(\sigma_y+\tau_y){\pi\over2}}K$ commutes with $U(1)_c$, the classification is still $\mathbb Z_2$ which is generated by $\eta_{x,y}=\pm1$. In the trivial class with $\eta_{x,y}=1$, the irrep reads
\[
D(C_{2z}T)K = e^{\pm i\tau_x{\pi\over2}}K,\ D(M_{x})=I,\ D(M_{y}T)K = e^{\pm i\tau_x{\pi\over2}}K,
\]
which can be realized by the representative $\phi_{C_{2z}T}=\phi_{M_{y}T}=e^{-i(\sigma_y\pm \tau_z){\pi\over2}}$.
For the nontrivial class with $\eta_{x,y}=-1$, the irrep is 4-dimensional, 
\beq\label{2'mm'_eta-1}
D(C_{2z}T)K \!=\! e^{-i(\sigma_x\pm \tau_x){\pi\over2}}\!K,\  
D(M_{x}) \!=\! e^{-i \sigma_y {\pi\over2}}\!,\  
D(M_{y}T)K \!=\! e^{-i(\sigma_z\pm \tau_x){\pi\over2}}\!K,
\eeq
which can be realized by the representative $\phi_{C_{2z}T}=e^{-i(\sigma_z\pm \tau_z){\pi\over2}}, \phi_{M_{y}T}=e^{-i(\sigma_x\pm \tau_z){\pi\over2}}$.

For $Z_2$ coefficient, there are also three invariants $(\eta_{x,y},\eta_x,\eta_y)$. The irrep of the class $(-1,1,1)$ is the same as (\ref{2'mm'_eta-1}) for the nontrivial class with $U(1)_c$ coefficient. The irrep in the class $(1,-1,1)$ reads
\Beq
D(C_{2z}T)K = e^{-i\tau_y{\pi\over2}}K,\ D(M_{y}T)K = e^{\pm i\tau_x{\pi\over2}}K,
\Eeq
which can be realized by the representative $\phi_{C_{2z}T}=e^{-i\sigma_y{\pi\over2}}, \phi_{M_{y}T}=e^{-i(\sigma_y\pm \tau_z){\pi\over2}}$; the irrep in the class $(1,1,-1)$ reads
\Beq
D(C_{2z}T)K = e^{\pm i\tau_x{\pi\over2}}K,\ D(M_{y}T)K = e^{-i\tau_y{\pi\over2}}K,
\Eeq
which can be realized by the representative $\phi_{C_{2z}T}=e^{-i(\sigma_y\pm \tau_z){\pi\over2}}, \phi_{M_{y}T}=e^{-i\sigma_y{\pi\over2}}$. The irreps in rest classes can be obtained by reducing the direct product of the irreps of the above three classes. 

For $Z_{4c}$ coefficient, there are also three invariants $(\eta_{x,y},\eta_x,\eta_y)$. The irrep of the class $(-1,1,1)$ is the same as (\ref{2'mm'_eta-1}) for the nontrivial class with $U(1)_c$ coefficient. The irrep in the class $(1,-1,1)$ reads
\Beq
D(C_{2z}T)K \!=\! e^{-i{1\over\sqrt2}(\tau_x\pm \tau_y){\pi\over2}}K,\ D(M_{y}T)K \!=\! e^{\pm i\tau_x{\pi\over2}}K,
\Eeq
(and $D(M_{x}) = e^{\pm i \tau_z{\pi\over4}}$) which can be realized by the representative $\phi_{C_{2z}T} = e^{-i(\sigma_y\pm{\tau_z\over2}){\pi\over2}}, \phi_{M_{y}T}=e^{-i(\sigma_y\pm\tau_z){\pi\over2}}$;  the irrep in the class $(1,1,-1)$ reads
\Beq
D(C_{2z}T)K = e^{\pm i\tau_x{\pi\over2}}K,\ D(M_{y}T)K =e^{-i{1\over\sqrt2}(\tau_x\pm \tau_y){\pi\over2}}K,
\Eeq
which can be realized by the representative $\phi_{C_{2z}T} =e^{-i(\sigma_y\pm \tau_z){\pi\over2}}, \phi_{M_{y}T}= e^{-i(\sigma_y\pm{\tau_z\over2}){\pi\over2}}$. The irreps in rest classes can be obtained by reducing the direct product of the irreps of the above three classes. 

{\bf irreps with nontrivial mapping $\varphi$}

Now we consider the case where $G_b$ has nontrivial action (taking inversion or complex conjugation) on the charge coefficient groups $U(1)_c$ and $Z_{4c}$. There are also several possibilities to realize this nontrivial action:\\
\indent (I) unitary group $G_b=\dot 2\dot mm=\{E, \dot C_{2z}, \dot M_{x}, M_{y}\}$ with the generator $C_{2z}, M_{x}$ associated by $e^{-i{\tau_x}{\pi\over2}}$. 

For $U(1)_c$ coefficient group, the twists are classified by
\Beq
H^2(\dot 2\dot mm, U(1)_s) = \mathbb Z_2\times \mathbb Z_2,
\Eeq
with invariants $\eta_{z} =\omega_2(\dot C_{2z}, \dot C_{2z}), \eta_{x} =\omega_2(\dot M_{x}, \dot M_{x})$. The projective rep of the class $(\eta_{z},\eta_{x})=(-1,1)$ reads
\Beq
D(\dot C_{2z}) = e^{\pm i{\tau_x}{\pi\over2}},\ D(\dot M_{x}) = ie^{\pm i{\tau_x}{\pi\over2}},
\Eeq
which can be realized by the representative $\phi_{C_{2z}} = e^{\pm i{\tau_x}{\pi\over2}}, \phi_{M_{x}} = e^{-i{(\sigma_z\pm \tau_x)}{\pi\over2}}$.
The projective rep of the class $(\eta_{z},\eta_{x})=(1, -1)$ reads
\Beq
D(\dot C_{2z}) = ie^{\pm i{\tau_x}{\pi\over2}},\ D(\dot M_{x}) = e^{\pm i{\tau_x}{\pi\over2}},
\Eeq
which can be realized by the representative $\phi_{C_{2z}} = e^{-i{(\sigma_z\pm \tau_x)}{\pi\over2}}, \phi_{M_{x}} = e^{\pm i{\tau_x}{\pi\over2}}$.
The reps of the rest calsses can be obtained in a similar way. 

For $Z_{4c}$ coefficient group, the classification 
\Beq
H^2(\dot 2\dot mm, Z_{4c}) = \mathbb Z_2\times \mathbb Z_2\times \mathbb Z_2,
\Eeq
is generated by the invariants $\eta_{z} =\omega_2(\dot C_{2z}, \dot C_{2z}),\ \eta_{x} =\omega_2(\dot M_{x}, \dot M_{x}), \ \eta_{y} =\omega_2(M_{y}, M_{y})^2$. The projective reps of the classes with $\eta_z=\pm1, \eta_x=\pm1, \eta_y=1$ are the same as the $U(1)_c$ coefficient group discussed above, while the rep for the class $(\eta_{z},\eta_{x}, \eta_{y})=(1,1,-1)$ reads
\Beq
&&D(\dot C_{2z}) = D(\dot M_{x}) =ie^{\pm i{\tau_x}{\pi\over2}},\\ 
&&D(M_{y}) = e^{\pm i{\tau_z}{\pi\over 4}},
\Eeq
which can be realized by the representative $\phi_{C_{2z}} = e^{-i{(\sigma_z\pm \tau_x)}{\pi\over2}}, \phi_{M_{x}} = e^{-i{(\sigma_z\pm \tau_x)}{\pi\over2}}, \phi_{M_{y}} = e^{\pm i{\tau_z}{\pi\over4}}$.
The irreps in rest classes can be obtained by reducing the direct product of the irreps of the known classes. 


The above discuss can be applied to the group $G_b = 2\dot m\dot m$ with slight modification and will not repeated here.

\indent (II) anti-unitary group $G_b=\dot 2'\dot m' m = \{E, \dot C_{2z}T, \dot M_{x}T, M_{y}\}$. \\
For $U(1)_c$ coefficient, the classification is 
$$H^2(\dot 2'\dot m'm, U(1)_c)=\mathbb Z_2\times \mathbb Z_2$$ 
generated by the invariants $\eta_{z} =\omega_2(\dot C_{2z}T, \dot C_{2z}T), \eta_{x} =\omega_2(\dot M_{x}T, \dot M_{x}T)$. The irrep of the class $(-1, 1)$ reads
\Beq
D(\dot C_{2z}T)K = e^{-i \sigma_y {\pi\over2}}K, \ \
D(\dot M_{x}T)K = IK, 
\Eeq
which can be realized by the representative $\phi_{C_{2z}T} = e^{- i{\tau_y}{\pi\over2}}, \phi_{M_{x}T} = e^{-i{(\sigma_y+ \tau_y)}{\pi\over2}}$; 
and the irrep of the class $(1, -1)$ reads
\Beq
D(\dot C_{2z}T)K = IK,\ \ 
D(\dot M_{x}T)K = e^{-i \sigma_y{\pi\over2}}K,
\Eeq
which can be realized by the representative $\phi_{C_{2z}T} = e^{-i{(\sigma_y+\tau_y)}{\pi\over2}}, \phi_{M_{x}T} = e^{- i{\tau_y}{\pi\over2}}$.  The reps of the rest classes can be obtained in a similar way. 

For $Z_{4c}$ coefficient group, the invariants are still $\eta_{z} =\omega_2(\dot C_{2z}T, \dot C_{2z}T),\ \eta_{x} =\omega_2(\dot M_{x}, \dot M_{x}), \ \eta_{y} =\omega_2(M_{y}, M_{y})^2$, the irreps in the classes $(\pm1,\pm1,1)$ are the same with $U(1)_c$ coefficient, while the irrep in the class $(1,1,-1)$ reads
\Beq
&&D(\dot C_{2z}T)K = D(\dot M_{x}T)K =IK,\\ 
&&D(M_{y}) = e^{\pm i{\tau_z}{\pi\over 4}},
\Eeq
which can be realized by the representative $\phi_{C_{2z}T} = e^{-i{(\sigma_y+ \tau_y)}{\pi\over2}}, \phi_{M_{x}T} = e^{-i{(\sigma_y+ \tau_y)}{\pi\over2}}, \phi_{M_{y}} = e^{\pm i{\tau_z}{\pi\over4}}$.
The irreps in rest classes can be obtained by reducing the direct product of the irreps of the known classes. 

The above discuss can be applied to the group $G_b = 2\dot m'\dot m'$ with slight modification and will not repeated here.


\indent (III) anti-unitary group $G_b=\dot 2'\dot mm'=\{E, \dot C_{2z}T, \dot M_{x},M_{y}T\}$. For $U(1)_c$ coefficient, one has $$H^2(\dot 2'\dot mm', U(1)_c)=\mathbb Z_2\times \mathbb Z_2$$ generated by the invariants $\eta_{z} =\omega_2(\dot C_{2z}T, \dot C_{2z}T), \eta_{x} =\omega_2(\dot M_{x}, \dot M_{x})$. The irrep of the class $(-1, 1)$ is 4-dimensional, such as
\Beq
D(\dot C_{2z}T)K = e^{-i \sigma_y {\pi\over2}}\otimes I K, \ \ 
D(\dot M_{x}) = e^{-i(\sigma_z\pm \tau_x){\pi\over2}}, 
\Eeq
which can be realized by the representative $\phi_{C_{2z}T} = e^{- i{\tau_y}{\pi\over2}}, \phi_{M_{x}} = e^{-i(\sigma_z\pm \tau_x){\pi\over2}}$; and the irrep of the class $(1, -1)$ reads
\Beq
D(\dot C_{2z}T)K = IK,\ \ 
D(\dot M_{x}) = e^{-i \tau_x{\pi\over2}},
\Eeq
which can be realized by the representative $\phi_{C_{2z}T} = e^{-i{(\sigma_y+\tau_y)}{\pi\over2}}, \phi_{M_{x}} = e^{- i{\tau_x}{\pi\over2}}$. 
The reps of the rest calsses can be obtained in a similar way. 

For $Z_{4c}$ coefficient group, the invariants are still $\eta_{z} =\omega_2(\dot C_{2z}T, \dot C_{2z}T),\ \eta_{x} =\omega_2(\dot M_{x}, \dot M_{x}), \ \eta_{y} =\omega_2(M_{y}, M_{y})^2$, the irreps in the classes $(\pm1,\pm1,1)$ are the same with $U(1)_c$ coefficient, while the irrep in the class $(1,1,-1)$ reads
\Beq
&&D(\dot C_{2z}T)K = ie^{\pm i \tau_z{\pi\over4}}K, \ D(\dot M_{x}) = ie^{\pm i\tau_x{\pi\over2}},\\ 
&&D(M_{y}T) = e^{-i{{1\over\sqrt2}(\tau_x\pm \tau_y)}{\pi\over 2}}K,
\Eeq
which can be realized by the representative $\phi_{C_{2z}T} = e^{-i{(\sigma_y+ \tau_y \pm{\tau_z\over2}+\sigma_z)}{\pi\over2}}, \phi_{M_{x}} = e^{-i{(\sigma_z\pm \tau_x)}{\pi\over2}}, \phi_{M_{y}T} = e^{-i(\sigma_y\pm {\tau_z\over2}){\pi\over2}}$.
The irreps in rest classes can be obtained by reducing the direct product of the irreps of the known classes. 

The above discuss can be applied to the groups $G_b = 2'\dot m' \dot m$ and $G_b= \dot 2\dot m' m'$ with slight modification and will not repeated here.

\subsubsection{Example-3: $G_b=mmm$ (3D)}

In the previous section,  we have shown that the irreps of each projective class can be realized by the homomorphism $\tilde \phi$ (or equivalently by the representative $\phi$). In this section we illustrate that this is not always the case. 

We consider the unitary point group $G_b=\dot m\dot m\dot m=\{E, C_{2x}, C_{2y}, C_{2z}\}\times \{E, \dot C_i\}$ which appears in 3 spatial dimensions and is beyond the 2-dimensional SCPGs discussed in Sec. \ref{sec: SCPG2D}. For $U(1)_c$ coefficient, where the elements $C_i, M_x=C_iC_{2x}, M_y=C_iC_{2y}, M_z=C_iC_{2z}$ have nontrivial action on $U(1)_c$. From Tabel.\ref{tab:antiunit_invar}, the classification is
\[
H^2(\dot m\dot m\dot m, U(1)_c) =\mathbb Z_2^4,
\]  
which is generated by the invariants $\eta_{x,y}={\omega_2(C_{2x},C_{2y}) \over \omega_2(C_{2y},C_{2x})}, \eta_{C_i}=\omega_2(C_{i},C_{i}), \eta_{M_x}=\omega_2(M_x,M_x), \eta_{M_y}=\omega_2(M_y,M_y)$.

Here we are not going to exhaust all the projective classes, instead, we will only focus on the class $(-1,-1,-1,-1)$. A 4-dimensional irrep can be constructed as
\Beq
&&D(C_{2x}) = e^{-i\sigma_x{\pi\over2}}\otimes I, \ D(C_{2y}) = e^{-i\sigma_y{\pi\over2}}\otimes I, \\
&& D(C_{i}) = I\otimes e^{-i\tau_x{\pi\over2}},\ D(C_{2z}) = e^{-i\sigma_z{\pi\over2}}\otimes I, \\
&& D(M_{x}) = ie^{-i(\sigma_x+\tau_x){\pi\over2}},  D(M_{y}) = ie^{-i(\sigma_y+\tau_x){\pi\over2}}.
\Eeq
Notice that the factor $i$ in $D(M_x)$ and $D(M_y)$ is beyond the $SO(4)$ operations and cannot be realized by $\phi$. 

Similarly, for the anti-unitary group $G_b={2'\over \dot m} {2'\over \dot m}{2\over \dot m}=\{E, C_{2x}T, C_{2y}T, C_{2z} \}\times \{E, \dot C_iT\}$  where the elements $C_iT, M_x, M_y, M_zT$ have nontrivial action on the coefficient group $U(1)_c$, the invariants are the same with the unitary group $\dot m\dot m\dot m$, and an 4-dimensional irrep of the class $(-1,-1,-1,-1)$ can be constructed as 
\Beq
&&D(C_{2x}T)K = e^{-i(\sigma_x+\tau_y){\pi\over2}}K, \\
&& D(C_{2y}T)K =  e^{-i(\sigma_z+\tau_y){\pi\over2}} K, \\
&& D(C_{i}T)K = e^{-i\sigma_y{\pi\over2}}\otimes IK,\\
&& D(M_{x}) =  i e^{-i (\sigma_z+\tau_y) {\pi\over2}},\\  
&& D(M_{y}) =  i e^{-i(\sigma_x+\tau_y){\pi\over2}}.
\Eeq
Again, the factor $i$ in $D(M_x)$ and $D(M_y)$ cannot be realized by $\phi$. 

Then we consider the anti-unitary group $G_b=\dot m'\dot m'\dot m'=\{E, C_{2x}, C_{2y}, C_{2z}\}\times \{E, \dot C_iT\}$ with $U(1)_c$ coefficient, where the elements $C_iT, M_xT, M_yT, M_zT$ have nontrivial action on $U(1)_c$. The invariants are the same with the unitary group $\dot m\dot m\dot m$, and an 4-dimensional irrep of the class $(-1,-1,-1,-1)$ can be constructed as 
\Beq
&&D(C_{2x}) = e^{-i(\mu_y +\sigma_x){\pi\over2}} , \ D(C_{2y}) =  e^{-i(\mu_y+\sigma_z){\pi\over2}} , \\
&& D(C_{i}T)K = I\otimes e^{-i\sigma_y{\pi\over2}} K,\ D(C_{2z}) = I\otimes e^{-i\sigma_y{\pi\over2}} , \\
&& D(M_{x}T)K =  e^{-i\mu_y{\pi\over2}}\otimes e^{-i \sigma_z {\pi\over2}} K,\\  
&&D(M_{y}T)K =  e^{-i\mu_y{\pi\over2}}\otimes e^{-i\sigma_x{\pi\over2}} K.
\Eeq
Here the operators $\mu_{x,y,z}$ are neither spin nor charge. Once again, this irrep cannot be realized by $\phi$. 

Certain classes of projective twist that cannot be realized via $\phi$ can be realized via site-dependent gauge sector $\gamma_l$, which have been illustrated via examples in the main text. It remains an open question how to judge if a projective class can be realized or not via gauging and how to construct the general solutions.


\section{Detailed derivations of results in the Main text}

\subsection{Spin-charge groups  for 3D fermionic superfluids}

\subsubsection{$G_b=O(3)\times Z_2^{T}$ and $ O_f=Z_2^f$}

In this section, we consider fermionic superfluid phase with $G_b=O(3)\times Z_2^{T}$, $ O_f=Z_2^f$. 

\underline{\it Global action:} The possible actions of $G_b$ on $O_f$ in spin-charge groups are characterized by the following continuous homomorphisms 
\begin{align}
\tilde{\phi}: O(3)\times Z_2^{T}\to SO(3)_s\times SO(3)_c. 
\end{align}
We now show that such continuous homomorphisms fall into exactly $36$ inequivalent class.

\begin{lemma}
Any continuous homomorphism $\rho:\SO(3)\to\SO(3)$ is either trivial or an automorphism.
\end{lemma}

\begin{proof}
Consider the differential at the identity,
\[
d\rho_e:\mathfrak{so}(3)\longrightarrow \mathfrak{so}(3),
\]
where $\mathfrak{so}(3)$ is the Lie algebra of $SO(3)$ which is simple, and $d\rho_e$ is either $0$ or injective. If $d\rho_e=0$, then $\rho$ has discrete image; since $SO(3)$ is connected, the image must be trivial. If $d\rho_e\neq 0$, then it is an isomorphism. Hence $\rho$ is a local diffeomorphism onto an open subgroup of the connected group $\SO(3)$, so it is surjective. Its kernel is a discrete normal subgroup of $\SO(3)$, hence central; but the center of $\SO(3)$ is trivial. Therefore the kernel is trivial and $\rho$ is an automorphism.
\end{proof}

\begin{lemma}
Every continuous automorphism of $\SO(3)$ is inner.
\end{lemma}

\begin{proof}
This is standard: $\mathrm{Aut}(\mathfrak{so}(3))\cong \SO(3)$, and the center of $\SO(3)$ is trivial. Hence every Lie-group automorphism of $\SO(3)$ is conjugation by some element of $\SO(3)$.
\end{proof}

\begin{proposition}
Up to conjugation equivalence, there are exactly three inequivalent continuous homomorphisms
\[
\tilde{\phi}:\SO(3)\times Z_2^{T}\longrightarrow \SO(3).
\]
Denoting $SO(3)\times Z_2^{T} =\big\{ R_{\bm{n}}(\theta) {T}^{m}; |\bm n|=1, 0\leq\theta<\pi, {m}\in\{0,1\} \big\}$, then the three homomorphisms are represented by
\[
\tilde{\phi}^{(a)}_{R_{\bm{n}}(\theta){T}^{m}}=\E,
\qquad
\tilde{\phi}^{(b)}_{R_{\bm{n}}(\theta) {T}^{m}}=[R_z(\pi)]^{{m}},
\qquad
\tilde{\phi}^{(c)}_{R_{\bm{n}}(\theta) {T}^{m}}=R_{\bm{n}}(\theta).
\]
\end{proposition}

\begin{proof}
Let $\tilde{\phi}$ be such a homomorphism and set
\[
\rho(R_{\bm{n}}(\theta))=\tilde{\phi}_{R_{\bm{n}}(\theta)},
\qquad
t=\tilde \phi_{T}.
\]
Because ${T}$ commutes with $SO(3)$, $t$ commutes with $\rho(\SO(3))$; moreover $t^2=\E$.

If $\rho$ is trivial, then $\tilde{\phi}$ is determined by the involution $t$. Up to conjugacy, there are only two possibilities for an involution: $\E$ and a $\pi$ rotation denote as $R_z(\pi)$. This gives $\tilde{\phi}^{(a)}$ and $\tilde{\phi}^{(b)}$.

If $\rho$ is nontrivial, then by the previous lemmas it is an inner automorphism. Up to conjugacy, we may assume $\rho(R_{\bm{n}}(\theta))=R_{\bm{n}}(\theta)$. Since $t$ commutes with all of $\SO(3)$ and the center of $\SO(3)$ is trivial, we must have $t=\E$. Hence
\[
\tilde{\phi}^{(c)}_{R_{\bm{n}}(\theta) {T}^{m}}=R_{\bm{n}}(\theta).
\]
These three classes are inequivalent because they have different restrictions to the identity component and, when the restriction is trivial, different images of ${T}$.
\end{proof}

We now consider $O(3)\times Z_2^{{T}}=SO(3)\times\{E,P\}\times Z_2^{{T}}$. For every $g\in\OO(3)$, 
denote $\det g=(-1)^{\eta(g)}$ with $\eta(g)=\frac{1-\det g}2\in\{0,1\}$. 
Then,
\begin{proposition}\label{Prop3}
Up to conjugation equivalence, there are exactly six inequivalent continuous homomorphisms
\[
\tilde{\phi}:\OO(3)\times Z_2^{{T}}\longrightarrow \SO(3),
\]
represented by
\begin{align*}
\tilde{\phi}^{(1)}_{g {T}^{m}} = \E,\qquad
\tilde{\phi}^{(2)}_{g {T}^{m}}=R_{z}(\pi)^{{m}},\qquad
\tilde{\phi}^{(3)}_{g {T}^{m}}=R_z(\pi)^{\eta(g)},\\
\tilde{\phi}^{(4)}_{g {T}^{m}}=R_z(\pi)^{\eta(g)+{m}},\qquad
\tilde{\phi}^{(5)}_{g {T}^{m}}=R_z(\pi)^{\eta(g)}R_x(\pi)^{{m}},\qquad
\tilde{\phi}^{(6)}_{g {T}^{m}}=(\det g)g.
\end{align*}
All exponents are understood modulo $2$.
\end{proposition}

\begin{proof}
Let $\tilde{\phi}$ be such a homomorphism. Restrict it to the connected subgroup $\SO(3)$, and denote the restriction by $\rho$. 
\bit
\item[I], If $\rho$ is nontrivial, then by the previous lemmas it is an inner automorphism. Then up to conjugacy, we can let
\[
\rho(R_{\bm{n}}(\theta))=R_{\bm{n}}(\theta).
\]
Now both $P$ and ${T}$ commute with the $SO(3)$. Hence $\tilde{\phi}_{P}$ and $\tilde{\phi}_{T}$ centralize all of $\rho(\SO(3))=\SO(3)$. Since the center of $\SO(3)$ is trivial, both $\tilde{\phi}_{P}$ and $\tilde{\phi}_{T}$ must be trivial. This yields the map $\tilde{\phi}^{(6)}$: 
\[
\tilde{\phi}^{(6)}_{g {T}^{m}}=(\det g)g.
\]
\item[II], If $\rho$ is trivial, namely $\rho=E$, then $\tilde{\phi}$ is determined by a commuting pair of involutions
\[
p=\tilde{\phi}_{P},\ \qquad t=\tilde{\phi}_{T}.
\]
\bit
\item[(1)] If $p=\E$, up to conjugacy one has $t^{(1)}=\E$ or $t^{(2)}=R_{z}(\pi)$, which respectively gives rise to $\tilde{\phi}^{(1)}$ and $\tilde{\phi}^{(2)}$;
\item[(2)] If $p\neq \E$, one can fix $p=R_{z}(\pi)$ up to conjugation equivalence. Then $t$ must lie in the centralizer $C_{\SO(3)}(R_{z}(\pi))$. Hence, there are exactly three possibilities:
\[
t^{(3)}=\E,\qquad t^{(4)}=R_{z}(\pi),\qquad t^{(5)}=R_{x}(\pi),
\]
which yield $\tilde\phi^{(3)}$, $\tilde\phi^{(4)}$, and $\tilde\phi^{(5)}$ respectively.
\eit
\eit
\end{proof}

\begin{theorem}
Let $\Phi=\{\tilde{\phi}^{(1)},\tilde{\phi}^{(2)},\tilde{\phi}^{(3)},\tilde{\phi}^{(4)},\tilde{\phi}^{(5)},\tilde{\phi}^{(6)}\}$ defined in Prop.~\ref{Prop3}. Up to the conjugation equivalence, the continuous homomorphisms
\[
O(3)\times Z_2^{T}\longrightarrow SO(3)_s\times SO(3)_c
\]
fall into exactly $6^2=36 $ inequivalent classes, represented by the ordered pairs $\big(\tilde\phi^{(\alpha)s}, \tilde\phi^{(\beta)c}\big)$ with $\alpha,\beta=1,2,...,6$. 
\end{theorem}

In the map $\tilde\phi^{(6)s}$, the spin rotation group is isomorphic to the spatial rotation group. The map $\tilde\phi^{(6)s}$ shows a nontrivial coupling between the orbital angular momentum and the spin angular momentum of Cooper pairs. This indicates that each Cooper pair in the corresponding superfluid has total spin $S=1$ (triplet-paring) and angular momentum $L=1$ ({\it i.e.} $p$-wave pairing symmetry).
For $\tilde\phi^{(1)s}\sim \tilde\phi^{(5)s}$, the spatial rotation is a symmetry by itself, indicating that the Cooper pairs have zero angular momentum $L=0$ ({\it i.e.} $s$-wave singlet pairing symmetry). \\

\underline{\it Invariants for projective twist:} Since $O_f$ has a nontrivial center $\mathcal Z(O_f)=Z_2^f$ (the action $\varphi$ is always trivial), the projective twist is classified by 
\Beq
H^2(O(3)\times Z_2^T, Z_2^f) =H^2(SO(3)\times Z_2\times Z_2, Z_2^f) = \mathbb Z_2^4
\Eeq
with invariants 
$$\eta_{x,y} = {\omega_2(C_{2x},C_{2y})\over \omega_2(C_{2y},C_{2x})},\ \eta_{P,T}={\omega_2(P,T)\over \omega_2(T,P)},\ \eta_P = \omega_2(P,P),\ \eta_T =\omega_2(T,T).$$\\

\underline{\it Relation to $^3$He superfluid:} Since the superfluid $^3$He B phase matches the `external' and `internal' symmetries $G_b=O(3)\times Z_2^{T}$, $ O_f=Z_2^f$, it falls in the class with coupling $\big(\tilde\phi^{(6)s}, \tilde\phi^{(4)c}\big)$ (since the Cooper pairs have $S=1$, $L=1$) and projective twist $\eta_{x,y}=-1, \eta_{P,T}=1, \eta_P=-1,\eta_T=-1$. The rest classes may correspond to new superfluids whose physical realization needs further study. 

\subsubsection{$G_b=SO(2)\times Z_2^P$ and $O_f  = U(1)_s\rtimes Z_2$ or $U(1)_{sc}\times Z_2$}

In this subsection, we consider $G_b=SO(2)\times Z_2^P$ and two different kinds of $O_f$, namely $O_f= \langle U(1)_s, e^ {-i( {\sigma_x\over2} + {\tau_z\over2}) \pi} \rangle =U(1)_s\rtimes Z_2$ or $O_f=\langle Z_2^f, e^{-i({\sigma_z\over2} + {\tau_z\over2})\theta}, \theta\in [0,2\pi) \rangle = U(1)_{sc}\times Z_2$. \\

\underline{\it Global action:} For the above two cases, the actions of $G_b$ on $O_f$ in spin-charge groups are characterized by the continuous homomorphisms $\tilde{\phi}:SO(2)\times Z_2^P\to O(2)$ (see Tab.~\ref{tab:Of_Infor} for information of $\mathscr N(O_f)/O_f$). 
The following proposition classifies the continuous homomorphisms $\tilde{\phi}:SO(2)\to O(2)$. 
\begin{proposition}
Every continuous homomorphism\ \( \tilde\phi:\SO(2)\to O(2)\) is conjugate equivalent to a unique representative of the form
\[
\tilde{\phi}^{(n)}_{R({\theta})} = R({n\theta}),\qquad n\in \mathbb{Z} {\rm\ with\ }{n\ge 0}.
\]
Hence the inequivalent classes are indexed by the nonnegative integer $n$.
\end{proposition}

\begin{proof}
Because $SO(2)$ is connected, its image under any continuous homomorphism is connected, hence lies in the identity component. Therefore
\[
\tilde\phi_{R({\theta})} = R({f(\theta)})
\]
for a continuous homomorphism $f:\mathbb R/2\pi\mathbb Z\to \mathbb R/2\pi\mathbb Z$. Such homomorphisms are exactly
\[
f^{(n)}(\theta)=n\theta,
\qquad n\in \mathbb{Z}.
\]
So every homomorphism is of the form $R({\theta})\mapsto R({n\theta})$. Conjugation by the reflection sends $n\mapsto -n$. Thus $n$ and $-n$ are equivalent. Distinct values of $|n|$ are inequivalent because they have different kernels (equivalently, different degrees on the circle). Therefore the classes are labeled by $n\in\mathbb{Z}$ with $n\ge0$.
\end{proof}

For each continuous homomorphism $\tilde{\phi}:SO(2)\times Z_2^P\to O(2)$, the restriction of $\tilde\phi$ to the connected component $SO(2)\subset O(2)$ has the form $\tilde\phi_{R({\theta})}=R({n\theta})) $ for a unique integer $n\ge 0$, up to conjugacy. Moreover, the relation
\begin{align}
\tilde\phi_{PR({\theta})P^{-1}} = \tilde{\phi}_{R({\theta})} = R({n\theta})
=\tilde{\phi}_{P} \tilde{\phi}_{ R({\theta})} \tilde{\phi}_{P}^{-1}=\tilde{\phi}_{P}R({n\theta})\tilde{\phi}_{P}^{-1}. 
\end{align}
implies that if $n> 0$ then $\tilde{\phi}_{P}\in SO(2)$ and if $n=0$ then $\tilde{\phi}_{P}\in O(2)$.  

\begin{theorem}
Up to conjugation equivalence, there are infinite inequivalent continuous homomorphisms
\[
\tilde{\phi}:SO(2)\times Z_2^P\longrightarrow O(2),
\]
represented by
\begin{align*}
&{\mbox{\rm type-I}}: \tilde{\phi}_{SO(2)}^{(0)_{a,b,c}}=E, {\rm\ and\ }\ \tilde{\phi}^{(0)_a}_{P}= E,\ \  \tilde{\phi}^{(0)_b}_{P}=R_{z}(\pi),\ \  \tilde{\phi}^{(0)_c}_{P}=R_{x}(\pi);\\
&{\mbox{\rm type-II}}: \tilde{\phi}^{(n)_{a,b}}_{R(\theta)} = R(n\theta),\ \ {\rm with\ }n\in\mathbb{Z}, n>0 {\ \rm and\ } \tilde{\phi}^{(n)_a}_{P}=E,\ \  \tilde{\phi}^{(n)_b}_{P} =R_{z}(\pi).\
\end{align*}
\end{theorem}
If the Cooper pair is parity-even under inversion (such as $s$-wave), then $\tilde \phi_{P}=R_z(\pi)$ (or $\tilde \phi_{P}=R_x(\pi)$ if $n=0$); otherwise if the Cooper pair has odd inversion parity (such as $p$-wave), then $\tilde \phi_{P}=E$.

The integer $n$ in type-II homomorphism is a {\it winding number} from spatial rotation $SO(2)$ to the target group $O(2)$. The winding number actually corresponds to the {\it angular momentum} of the Cooper pair whose gap function is proportional to $(p_x+ip_y)^n$. For instance, $n=1, 3$ stand for $p$- and $f$-wave pairing symmetry, respectively. 
For the type-I homomorphism, $n=0$, the gap function is proportional to $p_z^m$ or $\big(\sqrt{p_x^2+p_y^2}\big)^m$ with $m$ an integer. \\

\underline{\it Invariants for projective twist:} Now we consider the projective twist for two different $O_f$. 

(I) $O_f  = U(1)_s + e^{-i(\sigma_x+\tau_z){\pi\over2}}U(1)_s = U(1)_s\rtimes Z_2.$ In this case, $\mathcal Z(O_f)=Z_2^f$. Denoting $U(1)_s=\{R_\theta; \theta\in[0,4\pi)\}$, then the classification of projective twist reads
\Beq
H^2(SO(2)\times Z_2^P, Z_2^f) = \mathbb Z_2^2
\Eeq
with two invariants 
$$\eta_P^{Z_2}=\omega_2(P,P),\ \eta_\pi^{Z_2} =\omega_2(R_\pi, R_\pi).$$

(II) $O_f = U(1)_{sc}\times Z_2^f$. Since $O_f$ is abelian, the center is itself. Here we denote $U(1)_{sc}=\{R_\theta; \theta\in[0,2\pi)\}$. 

When the action is trivial (as $\tilde\phi$ belonging to type-II), then
\Beq
H^2(SO(2)\times Z_2^P, U(1)\times Z_2^f) = H^2(SO(2)\times Z_2^P, U(1)) \times H^2(SO(2)\times Z_2^P, Z_2^f) = H^2(SO(2)\times Z_2^P, Z_2^f) = \mathbb Z_2^2
\Eeq
with  two invariant 
$\eta_P^{Z_2}=\omega_2(P,P),\ \eta_\pi^{Z_2} =\omega_2(R_\pi, R_\pi).$

When the action is nontrivial (as $\tilde\phi$ belonging to type-I) with $\varphi^{(0)_c}(P)\cdot  e^{-i(\sigma_z/2+\tau_z/2)\theta} = e^{i(\sigma_z/2+\tau_z/2)\theta}$, then
\Beq
H^2_\varphi(SO(2)\times Z_2^P, U(1)\times Z_2^f) =  H^2_\varphi(SO(2)\times Z_2^P, U(1)) \times H^2(SO(2)\times Z_2^P, Z_2^f) =  \mathbb Z_2^4
\Eeq
with four invariant 
$$\eta_{T, R_\theta}^{U(1)} = {\omega_2^{U(1)}(T,R_\theta) \over\omega_2^{U(1)} (R_\theta, T)} \mbox{\ mod\ } R_{2\theta},\ \eta_P^{U(1)}=\omega_2^{U(1)} (P,P) \mbox{(Kramers type)}, \ \eta_P^{Z_2}=\omega_2^{Z_2}(P,P) , \ \eta_\pi^{Z_2} =\omega_2^{Z_2}(R_\pi, R_\pi).$$\\

\underline{\it Relation to $^3$He superfluid:} The $^3$He $A$ phase has internal group $O_f=U(1)_s\rtimes Z_2$ and the $A_1$ phase has $O_f=U(1)_{sc}\times Z_2^f$. The couplings $\tilde\phi$ for both phases are of type-II $\tilde\phi^{(1)_b}$ (of $(p_x+ip_y)$-wave), 
and the invariants of the projective twist read $\eta_{P}^{Z_2}=-1, \eta_{R_\pi}^{Z_2} =-1$. The rest classes predict the existence of more superfluid phases in nature, such as the $f$-wave triplet-pairing superfluids and so on. 


\subsection{Symmetry analysis of the AFM-flux model}

\begin{figure}[t]
  \centering
\includegraphics[scale=0.4]{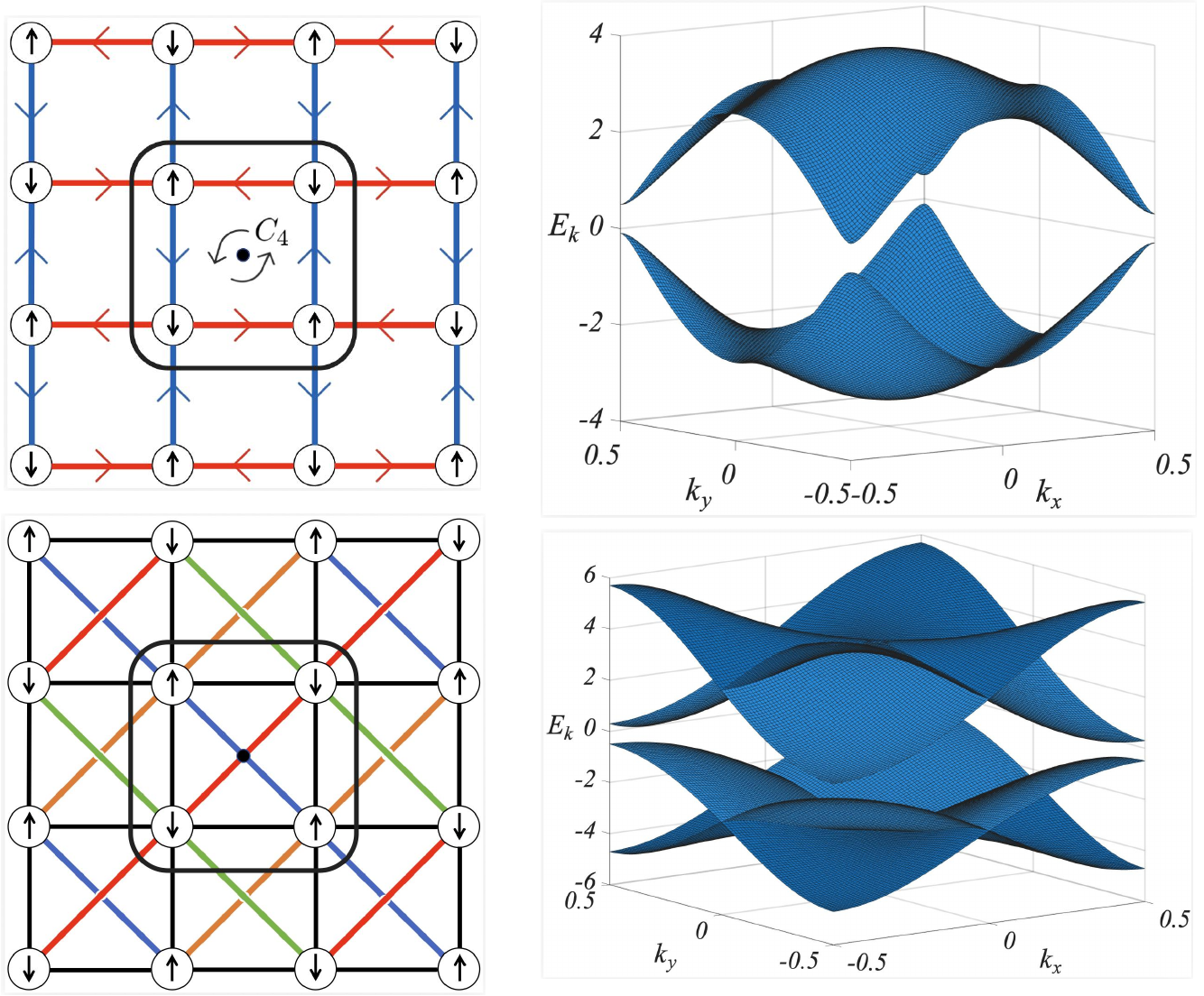}
\caption{Tight-binding model with AFM-flux. The black square indicates the enlarged unit cell. On the upper-left panel, the hopping matrix reads $\Lambda_{ij}=a+\mi b\sigma_z$ for the horizental arrows $i\to j$ and $\Lambda_{kl}=b+\mi a\sigma_z$ for vertical arrows $k\to l$, with $a,b$ real parameters. On the lower-left panel, the hopping matrix reads $\Lambda_{ij}=\alpha I+\beta\sigma_z$ for red bond, $\Lambda_{ij}=-\alpha I+\beta\sigma_z$ for blue bond, $\Lambda_{ij}=-\alpha I-\beta\sigma_z$ for green bond, and $\Lambda_{ij}=\alpha I-\beta\sigma_z$ for orange bond, with $\alpha,\beta$ real parameters. The upper-right panel shows the band structure of model~(\ref{SMmodel}) with $a=1$, $b=0.8$, $\lambda=0.3$, $J=0.2$ and $\alpha=\beta=0$. The lower-right panel shows the band structure including diagonal terms with $\alpha=0.6$ and $\beta=0.7$.}
\label{square-lattice}
\end{figure}

Here, we complete the information about the AFM-flux model discussed in the main text. The general Hamiltonian is given by
\begin{align}\label{SMmodel}
H_{\text{AFM-flux}} = \sum_{\langle ij\rangle} (C _i^\dagger \Lambda_{ij} C _j + \text{h.c.})  +  \sum_i \big[\lambda \hat N_i +  (-1)^i M \hat S^z_i \big] + \sum_{\langle\langle ij\rangle\rangle} (C _i^\dagger \Lambda_{ij} C _j + \text{h.c.}),
\end{align}
where \( \langle ij\rangle \) denotes nearest-neighbor sites and \( \langle\langle ij\rangle\rangle \) denotes next-nearest-neighbor sites. The corresponding hopping matrices and band structures are illustrated in Fig.~\ref{square-lattice}.

The four-fold degeneracy over the entire Brillouin zone, as shown in the upper-right panel of Fig.~\ref{square-lattice}, is protected by a four-dimensional irreducible representation. First, for a general momentum point \( (k_1,k_2) \), the corresponding band representation is generated by the following matrices:
\begin{small}
\begin{align}
U(1)=I_4\otimes e^{-i\theta\sigma_z} 
,\quad  &g_{T_x} =
\begin{pmatrix}
0_2 & e^{ik_1} i\sigma_x & 0_2 & 0_2 \\
i\sigma_x & 0_2 & 0_2 & 0_2 \\
0_2 & 0_2 & 0_2 & i\sigma_x \\
0_2 & 0_2 & e^{ik_1} i\sigma_x & 0_2
\end{pmatrix} \notag\\
  g_{T_y} =
\begin{pmatrix}
0_2 & 0_2 & 0_2 & e^{ik_2} i\sigma_x \\
0_2 & 0_2 & e^{ik_2} i\sigma_x & 0_2 \\
0_2 & i\sigma_x & 0_2 & 0_2 \\
i\sigma_x & 0_2 & 0_2 & 0_2
\end{pmatrix} 
,\quad  &g_T g_{C_2} =
\begin{pmatrix}
0_2 & 0_2 & -I_2 & 0_2 \\
0_2 & 0_2 & 0_2 & -e^{-2i\theta_0\sigma_z} \\
I_2 & 0_2 & 0_2 & 0_2 \\
0_2 & e^{-2i\theta_0\sigma_z} & 0_2 & 0_2
\end{pmatrix}.
\end{align}
\end{small}
Introducing unitary transformation 
{\small $U=\Bmat 
0 & 0& 0 &0 &1 &0 & 0 &0 \\
1 & 0& 0 &0 &0 &0 & 0 &0 \\
0 & 1& 0 &0 &0 &0 & 0 &0 \\
0 & 0& 0 &0 &0 &1 & 0 &0 \\
0 & 0& 0 &0 &0 &0 & 1 &0 \\
0 & 0& 1 &0 &0 &0 & 0 &0 \\
0 & 0& 0 &1 &0 &0 & 0 &0 \\
0 & 0& 0 &0 &0 &0 & 0 &1 \\
\Emat,
$}
then the above representation can be transformed into a direct sum of two four-dimensional irreducible representations: 
\begin{small}
\begin{align}
U^\dag U(1) U = I_2\otimes e^{i\theta\sigma_z} \oplus I_2 \otimes e^{-i\theta\sigma_z}
,\quad  &U^\dag g_{T_x} U=
i\begin{pmatrix}
0& e^{ik_1} & 0 & 0 \\
1 & 0 & 0 & 0 \\
0 & 0 & 0 & 1 \\
0 & 0 & e^{ik_1} & 0
\end{pmatrix}\oplus i\begin{pmatrix}
0& e^{ik_1} & 0 & 0 \\
1 & 0 & 0 & 0 \\
0 & 0 & 0 & 1 \\
0 & 0 & e^{ik_1} & 0
\end{pmatrix} \notag\\
U^\dag  g_{T_y}U =
i\begin{pmatrix}
0 & 0 & 0 & e^{ik_2} \\
0 & 0 & e^{ik_2} & 0 \\
0 & 1 & 0 & 0 \\
1 & 0 & 0 & 0
\end{pmatrix} \oplus i\begin{pmatrix}
0 & 0 & 0 & e^{ik_2} \\
0 & 0 & e^{ik_2} & 0 \\
0 & 1 & 0 & 0 \\
1 & 0 & 0 & 0
\end{pmatrix} 
,\quad  &U^\dag g_T g_{C_2} U^*=
\begin{pmatrix}
0 & 0 & -1 & 0 \\
0 & 0 & 0 & -e^{-2i\theta_0} \\
1 & 0 & 0 & 0 \\
0 & e^{-2i\theta_0} & 0 & 0
\end{pmatrix} \oplus \begin{pmatrix}
0 & 0 & -1 & 0 \\
0 & 0 & 0 & e^{-2i\theta_0} \\
1 & 0 & 0 & 0 \\
0 & -e^{-2i\theta_0} & 0 & 0
\end{pmatrix}.
\end{align}
\end{small}

Since the group contains the continuous subgroup \( U(1) \), the most convenient way to prove this fact is the so-called Hamiltonian approach\cite{yang2021hamiltonian,song2025constructions}. The Hamiltonian approach is a numerical method for decomposing linear or projective representations into irreducible ones: one constructs a random complex Hermitian matrix that is invariant under the conjugation actions of the representation matrices, and each distinct eigenvalue of this matrix labels an irreducible representation, with its degeneracy equal to the dimension of the corresponding irreducible representation. For our case, we first select the discrete subgroup \( \mathbb{Z}_8 \subset U(1) \), generated by \( z_0 = I_4 \otimes e^{-i\pi\sigma_z/4} \). Then, we construct the desired matrix according to Algorithm~\ref{algorithm1}. As a result, the distribution of the eigenvalues of the random Hermitian matrix \( H \) is \( 4+4 \). The corresponding irreducible representation is the following:

When adding next-nearest-neighbor hopping terms, the $g_T$ symmetry is broken (so $g_{C_2}g_{T}$ is) and the $g_{T}g_{M_{x,y}}$ symmetry is preserved. From the symmetry point of view, the band rep for $U(1)$, $g_{T_x}$ and $g_{T_y}$ at an arbitrary momentum point contains four two-dimensional irreducible representations, which can also be proven by the Hamiltonian approach. However, along the high-symmetry lines $k_x=0$, there is an extra symmetry
\[
g_{T} g_{M_{y}}=
\begin{pmatrix}
0_2 & 0_2 & 0_2 & -i\sigma_x \\
0_2 & 0_2 & -i\sigma_xe^{-2i\theta_0\sigma_z} & 0_2 \\
0_2 & i\sigma_x & 0_2 & 0_2 \\
i\sigma_xe^{-2i\theta_0\sigma_z} & 0_2 & 0_2 & 0_2
\end{pmatrix}.
\] 
Under the unitary transformation 
{\small $U=\Bmat 
{1\over\sqrt2}& 0& 0 &0 &{1\over\sqrt2} &0 & 0 &0 \\
{1\over\sqrt2} & 0& 0 &0 &-{1\over\sqrt2} &0 & 0 &0 \\
0 & {1\over\sqrt2}& 0 &0 &0 &{1\over\sqrt2} & 0 &0 \\
0 & {1\over\sqrt2}& 0 &0 &0 &-{1\over\sqrt2} & 0 &0 \\
0 & 0& {1\over\sqrt2} &0 &0 &0 & {1\over\sqrt2} &0 \\
0 & 0& {1\over\sqrt2} &0 &0 &0 & -{1\over\sqrt2} &0 \\
0 & 0& 0 &{1\over\sqrt2} &0 &0 & 0 &{1\over\sqrt2} \\
0 & 0& 0 &{1\over\sqrt2} &0 &0 & 0 &-{1\over\sqrt2} \\
\Emat,
$} 
the 8-dimensional representation of $U(1), g_{T_x}, g_{T_y}$ and $g_{T}g_{M_y}$ can be reduced into a direct sum of two 4-dimensional
irreducible representations. 
The band degeneracy shown in the lower-right panel coincides with the above symmetry analysis.

\begin{center}
\refstepcounter{algorithm}
\begin{minipage}{0.96\columnwidth}
\hrule
\vspace{0.5em}
\textbf{Algorithm \thealgorithm.} The Hamiltonian approach\cite{yang2021hamiltonian} to obtain irreducible representations
\label{algorithm1}
\vspace{0.3em}
\hrule
\vspace{0.6em}
\end{minipage}
\end{center}

\subsection{Details of spin supercurrent calculations}

Consider a spinful pairing matrix
\begin{align}
\Delta_{mn}(\bm r)=\langle c_m(\bm r)c_n(\bm r)\rangle,
\end{align}
where \(m,n\) label the spin \(S_z\) and orbital quantum numbers. The gradient free energy is written as
\begin{align}
F_{\mathrm{grad}}=\int d^2 r \,
K_{ij}\,
\operatorname{Tr}\!\left[(D_i\Delta)^\dagger (D_j\Delta)\right],
\label{eq:Fgrad}
\end{align}
where \(K_{ij}\) is the stiffness tensor, and in general, the covariant derivative is
\begin{align}
D_i\Delta=\partial_i\Delta- \mi A_i^{\tau_z}\Delta- \mi a_i^z S_z\Delta- \mi \omega_i L_z\Delta ,
\label{eq:covD}
\end{align}
with \(A_i^{\tau_z}\) the charge source, \(a_i^z\) the spin source, and \(\omega_i\) the orbital-rotation source. Here the spin generator acts on the pairing matrix as
\begin{align}
S_z\Delta = \frac{1}{2}\left(\sigma_z\Delta+\Delta\sigma_z^T\right).
\label{eq:SzDelta}
\end{align}
Varying Eq.~\eqref{eq:Fgrad} with respect to the spin source \(a_i^z\) gives the spin supercurrent
\begin{align}
J_i^{z,\mathrm{SC}}= - \frac{\delta F_{\mathrm{grad}}}{\delta a_i^z}= 2K_{ij}\,\operatorname{Im} \operatorname{Tr}\!\left[(S_z\Delta)^\dagger D_j\Delta\right].
\label{eq:Jsc_def}
\end{align}
Notice that Eq.~\eqref{eq:Jsc_def} is bilinear in \(\Delta^\dagger D\Delta\). It is therefore a condensate current associated with the Cooper pair field. 

We first examine the symmetry behavior of the current $J_i^{z}$ defined in Eq.~(\ref{eq:Jsc_def}). Consider a general unitary symmetry operation \(g\).  Let $l_g$ denote its action on spatial degrees of freedom. Then the pairing order parameter transforms as
\begin{align}
\Delta(\bm r)\to \Delta'(\bm r')=  U_s(g)\,U_L(g)\, \Delta(\bm r)\, U_s^T(g),
\label{eq:Delta_transform}
\end{align}
where \(U_s(g)\) denotes the spin representation of $g$ and \(U_L(g)\) denotes the combination of the orbital and pairing representations of \(g\). The covariant derivative transforms in the corresponding covariant manner,
\begin{align}
D_i\Delta\to D_i'\Delta'(\bm r') =  U_s(g)\,U_L(g)\,\bigl[(l_g)_{ij}D_j\Delta(\bm r)\bigr] \,U_s^T(g).
\label{eq:DiDelta_transform}
\end{align}
Since the system has spin $U(1)_s$ symmetry, the spin sector of the symmetry $g$ must satisfy
\begin{align}
U_s^\dagger(g)\,\sigma_z\,U_s(g)=\eta_J(g)\,\sigma_z,\qquad\eta_J(g)=\pm1.
\label{eq:eta_def}
\end{align}
Then Eq.~\eqref{eq:SzDelta} implies
\begin{align}
S_z\Delta'=\eta_J(g)\,U_s(g)\,U_L(g)\,\bigl[S_z\Delta\bigr]\,U_s^T(g).
\label{eq:SzDelta_transform}
\end{align}
Substituting Eqs.~\eqref{eq:DiDelta_transform} and \eqref{eq:SzDelta_transform} into Eq.~\eqref{eq:Jsc_def}, we obtain
\begin{align}
J_i^{z,\mathrm{SC}}(\bm r')\to {J'}_i^{z,\mathrm{SC}}(\bm r')&=2K_{ij}\,\operatorname{Im} \operatorname{Tr}\!\left[(S_z\Delta')^\dagger D_j'\Delta'\right]\nonumber\\
&=2K_{ij}\,\operatorname{Im} \operatorname{Tr}\!\left[\eta_J(g)\,\bigl(U_sU_L(S_z\Delta)U_s^T\bigr)^\dagger\bigl(U_sU_L[(l_{g})_{j\ell} D_\ell\Delta]U_s^T\bigr)\right].
\end{align}
Using the unitarity of \(U_s\) and \(U_L\), together with cyclicity of the trace, all internal rotations cancel. One therefore finds
\begin{align}
{J'}_i^{z,\mathrm{SC}}(\bm r')=2\,\eta_J(g)\,K_{ij}\,(l_{g})_{j\ell}\,\operatorname{Im} \operatorname{Tr}\!\left[(S_z\Delta)^\dagger D_\ell\Delta\right].
\label{eq:J_intermediate}
\end{align}

Because \(g\) is a symmetry of the gradient functional in Eq.~\eqref{eq:Fgrad}, the tensor \(K_{ij}\) must satisfy $K_{ij}=(l_{g})_{ik}(l_{g})_{j\ell}K_{k\ell}$, or equivalently
\begin{align}
K l_g = l_g K.
\label{eq:K_commute}
\end{align}
Using Eq.~\eqref{eq:K_commute}, Eq.~\eqref{eq:J_intermediate} may be rewritten as
\begin{align}
{J'}_i^{z,\mathrm{SC}}(\bm r')=\eta_J(g)\,(l_{g})_{ij}\,J_j^{z,\mathrm{SC}}(\bm r).
\label{eq:J_final_component}
\end{align}
In vector form,
\begin{align}
\bm J^{z,\mathrm{SC}}
\to
\eta_J(g)\,l_g\,\bm J^{z,\mathrm{SC}}.
\label{eq:J_final_vector}
\end{align}
Equation~\eqref{eq:J_final_vector} implies that the Cooper pair spin supercurrent transforms as an ordinary spatial vector, multiplied by the sign \(\eta_J(g)\) associated with the action of the symmetry on \(S_z\).

We now consider the following linear response 
\begin{align}\label{eq:lrspds}
J_i^{z}=\sum_{j}\rho_{ij} A_j^{\tau_z},
\end{align}
where $A_j^{\tau_z}$ is the charge source with $\tau_z$ sector. Hence, we have $e^{-\mi\tau_x\pi/2}\ A_j^{\tau_z}\ e^{\mi\tau_x\pi/2}=-A_j^{\tau_z}$. Thus, similar to Eq.~(\ref{eq:J_final_vector}), the action of an arbitrary symmetry $g$ on the vector $\bm{A}^{\tau_z}$ can be denoted by
\begin{align}
\bm{A}^{\tau_z}\to \eta_{A}(g)l_g \bm{A}^{\tau_z}. 
\label{eq:A_final}
\end{align}

According to (\ref{eq:J_final_vector}) and (\ref{eq:A_final}),  the tensor $\bm{\rho}$ defined in Eq.~(\ref{eq:lrspds}) must satisfy
\begin{empheq}[box=\fbox]{align}
\bm{\rho}={\eta_{J}(g)\over \eta_{A}(g)}\, l_g\cdot \bm{\rho}\cdot l_g^{-1},\quad \forall g.
\end{empheq}

\paragraph*{For Group $G_1$: collinear magnet without spin flux.} Since $g_{C_4}=(e^{-\mi(\sigma_x+\tau_x){\pi\over 2}}\mid C_4)$ and $g_{M_x}=(e^{-\mi\sigma_x{\pi\over 2}}\mid M_x)$ are global operations, the tensor $\bm{\rho}$ is uniform, and satisfies 
\begin{align}
\bm{\rho}=+ l_{g_{C_4}}\bm{\rho} l_{g_{C_4}}^{-1}\qquad \bm{\rho}=- l_{g_{M_x}}\bm{\rho} l_{g_{M_x}}^{-1}\notag
\end{align}
The only possible solution for uniform $\rho$ is 
\begin{equation}
\rho=\beta\begin{pmatrix}0&1\\-1&0\end{pmatrix},\quad \beta\in\mathbb{C}.
\label{eq:g5final2}
\end{equation}
This is a Hall-like spin supercurrent response.

\paragraph*{For Group $G_2$: collinear magnet with spin flux.}
Since $g_{C_4}=(\gamma_{C_4}(r)e^{-\mi(\sigma_x+\tau_x){\pi\over 2}}\mid C_4)$ is not a global symmetry but $g_{C_4}^2$ is, the tensor $\bm{\rho}$ is \emph{non-uniform} and can be devided into two sectors related by $g_{C_4}$. In other words,  $\bm{\rho}$ is sub-lattice dependent. Since $\gamma_{C_4}(r)$ commutes with $\sigma_z$, the effective symmetry constrains come from $e^{-\mi\sigma_x{\pi\over 2}}\sigma_z(e^{-\mi\sigma_x{\pi\over 2}})^{-1}=-\sigma_z$ and $e^{-\mi\tau_x{\pi\over 2}}\tau_z(e^{-\mi\tau_x{\pi\over 2}})^{-1}=-\tau_z$.

Writing $\bm{\rho}_A=\bigl(\begin{smallmatrix}a&b\\c&d\end{smallmatrix}\bigr)$, the action of $g_{M_x}$ and $g_{C_4}$ imply
\begin{equation}
\bm{\rho}_B=-M_x\rho_AM_x^T=
\begin{pmatrix}-a&b\\c&-d\end{pmatrix},
\qquad
\bm{\rho}_B=C_4\rho_AC_4^T=
\begin{pmatrix}d&-c\\-b&a\end{pmatrix},
\label{eq:g2eq2}
\end{equation}
which yields $d=-a$ and $c=-b$. Therefore
\begin{equation}
\bm{\rho}_A=\alpha\begin{pmatrix}1&0\\0&-1\end{pmatrix}+\beta\begin{pmatrix}0&1\\-1&0\end{pmatrix},
\quad
\bm{\rho}_B=-\alpha\begin{pmatrix}1&0\\0&-1\end{pmatrix}+\beta\begin{pmatrix}0&1\\-1&0\end{pmatrix},\quad \alpha,\beta\in\mathbb{C}.
\label{eq:g2final2}
\end{equation}
Coarse graining over the magnetic unit cell gives
\begin{equation}
\bar{\bm{\rho}}\equiv \frac{\bm{\rho}_A+\bm{\rho}_B}{2}=\beta\begin{pmatrix}0&1\\-1&0\end{pmatrix},
\qquad
\Delta\bm{\rho}\equiv\frac{\bm{\rho}_A-\bm{\rho}_B}{2}=\alpha\begin{pmatrix}1&0\\0&-1\end{pmatrix}.
\label{eq:g2bulk2}
\end{equation}
Thus $G_2$ has a bulk uniform Hall-like tensor plus a hidden staggered nematic sector. 

\paragraph*{For Group $G_3$: collinear magnet with SPG-I.}
In this case, the symmetry operations are all global: $g_{C_4}=(e^{-\mi\sigma_x{\pi\over 2}}\mid C_4)$,  \( g_{M_x}=(e^{-\mi\sigma_x{\pi\over 2}}\mid M_x) \), and \( g_{T}=K \). The complex conjugation $K$ reverses the sign of condensate phase velocity and $e^{-i\sigma_z\pi/2}$ acts trivially on $\sigma_z$. Hence $g_{T}$ reverses the sign of $\bm{J}^{z,SC}$. Furthermore, the vector $\bm{A}$ is time-reversal odd.  Therefore, we have
\begin{align}
\bm{\rho}=a\begin{pmatrix}0&1\\1&0\end{pmatrix},\quad a\in\mathbb{R}
\end{align}
This is a symmetric off-diagonal spin supercurrent response.

\paragraph*{For Group $G_4$: collinear magnet with SPG-II.}
In this case, the symmetry operations are \( g_{C_4}=(e^{-\mi\sigma_x{\pi\over 2}}\mid C_4) \), \( g_{M_x}=M_x \), and \( g_{T}=K \). We have
\begin{equation}
\bm{\rho}=b \begin{pmatrix}1&0\\0&-1\end{pmatrix},\quad b\in\mathbb{R}.
\label{eq:g8final2}
\end{equation}
This is a purely nematic spin supercurrent response.

\bibliography{SO(4)_SM}